\documentclass[structabstract]{aa}  
\usepackage[toc,page]{appendix}
\usepackage{graphicx}
\usepackage{natbib}
\bibpunct{(}{)}{;}{a}{,}{,}
\usepackage{txfonts}
\begin{document}
   \title{Gas phase Elemental abundances in Molecular cloudS (GEMS)}
   \subtitle{ VII. Sulfur elemental abundance}

   \author{ A.~Fuente\inst{1}  
     \and
     P.~Rivi\`ere-Marichalar\inst{1}
     \and
     L. Beitia-Antero\inst{1,2} 
     \and
     P.~Caselli\inst{3}       
    \and
   V.~Wakelam\inst{4}
   \and
   G.~Esplugues\inst{1}
   \and
   M.~Rodr\'{\i}guez-Baras\inst{1}
   \and
   D.~Navarro-Almaida\inst{5}
    \and
     M.~Gerin\inst{6,7}
    \and
     C.~Kramer\inst{8,9}
    \and         
    R.~ Bachiller \inst{1}
   \and   
   J.~ R.~Goicoechea \inst{10}
    \and
    I.~Jim\'enez-Serra\inst{11}     
     \and
    J.~C.~Loison\inst{12}
     \and
    A.~Ivlev\inst{3}
     \and
     R.~ Mart\'{\i}n-Dom\'enech\inst{13}
      \and
     S.~ Spezzano\inst{3}
     \and
     O.~Roncero\inst{10}
     \and
     G.~Mu\~noz-Caro\inst{11}
      \and      
   S.~Cazaux \inst{14,15}
         \and      
   N.~Marcelino\inst{1}
               }
              
   \institute{ Observatorio Astron\'omico Nacional (OAN), Alfonso XII, 3
             28014, Madrid, Spain
             \and
             Departamento de F\'{\i}sica de la Tierra y Astrof\'{\i}sica, Facultad de CC. Matem\'aticas, Universidad Complutense de Madrid Plaza de las Ciencias 3, E-28040, Madrid, Spain
\and
Centre for Astrochemical Studies, Max-Planck-Institute for Extraterrestrial Physics, Giessenbachstrasse 1, 85748, Garching, Germany    
\and
Laboratoire d'astrophysique de Bordeaux, Univ. Bordeaux, CNRS, B18N, all\'ee Geoffroy Saint-Hilaire, 33615, Pessac, France  
\and
Universit\'e Paris-Saclay, CEA, AIM, D\`epartement d'Astrophysique (DAp), F-91191 Gif-sur-Yvette, France 
\and
Observatoire de Paris, PSL Research University, CNRS, \'Ecole Normale Sup\'erieure, Sorbonne Universit\'e, UPMC Univ. Paris 06, 75005, Paris, France
\and
LERMA, Observatoire de Paris, PSL Research University, CNRS, UMR8112, Place Janssen, 92190, Meudon Cedex, France 
\and
Institut de Radioastronomie Millim\'etrique (IRAM), 300 rue de la Piscine, 38406 Saint Martin d'H\`eres, France
\and
Instituto Radioastronom\'{\i}a Milim\'etrica (IRAM), Av. Divina Pastora 7, Nucleo Central, 18012, Granada, Spain
\and
Instituto de F\'{\i}sica Fundamental (CSIC), Calle Serrano 121-123, 28006, Madrid, Spain
\and
Centro de Astrobiolog\'{\i}a (CSIC-INTA), Ctra. de Ajalvir, km 4, Torrej\'on de Ardoz, 28850, Madrid, Spain
\and
ISM, UMR 5255, CNRS - Univ. Bordeaux, F-33400 Talence, France
\and
 Center for Astrophysics $|$ Harvard \& Smithsonian, 60 Garden St., Cambridge, MA 02138, USA
\and
Faculty of Aerospace Engineering, Delft University of Technology, Delft, The Netherlands  
\and
University of Leiden, P.O. Box 9513, NL, 2300 RA, Leiden, The Netherlands 
}
 
 \abstract 
   { Gas phase Elemental abundances in molecular CloudS (GEMS) is an IRAM 30m large program aimed at determining the elemental abundances of carbon (C), oxygen (O), nitrogen (N), and sulfur (S) in a selected set of prototypical star-forming filaments. In particular, the elemental abundance of S remains uncertain by several orders of magnitude, and its determination is one of the most challenging goals of this program.}
   { This paper aims to constrain the sulfur elemental abundance in Taurus, Perseus, and Orion A based on the GEMS molecular database. The selected regions are prototypes of low-mass, intermediate-mass, and high-mass star-forming regions, respectively, providing useful templates for the study of interstellar chemistry.}
   {We have carried out an extensive chemical modeling of the fractional abundances of CO, HCO$^+$, HCN, HNC, CS, SO, H$_2$S, OCS, and HCS$^+$  to determine the sulfur depletion toward the 244 positions in the GEMS database. These positions sample visual extinctions from A$_V$ $\sim$ 3 mag to $>$50 mag, molecular hydrogen densities ranging from a few $\times$10$^3$~cm$^{-3}$ to 3$\times$10$^6$~cm$^{-3}$, and T$_k$ $\sim$ 10$-$35~K. We investigate the possible relationship between sulfur depletion and the grain charge distribution in different environments.}
  {Most of the positions in Taurus and Perseus are best fitted assuming early-time chemistry, t=0.1 Myr, $\zeta_{H_2}$$\sim$ (0.5$-$1)$\times$10$^{-16}$ s$^{-1}$, and [S/H]$\sim$1.5$\times$10$^{-6}$. On the contrary, most of the positions in Orion are fitted with t=1~Myr and $\zeta_{H_2}$$\sim$ 10$^{-17}$ s$^{-1}$. Moreover,  $\sim$40\% of the positions in Orion are best fitted  assuming the undepleted sulfur abundance, [S/H]$\sim$1.5$\times$10$^{-5}$. We find a tentative trend of  sulfur depletion increasing with density. 
}
 {Our results suggest that sulfur depletion depends on the environment. While the abundances of sulfur-bearing species are consistent with undepleted sulfur in Orion,  a depletion factor of $\sim$20 is required to explain those observed in Taurus and Perseus.  We propose that differences in the grain charge distribution might explain these variations. Grains become negatively charged at a visual extinction of A$_V$$\sim$ 3.5 mag in Taurus and Perseus. At this low visual extinction, the S$^+$ abundance is high, X(S$^+$)$>$10$^{-6}$,  and the electrostatic attraction between S$^+$ and negatively charged grains could contribute to enhance sulfur depletion. In Orion, the net charge of grains remains approximately zero until higher visual extinctions (A$_V$$\sim$ 5.5 mag), where the abundance of S$^+$ is already low because of the higher densities, thus reducing sulfur accretion. The shocks associated with past and ongoing star formation could also contribute to enhance [S/H].
 }
  
   \maketitle


\section{Introduction}
\label{intro}
In recent years, space telescopes such as Spitzer and Herschel have revolutionized 
our view of star-forming regions. Images of giant molecular clouds 
and dark cloud complexes have revealed  spectacular networks of filamentary structures where stars 
are born \citep{Andre2010}.
Interstellar filaments are almost everywhere in the Milky Way. Now we believe that 
filaments precede the onset of star formation, funneling interstellar gas and dust into 
increasingly denser concentrations that will contract and fragment leading to gravitationally bound 
prestellar cores that will eventually  form stars.

Gas chemistry  has a key role in the star formation process 
by determining aspects such as the gas cooling and the gas ionization degree. Molecular filaments can
fragment to prestellar cores to a large extent because molecules cool the gas, thus diminishing the thermal
support relative to self-gravity. The ionization fraction controls the coupling of magnetic fields 
with the gas driving the dissipation of turbulence and angular momentum transfer. Therefore, chemistry
plays a crucial role in the cloud collapse (isolated vs. clustered star formation) and the dynamics 
of accretion disks (see \citealp{Zhao2016, Padovani2013}). 
In the absence of other ionization agents (X-rays, UV photons, and J-type shocks), 
the ionization fraction is proportional to $\sqrt{\zeta _{H_2}}$, where $\zeta _{H_2}$ is the 
cosmic-ray ionization rate for H$_2$ molecules, which becomes a key parameter in the molecular 
cloud evolution \citep{Kee1989, Caselli2002}. In addition to  $\zeta _{H_2}$, 
the gas ionization fraction, X(e$^-$), depends on the elemental 
depletion factors \citep{Caselli1998}. In particular, carbon (C) is the main donor of electrons 
in the cloud surface (A$_v$$<$4 mag). As long as 
it is not heavily depleted, sulfur (S) is the main donor in the range of $\sim$3.7$-$7 magnitudes, which
encompasses a large fraction of the molecular cloud mass. Depletions of C 
and O determine the cooling gas rate since CO, [CII],  and [OI] are main coolants in molecular clouds.

Gas phase Elemental abundances in Molecular CloudS (GEMS) is an IRAM 30m large program aimed at determining the S, C, N, and O depletions, and X(e$^-$) 
in a set of selected prototypical star-forming filaments. The observations corresponding to this program and the resulting molecular database were presented
by \citet{Rodriguez-Baras2021}. In that paper, we explored the relationship between the abundances of the different molecular species and the local 
physical parameters, concluding that density is the main parameter determining the chemical abundances in dark clouds. The determination of the elemental abundances is
a challenging task, especially in the case of sulfur for which the main potential sulfur reservoirs (atomic S, and H$_2$S ice) cannot be easily observed, 
The fine-structure line emission of atomic sulfur at $\sim$25~$\mu$m has been detected in bipolar outflows using the Spitzer space telescope \citep{Anderson2013}. However,
the high excitation conditions of this line (E$_u$=570~K, n$_{\rm crit}$ = 1.5 $\times$10$^6$ cm$^{-3}$) prohibits its detection in dark clouds. 
Alternatively, the S$^+$ abundance can be derived from the emission of the sulfur radio recombination lines \citep{Pankonin1978, Goicoechea2021a}. However, these lines are very weak, thus hindering their use as routine S$^+$ tracers  in molecular clouds. The detection of H$_2$S in the ice is hampered by the strong overlap between the 2558 cm$^{-1}$ band and the methanol bands at 2530 and 2610 cm$^{-1}$, and upper limits have been derived thus far \citep{Smith1991, Jimenez-Escobar2011}.
Therefore, the estimate of sulfur elemental abundance in dark clouds must rely on the observation of minor sulfur-bearing species (e.g., CS, and SO) and the predictive power of chemical models. The
GEMS project relies on a collaborative effort among astronomers, modelers,  theoretical chemists, and experimentalists, working in  a coordinated way to improve the sulfur 
chemical network and, therefore, the reliability of model predictions. Theoretical ab initio calculations have been carried out to
obtain more accurate estimations of  the rates if important reactions associated with the SO and CS chemistry (\citealp{Fuente2019}, \citealp{Bulut2021}, and Bulut et al., in prep). 
An important upgrade of the sulfur surface chemistry was carried out by \citet{Laas2019}.
 \citet{Navarro-Almaida2020} and \citet{Navarro-Almaida2021} used a progressively updated chemical network to investigate the sulfur chemistry in the
 prototypes of cold dense cores, TMC~1 (CP), TMC~1 (C), and Barnard 1b.
 \citet{Goicoechea2021b} also used an updated network to account for the chemistry of sulfur hydrides in the Orion Bar.
Based on the GEMS database, \citet{Spezzano2022} and \citet{Esplugues2022} carried out a detailed study of the CH$_3$OH and H$_2$CS chemistry. These two species are
crucial to constrain the formation of complex organic molecules (COMs) and organo-sulfur compounds in starless cores. 

Elemental abundance, as referred in this paper, is the amount of a given atom in volatiles, that is to say in the gas and in the icy grain mantles. The elemental abundances are
given as initial condition to most chemical models and remain constant during the calculations, since the atoms are not expected to incorporate to the refractory part of grains in molecular clouds. Usually, the adopted initial abundances of C, N, and O correspond to the values derived in the diffuse cloud $\zeta$ Oph by \citet{Jenkins2009},  which result in a C/O abundance ratio of 0.55, giving reasonable prediction for dark clouds \citep{Agundez2013}. However, the elemental abundance of S is uncertain by several orders of magnitude. While the observed gaseous sulfur accounts for its total solar abundance ([S/H]$\sim$1.5$\times$10$^{-5}$, \citealp{Asplund2009})
in diffuse clouds \citep{Neufeld2015} and highly irradiated photon-dominated regions \citep{Goicoechea2021a}, a sulfur depletion 
of a factor of $>$100 is usually needed to account for the observed
abundances of sulfur-bearing species in starless cores \citep{Ruffle1999, Vastel2018} and proto-planetary disks \citep{Dutrey2011, LeGal2019, Riviere-Marichalar2020, LeGal2021}. Sulfur depletions of
a factor of $\sim$10 have been estimated in the translucent part of molecular clouds \citep{Fuente2019}, hot cores \citep{Blake1996, Wang2013, Fuente2021, Bouscasse2022}, low irradiated photon-dominated  regions \citep{Goicoechea2006}, and bipolar outflows \citep{Anderson2013,Feng2020,Taquet2020}. \citet{Vidal2017} were able to reproduce the abundances of the sulfur bearing species detected in TMC~1-CP with undepleted sulfur abundance. Thus far, there is no unified scheme accounting for the observed abundances of sulfur species in different regions of the interstellar medium. The determination of the sulfur depletion is the objective of this paper. The large molecular GEMS database allows us to carry out a uniform and systematic study in regions located in different environments, avoiding the uncertainties coming from the use of different observations and methodology. 

\begin{table*}
\caption{Cores included in the GEMS sample and observation cuts associated with them. N$_{\rm o}$ indicates the total number of points observed in 
the corresponding cut. N$_{\rm v}$ indicates the number of points where the molecular hydrogen density could be derived.}
\label{Table: GEMS sample}
\centering
\begin{tabular}{lllrlllll}\\
\hline\hline
\noalign{\smallskip}
Cloud & Cloud & Core & \multicolumn{2}{c}{Coordinates} & Other names & Cut & N$_{\rm o}$ & N$_{\rm v}$ \\
complex & ID & RA (J2000) & Dec (J2000) & & & & \\
\hline
\noalign{\smallskip}                          
%
Taurus &  TMC~1        &  CP   & 04:41:41.90   &  $+$25:41:27.1    & & C1          &    6& 6 \\
&            &  NH3  &  04:41:21.30   &  $+$25:48:07.0    & &  C2           &   6& 6  \\
&           &  C    &  04:41:38.80   &  $+$25:59:42.0   & &   C3          &  6 & 6  \\
&  B\,213-N$^1$  &  \#1  &  04:17:41.80   &     $+$28:08:47.0     & 5   &  C1    &   9 & 9  \\
&           &  \#2  &   04:17:50.60   &     $+$27:56:01.0     &  $-$ &  C2   &   9 & 7  \\
&            &  \#5  &  04:18:03.80   &     $+$28:23:06.0     & 7  &   C5   &  9 & 9  \\
&            &  \#6  &  04:18:08.40   &     $+$28:05:12.0     &  8  &  C6   &   9 &  5  \\
&            &  \#7  &  04:18:11.50   &     $+$27:35:15.0     & 9  &  C7     &   9 & 8  \\
& B\,213-S$^1$  &  \#10 &  04:19:37.60   &     $+$27:15:31.0     & 13a &  C10    &   9 &  8  \\
&            &  \#12 &  04:19:51.70   &     $+$27:11:33.0     & $-$ &   C12    &  9 & 6  \\
&            &  \#16 &  04:21:21.00   &     $+$27:00:09.0     & $-$ &  C16    &  9  &  9  \\
&           &  \#17 &   04:27:54.00  &    $+$26:17:50.0      & 26b &   C17    &  9 & 5  \\
Perseus &  L1448$^2$   &  \#32 &  03:25:49.00   &     $+$30:42:24.6      &   &    C1   & 8 & 7 \\
& NGC\,1333$^2$   &       &   03:29:18.26   &      $+$31:28:02.0     &  &   C1    &   21 & 18 \\
&            &       &  03:28:41.60   &      $+$31:06:02.0     &   &   C2  &  17 & 13 \\
&            &  \#46 &  03:29:11.00   &   $+$31:18:27.4     &   SK20$^3$  &  C3  & 17 & 11 \\
&            &  \#60 & 03:28:39.40    &      $+$31:18:27.4     &   &      & & \\
&            &  \#51 &  03:29:08.80   &      $+$31:15:18.1     &   SK16    &    C4  & 16  & 13  \\
&            &  \#53 & 03:29:04.50  &      $+$31:20:59.1     &   SK26   &   C5  & 11 &  11  \\
&            &  \#57 &  03:29:18.20     &      $+$31:25:10.8     &    SK33  &   C6  & 9  & 9  \\
&            &  \#64 &  03:29:25.50     &      $+$31:28:18.1     &  &   C7      & 1 & 1  \\
& Barnard 1$^2$   &  1b   &  03:33:20.80 & $+$31:07:34.0  & &  C1   & 18  & 18 \\
&            &       &  03:33:01.90     &    $+$31:04:23.2    &  &    C2      & 8   & 6  \\
&Barnard 5$^2$   &  \#79 &  03:47:38.99    &      $+$32:52:15.0    &   &    C1     &  13 &  9 \\
&IC\,348$^2$     &  \#1  &    03:44:01.00    &   $+$32:01:54.8   &  &  C1  & 14 & 11 \\
&           &  \#10 &  03:44:05.74   &       $+$32:01:53.5      &    &     &   &  \\
Orion &Orion A   & OMC~3        &   05:35:19.54   &      $-$05:00:41.5        & &   C1         & 20 & 10  \\
&            & OMC~4      &  05:35:08.15    &      $-$05:35:41.5        & &   C2         & 20 & 14  \\
&            &  OMC~2     & 05:35:23.68  &     $-$05:12:31.8     &  &    C3         & 13 & 9  \\

\hline
\noalign{\smallskip}
TOTAL       &              &    & &  &  & 305 & 244 \\                  
\hline \hline
\end{tabular}

\noindent
$^1$B~213 core IDs are from \citet{Hacar2013}. IDs indicated in "Other names" column are from \citet{Onishi2002}. \\

\noindent
$^2$Perseus core IDs (L 1448, NGC 1333, Barnard 1, Barnard 5, IC348) are from \citet{Hatchell2007a}. \\

\noindent
$^3$NGC\,1333 core IDs indicated in $"$Other names$"$ column are from \citet{Sandell2001}.
\end{table*}

\section{Observations and molecular database}
\label{mol}

\subsection{Observations}
GEMS observations were carried out using the IRAM 30-m telescope at Pico Veleta (Spain) during several observing periods from July 2017 to December 2020. 
In addition, complementary observations of  the CS 1$\rightarrow$0 line were carried out with the Yebes 40m radiotelescope \citep{Tercero2021} toward TMC~1 and Barnard 1b. 
All these observations have been described in detail in the first GEMS paper \citep{Fuente2019}. The complete database is public and available through the IRAM Large Program webpage \footnote{https://www.iram-institute.org/EN/}, or directly from the GEMS webpage \footnote{https://www.oan.es/gems/}. 

\subsection{Sample}
Our project focuses on the nearby star-forming regions Taurus, Perseus, and  Orion,  which are considered as prototypes of low-mass, intermediate-mass, and massive 
star-forming regions. Moreover, these molecular cloud complexes have been observed with Herschel and SCUBA as part of the  Gould Belt Survey \citep{Andre2010}, and accurate visual extinction (A$_V$) and dust temperature (T$_d$) maps are available \citep{Malinen2012, Palmeirim2013, Lombardi2014, Zari2016}. The angular resolution of the A$_{\rm V}$ and T$_d$ maps ($\sim$36$''$) is  similar to that provided by the 30m telescope at 3mm allowing a direct comparison of continuum and spectroscopic data.

Chemical calculations show that the determination of elemental abundances requires the observation of more than a dozen of molecular compounds, some of them presenting weak emission \citep{Fuente2016, Fuente2019}. Full-sampling mapping of these regions in such a large number of species would have been unrealistic in terms of telescope observing time. An alternative strategy was adopted in GEMS. Our project observed toward 305 positions distributed in 29 cuts roughly perpendicular to a set of selected filaments. These cuts were defined to intersect the filament along one of the embedded starless cores, avoiding the position of protostars, HII regions and bipolar outflows. The final set of observed positions probes visual extinctions from A$_V\sim$3 mag up to A$_V\sim$200 mag. The separation between one position and another in a given cut was selected to sample regular bins of A$_V$. This is, however, not  possible close to the visual extinction peaks where the surface density gradient is steep. 
In Table~\ref{Table: GEMS sample}, we show a summary of the cuts forming our sample. A more detailed description of the cuts and their location within the clouds was given by \citet{Rodriguez-Baras2021}. 

\subsection{Molecular database}
Carbon sulfide (CS) has been largely used as density and column density tracer in the interstellar medium because it is an abundant species, with a simple rotational spectrum, and its excitation is well-known \citep{Linke1980, Zhou1989, Tatematsu1993, Zinchenko1995, Anglada1996, Bronfman1996, Laundhardt1998, Shirley2003, Bayet2009, Wu2010, Zhang2014, Scourfield2020}. In order to calculate the gas density toward the observed positions, we fit the rotational lines of CS and its isotopologues using the non-LTE molecular excitation and radiative transfer code RADEX \citep{Tak2007}, and the collisional coefficients calculated by \citet{Denis2018}.

Several assumptions have been made in these calculations. First, we adopted fixed isotopic ratios,  $^{12}$C/$^{13}$C=60, $^{32}$S/$^{34}$S= 22.5, and
$^{16}$O/$^{18}$O=550  \citep{Wilson1994, Savage2002, Gratier2016}. 
The assumption of fixed isotopic ratios is justified in the case of CS because isotopic fractionation is not expected to be important for sulfur. The CS molecule is enriched in $^{34}$S at early time because of the $^{34}$S$^+$ + CS reaction but not at the characteristic chemical ages of dense clouds \citep{Loison2019b}. More controversial could be the $^{12}$CS/$^{13}$CS ratio; the adopted value is consistent with the results of \citet{Gratier2016} in TMC~1 and \citet{Agundez2019} in L~483. This value is also consistent with chemical predictions for typical conditions and chemical ages in dark clouds \citep{Colzi2020, Loison2020}. 
Isotopic chemical fractionation might be important for HCN \citep{Loison2020}, but the HCN/H$^{13}$CN does not differ from the assumed value in more than a factor of $\sim$2.
Oxygen isotopic fractionation is expected in some species such as NO, SO, O$_2$, and SO$_2$ \citep{Loison2019b}. However, it is not expected to be relevant for C$^{18}$O and HC$^{18}$O$^+$, the two oxygen isotopologues in the GEMS database.
We assumed a beam filling factor of 1 for all transitions, that is to say that the emission is more extended than the beam size, which is reasonable for most positions.  

The observed CS, C$^{34}$S, and $^{13}$CS line intensities were fitted by varying only two parameters, n(H$_2$) and  N(CS). The gas kinetic temperature was assumed to be equal to the dust temperature as derived from the Herschel data. 
The parameter space was explored following the Monte Carlo Markov Chain (MCMC) methodology with a Bayesian inference approach. In particular, we use the \emph{emcee} \citep{Foreman2012} implementation of the Invariant MCMC Ensemble sampler methods by \citet{Goodman2010}. At the very low visual extinctions, not all the observed CS lines were detected, and we could not determine the gas density  (see Table~\ref{Table: GEMS sample}). 
Following this methodology, we estimated the gas density toward the 244 positions out of the whole sample. The abundances observed toward these positions constitutes the observational basis of this work.

Once the density was determined from the CS line fitting, we assumed this value to estimate the column densities and abundances of the rest of species: $^{13}$CO, C$^{18}$O, HCO$^+$, H$^{13}$CO$^+$, HC$^{18}$O$^+$, H$^{13}$CN, HNC, HCS$^+$, SO, $^{34}$SO, H$_2$S, and OCS. To do so,
we used the RADEX code and the collisional coefficients from the references listed in Table~\ref{Table:collisional coefficients}. The lines of the most abundant species, CO, HCO$^+$, HCN and HNC are expected to be optically thick and the abundances derived from them are indeed lower limits to the real ones. In order to minimize the effect of high opacities, in this paper we use rare isotopologues to calculate the abundance of the main isotopologue as follows, X(CO)=X(C$^{18}$O) $\times$ 550, X(HCO$^+$)=X(H$^{13}$CO$^+$) $\times$ 60, and X(HCN)=X(H$^{13}$CN) $\times$ 60. In the case of HNC, we did not observe HN$^{13}$C in GEMS, and the estimated abundance might be underestimated in positions with A$_V$$>$8 mag. A more detailed description of the molecular database and a first statistical analysis was published by \citet{Rodriguez-Baras2021}. 

\begin{table}
\begin{centering}
\caption{Chemical model parameters}
\label{Table:grid}
\begin{tabular}{l |l}
 \hline \hline
{\bf Taurus}  & \\
n$_{\rm H}$  & 3.16$\times$10$^3$ to 3.16$\times$10$^6$ cm$^{-3}$ in steps of $\times$3.16\\ 
A$_V$  & 1.5, 3.5, 5.5, 7.5, 9.5, 11.5 mag  \\ 
T  & 10, 15, 20, 25, 30, 35, 45 K  \\ 
CRIR$^1$   & (1.0, 5.0, 10, 50) x 10$^{-17}$ s$^{-1}$  \\ 
$\chi_{UV}$ (Draine) &   5.0$^2$  \\ 
$[S/H]$    & (1.5, 0.15, 0.008) $\times$10$^{-5}$ \\ \hline
{\bf Perseus} & \\
n$_{\rm H}$  & 3.16$\times$10$^3$ to 3.16$\times$10$^6$ cm$^{-3}$ in steps of $\times$3.16\\ 
A$_V$  & 1.5, 3.5, 5.5, 7.5, 9.5, 11.5, 13.5 mag  \\ 
T  & 10, 15, 20, 25, 30, 35, 45 K  \\ 
CRIR$^1$   & (1.0, 5.0, 10, 50) x 10$^{-17}$ s$^{-1}$  \\ 
$\chi_{UV}$ (Draine) &   25.0$^2$     \\ 
$[S/H]$    &   (1.5, 0.15, 0.008) $\times$10$^{-5}$\\ \hline 
{\bf Orion}  & \\
n$_{\rm H}$  & 3.16$\times$10$^3$ to 3.16$\times$10$^6$ cm$^{-3}$ in steps of $\times$3.16\\ 
A$_V$  & 1.5, 3.5, 5.5, 7.5, 9.5, 11.5, 15.5, 20 mag  \\ 
T  & 10, 15, 20, 25, 30, 35, 45 K  \\ 
CRIR$^1$   & (1.0, 5.0, 10, 50) x 10$^{-17}$ s$^{-1}$  \\ 
$\chi_{UV}$ (Draine) &   50$^3$   \\ 
$[S/H]$    & (1.5, 0.15, 0.008) $\times$10$^{-5}$  \\ 
\hline \hline
\end{tabular}

\noindent
$^1$Molecular hydrogen cosmic-ray ionization rate ($\zeta_{H_2}$).
\end{centering}

\noindent
$^2$ Estimated by \citet{Navarro-Almaida2020} based on the dust temperature and the analytical expression derived by \citet{Hocuk2017}.

\noindent
$^3$ Assumed value based on  UV field estimated toward the photon-dominated region in the Horesehead nebula \citep{Pety2005}.
\end{table}

\section{Chemical network}
\label{Sect:chem}

A great effort has been done in the last decade to improve the gas-phase \citep{Fuente2016, Fuente2017,Vidal2017,Fuente2019}, and surface \citep{Laas2019} sulfur chemical network. 
\citet{Navarro-Almaida2020} incorporated all these novelties to build the chemical network used in their paper. In addition, we have implemented the reaction rate for
the  S + CH $\rightarrow$ CS + H recently estimated by Bulut et al. (in preparation). Using this updated chemical network, we performed the abundance calculations using \textsc{Nautilus 1.1} \citep{Ruaud2016}, which is a numerical model suited to the study chemistry in astrophysical environments. It is a three-phase model, in which gas, grain surface and grain mantle phases, and their interactions, are considered. It solves the kinetic equations for the gas-phase and the solid species at the surface of interstellar dust grains. 

In dark clouds, where the temperature of grain particles is below the sublimation temperature of most species, non-thermal desorption processes are needed to maintain significant abundances of molecules in gas phase. In Nautilus, desorption into the gas phase is only allowed for the surface species, considering both thermal and non-thermal mechanisms. 
The latter include desorption induced by cosmic rays (Hasegawa \& Herbst 1993), direct (UV field) and indirect (secondary UV field induced by the cosmic-ray flux) photo-desorption, and reactive chemical desorption \citep{Garrod2007, Minissale2016}. 
In the interior of dark clouds where the UV field is highly attenuated by dust grains and photo-desorption is not efficient,  desorption induced by cosmic rays and reactive chemical desorption become the main desorbing mechanisms. Since we are considering molecular gas shielded from the UV radiation, we use the prescription proposed by \citet{Minissale2016} for ice coated grains to calculate the reactive chemical desorption. 

\begin{figure*}
\includegraphics[angle=0,scale=.3]{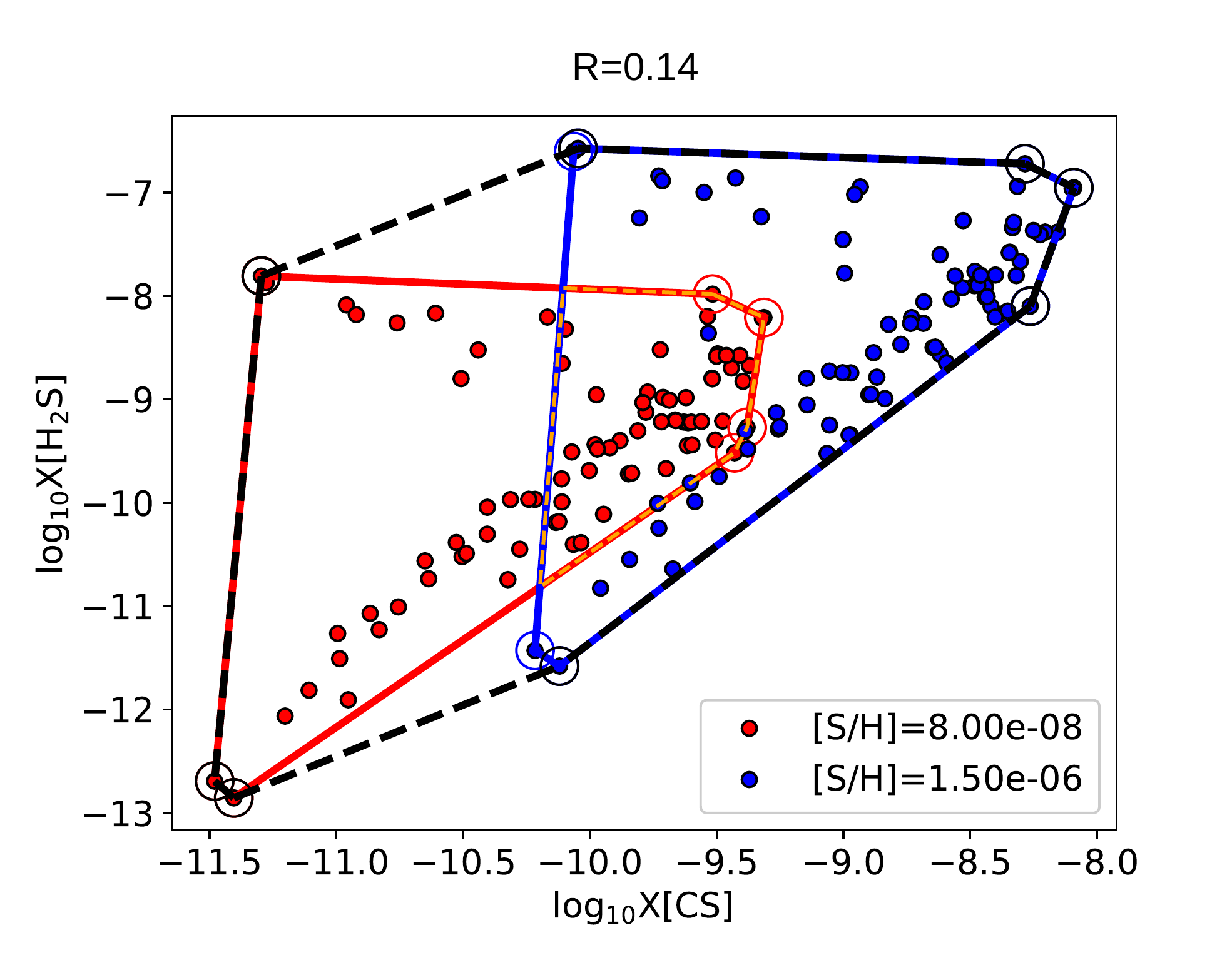}
\includegraphics[angle=0,scale=.3]{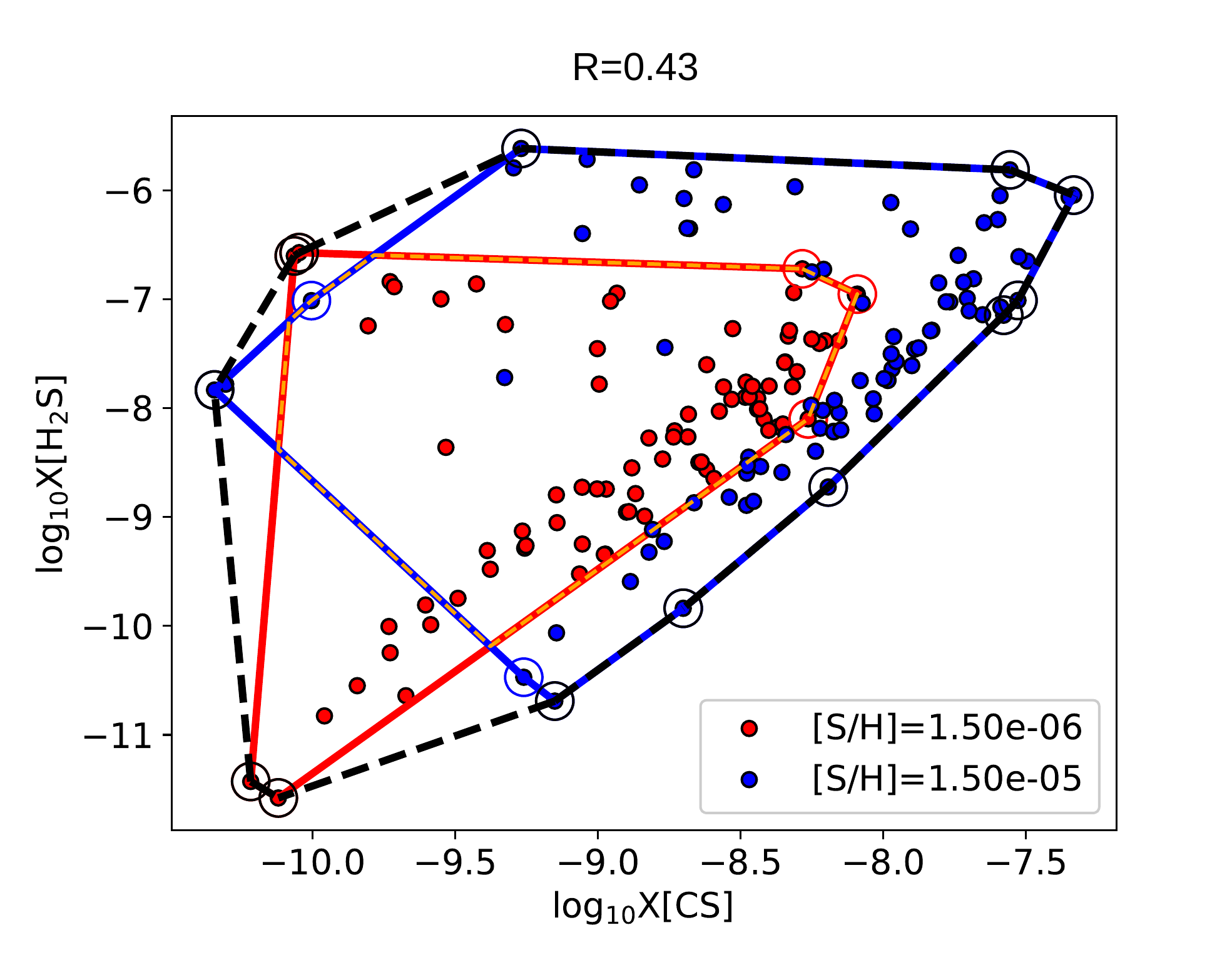}
\includegraphics[angle=0,scale=.3]{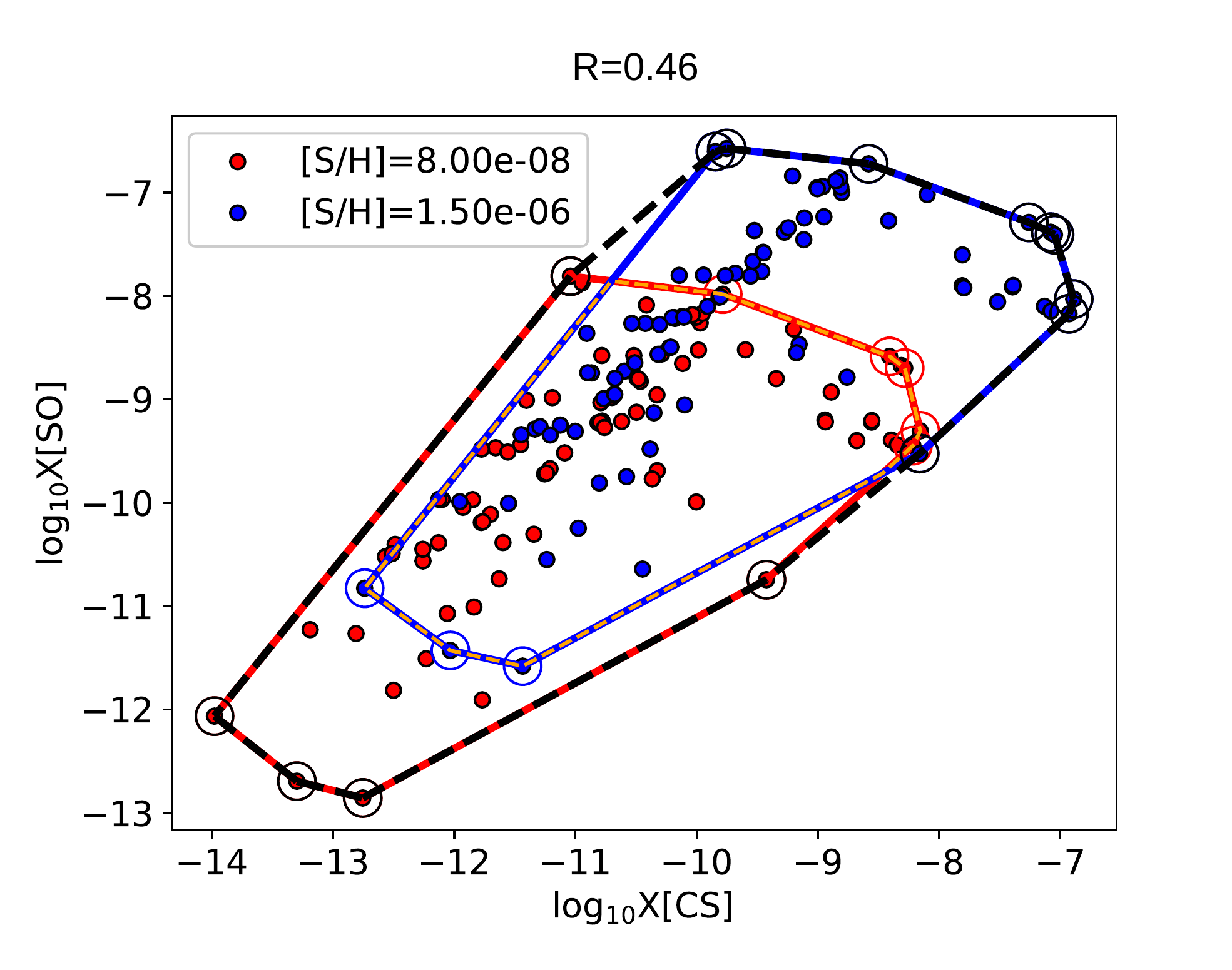} \\
\includegraphics[angle=0,scale=.3]{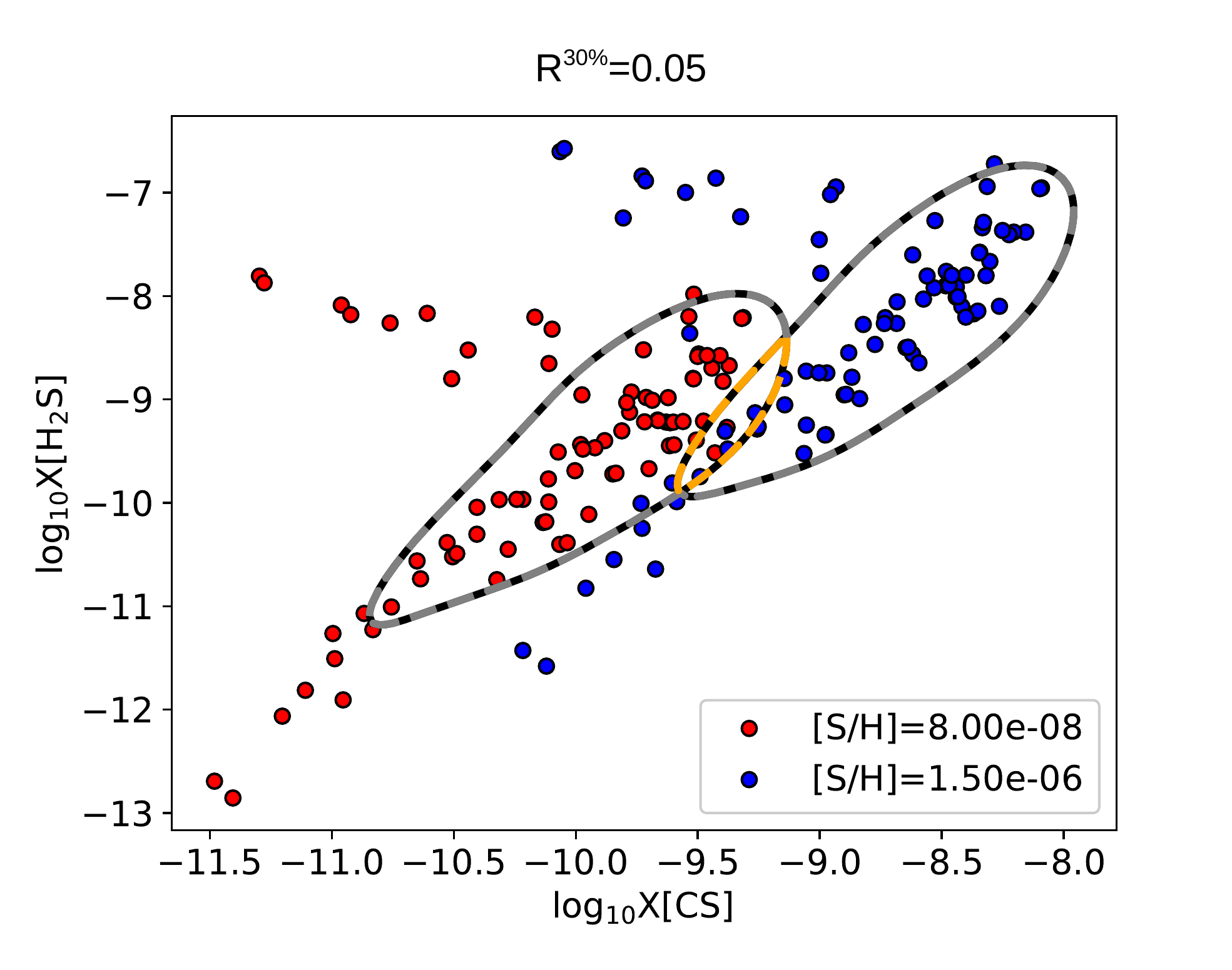}
\includegraphics[angle=0,scale=.3]{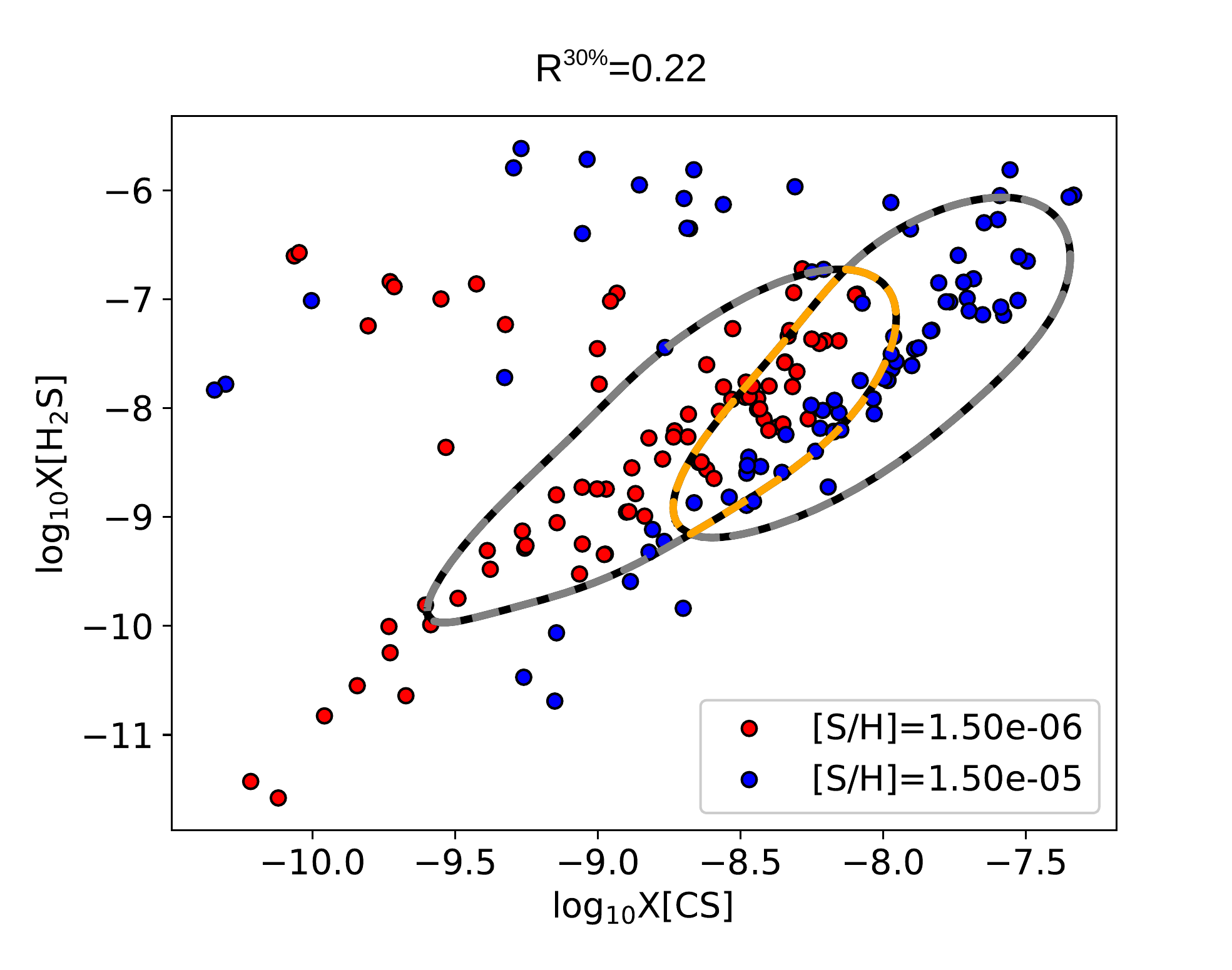}
\includegraphics[angle=0,scale=.3]{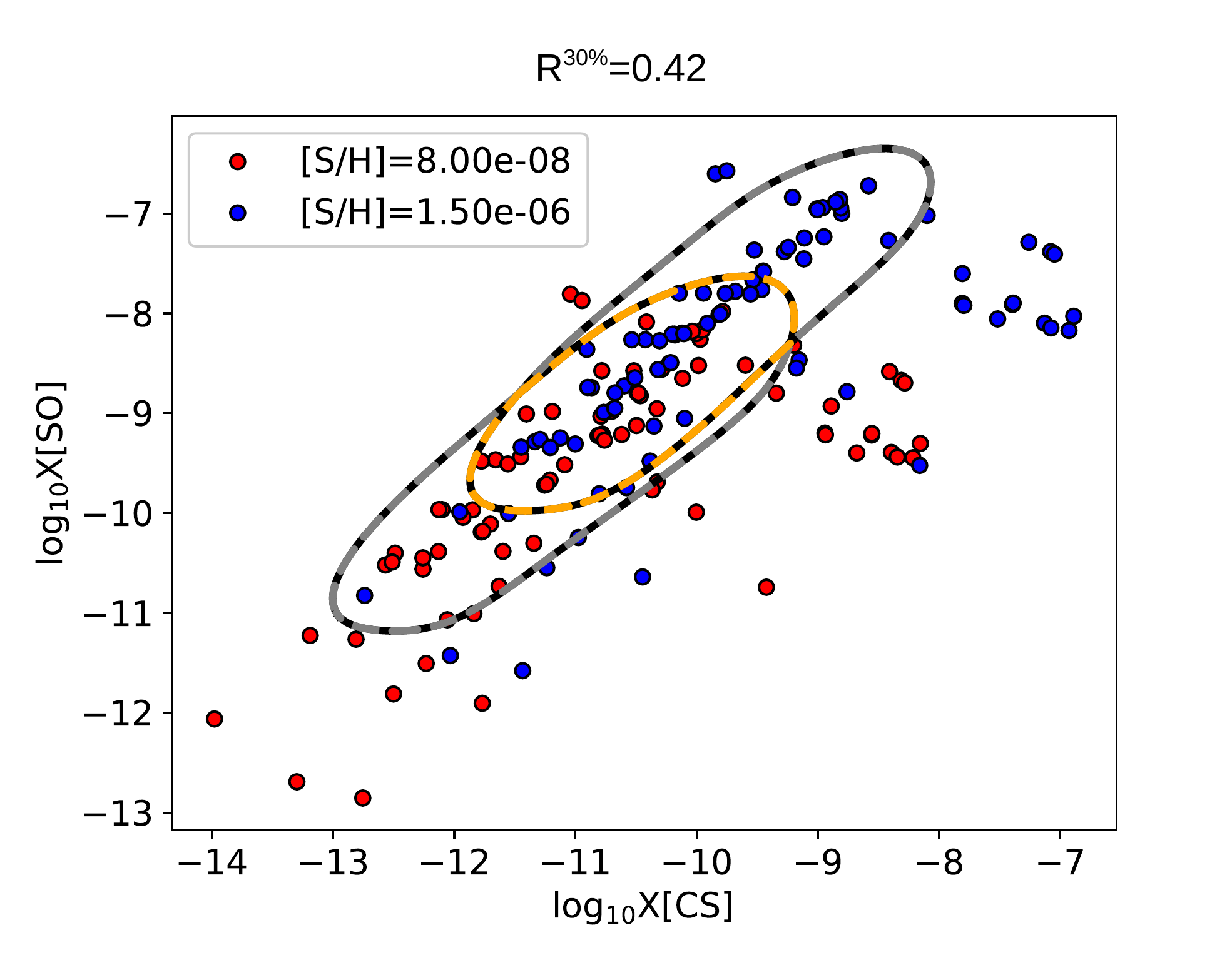}
\caption{Model clouds formed by selecting all the models with T=10~K and A$_V$=11.5 mag  from the output of the grid listed in Table~\ref{Table:grid} for Taurus. These models are used to illustrate  the calculation of the parameter  $R$ (upper panels) and $R^{30\%}$ (bottom panels) (see Sect.~\ref{Sect:exploration} and Table~\ref{Table:R}). {\it Upper panels:} red and  blue lines show the polygons obtained by connecting the extreme values of each [S/H] model cloud. The  parameters $R$ is then calculated  as the ratio of the red and blue contours intersection area between over the total area.  
 {\it Bottom panels:} grey lines are iso-density contours corresponding to 30\% the point density peak in each model cloud. We show in orange the intersection between the grey contours corresponding to the two values of [S/H]. The  parameters $R^{30\%}$ is defined  as the ratio of the intersection area (area enclosed within the orange contour) over the total area.
}
\label{Rpar}
\end{figure*}

\begin{table*}
\caption{$R$ and $R^{30\%}$ factors}
\label{Table:R}
\begin{tabular}{l |l |l |l |l | |l  |l  |l}
\hline \hline
\multicolumn{1}{l|}{Fixed parameter} &
\multicolumn{1}{l|}{D$_S$$^1$} &
\multicolumn{1}{l|}{R$_{min}$} &
\multicolumn{1}{l|}{R$_{max}$} & 
\multicolumn{1}{l| | }{R$_{mean}$} &
\multicolumn{1}{l|}{R$^{30\%}$$_{min}$} &
\multicolumn{1}{l|}{R$^{30\%}$$_{max}$} & 
\multicolumn{1}{l}{R$^{30\%}$$_{mean}$} 
 \\  \hline 
T=10~K &   187/10 &  0.14(H$_2$S-CS) & 0.46(CS-SO) &  0.31 &  
                                    0.05(H$_2$S-CS) & 0.51(OCS-SO) &  0.25  \\
        &  10/1   &   0.43(H$_2$S-CS)     & 0.62(CS-SO)    &  0.53 & 
                           0.22(H$_2$S-CS)    & 0.64(OCS-SO)  & 0.44   \\
        &    187/1  &  0.06(H$_2$S-CS)   & 0.29(OCS-SO) &  0.18 &  
                              0.00(H$_2$S-CS)   & 0.41(OCS-SO)  &  0.13    \\ \hline
T=25~K &   187/10  &  0.39(H$_2$S-SO) & 0.55(OCS-SO) &  0.49 &   
                                    0.24(H$_2$S-CS) & 0.42(OCS-SO)  & 0.31    \\ 
             &  10/1  &   0.66(H$_2$S-CS)    & 0.83(OCS-SO) &  0.71 &  
                               0.58(H$_2$S-OCS) & 0.66(H$_2$S-CS) & 0.62    \\
             &    187/1  &  0.27(H$_2$S-SO)  & 0.54(OCS-SO) &  0.37 &  
                                   0.21(H$_2$S-CS)  & 0.46(OCS-SO)  & 0.35      \\ \hline     
T=35~K &   187/10 &  0.32(OCS-H$_2$S) & 0.58(OCS-CS) &  0.47 & 
                                    0.04(OCS-CS)        & 0.69(OCS-H$_2$S)   &  0.31   \\
        &    10/1  &  0.56(OCS-H$_2$S) & 0.65(OCS-SO)        &  0.60 &   
                             0.21(OCS-SO)        &  0.46(H$_2$S-CS)  &  0.35   \\ 
        &   187/1  &  0.28(OCS-SO)   &    0.43(OCS-CS)            &  0.34 &    
                            0.03(OCS-CS)   &    0.40(OCS-H$_2$S)     &  0.20    \\ \hline
n$_{\rm H}$=10$^4$~cm$^{-3}$  &    187/10.   &  0.37(OCS-CS)     &  0.58(H$_2$S-CS) & 0.46  &  
                                                                             0.01(H$_2$S-CS) &  0.25(OCS-SO)      & 0.14   \\
                      &   10/1 &  0.48(H$_2$S-SO)   & 0.62(OCS-SO) & 0.56 &   
                                        0.25(OCS-CS)         & 0.59(OCS-SO) & 0.40  \\
                      &    187/1  &  0.18(H$_2$S-CS)  & 0.30(OCS-SO) & 0.25 &    
                                           0.00(OCS-CS)      &  0.21(OCS-SO) & 0.07   \\ \hline
n$_{\rm H}$=10$^5$~cm$^{-3}$  &   187/10 &  0.34(OCS-H$_2$S)  & 0.57(OCS-SO) & 0.45 &  
                                                                            0.32(OCS-CS)           & 0.58(OCS-SO)  & 0.43 \\ 
                     &  10/1    &    0.33(OCS-H$_2$S)  & 0.60(CS-SO)     & 0.41 &   
                                       0.29(H$_2$S-CS)     & 0.63(OCS-SO)  & 0.43 \\  
                 &   187/1  &    0.09(OCS-H$_2$S)  & 0.32(CS-SO)     & 0.19 &   
                                       0.00(OCS-CS)          &  0.34(OCS-SO)  & 0.16     \\ \hline
n$_{\rm H}$=10$^6$~cm$^{-3}$  &   187/10  &  0.40(H$_2$S-SO)   & 0.65(OCS-SO) & 0.53  &  
                                                                            0.38(H$_2$S-SO)   & 0.63( CS-SO)   & 0.51 \\ 
                      &   10/1 &  0.49(H$_2$S-SO)   & 0.71(CS-SO)  & 0.58 &    
                                       0.27(OCS-SO)        & 0.71(CS-SO)   & 0.48   \\
                      &   187/1  &  0.19(H$_2$S-SO)   & 0.50(CS-SO)  & 0.35 &   
                                           0.23(OCS-CS)         & 0.53(CS-SO)  & 0.35 \\  \hline
t=0.1~Myr  &  187/10   &   0.48(H$_2$S-CS)  & 0.63(CS-SO)           &  0.53 &    
                                         0.28(SO-CS)         & 1.0 (OCS-H$_2$S)  &  0.47  \\
             &    10/1  &  0.55(OCS-CS)  &  0.68(CS-SO)             &  0.64 &    
                                0.43(OCS-CS)  &  0.63(OCS-H$_2$S)  &  0.55    \\
             &    187/1  &  0.32(H$_2$S-CS) & 0.42(CS-SO)            &  0.37 &   
                                  0.06(SO-CS)         & 1.0 (OCS-H$_2$S)  &  0.40    \\  \hline
t=1~Myr    &  187/10  &  0.41(H$_2$S-SO) & 0.59(CS-SO)  &  0.51 &   
                                       0.29(OCS-CS)      & 1.0(CS-SO)    &  0.50   \\
             &   10/1  &  0.54(H$_2$S-SO) & 0.75(CS-SO)  &  0.64 &  
                                0.42(H$_2$S-SO) & 0.61(CS-SO)  & 0.52    \\ 
             &   187/1  &  0.26(H$_2$S-SO) & 0.52(OCS-H$_2$S) &  0.41  &   
                   0.04(OCS-CS)       & 0.34(OCS-SO)         &  0.24  \\  \hline
t=10~Myr    &  187/10  &   0.51(OCS-CS) & 0.61 (CS-SO)    &  0.58 &   
                          0.24(OCS-CS) & 0.61(OCS-SO)  & 0.47   \\
             &  10/1  &  0.56(H$_2$S-SO)    & 0.70 (OCS-SO) &  0.65 &    
                   0.41(OCS-H$_2$S)  & 0.64(CS-SO)    &   0.53   \\
             &   187/1  &  0.32(H$_2$S-SO) & 0.48 (SO-CS)      &  0.41 &   
                   0.16(OCS-CS)       & 0.42(OCS-SO)    & 0.24  \\  \hline
$\zeta_{H_2}$= 1 $\times$ 10$^{-17}$ s$^{-1}$ &  187/10  & 0.48(H$_2$S-CS) & 0.61(CS-SO) &  0.55 &   
                                                                               0.47(H$_2$S-CS) & 0.61(CS-SO) &  0.54    \\
                                         &    10/1  &  0.57(H$_2$S-CS) & 0.72(CS-SO) &  0.65 &    
                                                  0.58(OCS-SO)      & 0.75(CS-SO)  & 0.62   \\
                                          &  187/1   &  0.27(H$_2$S-CS) & 0.44(CS-SO) &  0.37 &    
                                                  0.31(H$_2$S-CS) & 0.50(CS-SO)  & 0.37   \\ \hline
$\zeta_{H_2}$= 5 $\times$ 10$^{-17}$ s$^{-1}$  &187/10  & 0.45(H$_2$S-CS) &  0.67(OCS-SO) & 0.57 &  
                                                                               0.36(OCS-CS)      &  0.55(OCS-SO)  & 0.46   \\
                                             &  10/1  & 0.50(H$_2$S-SO) &  0.70(CS-SO)    & 0.59 &   
                                                   0.49(H$_2$S-SO) & 0.62(OCS-SO)  & 0.54 \\ 
                                             & 187/1   & 0.27(H$_2$S-SO) &  0.46(CS-SO)  & 0.35 &   
                                                   0.26(H$_2$S-SO) &  0.29(OCS-CS) & 0.29  \\  \hline
$\zeta_{H_2}$= 1 $\times$ 10$^{-16}$  s$^{-1}$ & 187/10  & 0.41(H$_2$S-CS) &  0.64(OCS-SO) & 0.52 &  
                                                                             0.31(OCS-CS)       & 0.54(OCS-SO) &  0.42  \\
                                            &   10/1 & 0.51(H$_2$S-SO) &  0.65(CS-SO)    & 0.57 &    
                                                  0.48(H$_2$S-SO) &  0.63(OCS-SO)  & 0.56  \\ 
                                           &  187/1    & 0.24(H$_2$S-SO) &  0.40(OCS-SO) & 0.31 &  
                                                  0.19(OCS-CS)       &  0.33(OCS-SO) & 0.26  \\ \hline\hline
\end{tabular}

\noindent
$^1$ D$_{\rm S}$= [S/H]/[S/H]$_{\rm solar}$, where [S/H]$_{\rm solar}$=1.5$\times$10$^{-5}$.
\end{table*}  

\section{Exploration of the parameter space}
\label{Sect:exploration}

In this section, we aim to select the best strategy to determine sulfur depletion (D$_{\rm S}$) based on GEMS database.
We ran a grid of models with typical conditions in dark clouds, in particular the physical conditions listed for Taurus in Table~\ref{Table:grid}. The parameter [S/H] is allowed to take three discrete values that represent the cases of no depletion,  D$_{\rm S}$=1 ([S/H]=1.5$\times$10$^{-5}$),  moderate depletion, D$_{\rm S}$=10  ([S/H]=1.5$\times$10$^{-6}$), and high depletion, D$_{\rm S} \sim$ 187  ([S/H]=8.0$\times$10$^{-8}$). These values correspond to different estimates of D$_{\rm S}$ in star-forming regions \citep{Vastel2018, Navarro-Almaida2020,  Navarro-Almaida2021,  Hily-Blant2022}.  The value of $\zeta_{H_2}$ varied from $\zeta_{H_2}$=10$^{-17}$~s$^{-1}$ which is typically found in dense and evolved cores \citep{Caselli2002} to  $\zeta_{H_2}$=5$\times$10$^{-16}$~s$^{-1}$, expected in low visual extinction regions \citep{Neufeld2017}.
Finally, we considered three chemical ages,  $t$, that represent early chemistry (0.1 Myr), late chemistry (1 Myr), and steady-state chemistry (10 Myr). For this exploratory work, we only consider models with A$_V$ = 11.5 mag, thus neglecting the effect of the UV field.
Using these chemical calculations, we performed corner diagrams of the abundances of the S-bearing species included in GEMS database, CS, SO, H$_2$S, and OCS, for different values of D$_{\rm S}$ (see. Fig.~\ref{modelos-Tg-1} to Fig.~\ref{modelos-age}). In each diagram, one of the input parameters is kept fixed while the others are allowed to vary in the whole range. The fixed parameter is the one indicated in Fig.~\ref{modelos-Tg-1} to Fig.~\ref{modelos-age}. Models with different sulfur depletion are differentiated using a color code. 
When the model clouds corresponding to different values of  D$_{\rm S}$ appear grouped in bands with little overlap among them, we can conclude that the considered molecules are robust tracers of  D$_{\rm S}$. If model clouds largely overlap then the solution is degenerate and we need to fix additional parameters to determine  D$_{\rm S}$.

In order to base our discussion on quantitative grounds,
we define the parameters $R$ and $R^{30\%}$, which measure the overlap between model clouds corresponding to  different values of D$_{\rm S}$. Fig.~\ref{Rpar} illustrates how these parameters are calculated. 
In the upper panels of  Fig.~\ref{Rpar}, the extreme points of models with the same D$_{\rm S}$ are connected to construct polygons. 
The $R$ parameter is calculated as, $R = (Intersection)/(Total)$, where {\it Intersection} refers to the intersection between the areas of the polygons corresponding to two values of D$_{\rm S}$ (intersection between the areas defined by the blue and red contours in Fig.~\ref{Rpar}), and {\it Total} indicates the area defined by the whole cloud of points, which mathematically would be $Area1 + Area2 - Intersection$. With this definition, $R$=1 indicates that the two clouds of points fully overlap and it is not possible to differentiate between the two values of D$_{\rm S}$. On the contrary, if $R$=0, the intersection is null and we have a clear diagnostic. This parameter gives a good measure of the overlapping when the points are uniformly distributed. However,  it is not a good descriptor when the points are not evenly distributed, with some located very far from the bulk. In order to account for these cases, we have defined another parameter, R$^{30\%}$, that uses density curves to define the area corresponding to a given value of D$_{\rm S}$. In particular, $Area1^{30\%}$ and $Area2^{30\%}$ are defined as the areas enclosed by the 30\% point density peak contour, and $R^{30\%} = (Intersection^{30\%} )/(Total^{30\%} )$ (see bottom panels of  Fig.~\ref{Rpar}). As expected, we find $R^{30\%} < R$ in the case of a peaky distribution of points.

Table~\ref{Table:R} shows the minimum, maximum, and mean values of $R$ and $R^{30\%}$ for the panels forming the corner diagrams shown in Fig.~\ref{modelos-Tg-1} to Fig.~\ref{modelos-age}. We remind that we have only used  the pairs of molecules,  OCS-H$_2$S, OCS-SO, OCS-CS, H$_2$S-CS, H$_2$S-SO, and CS-SO, in our analysis.  Thus, the minimum, maximum, and mean R and $R^{30\%}$ values are estimated taking into account  only these six combination of sulfur species.  Although the exact value of  $R$ and $R^{30\%}$ are slightly different, we find the same trends in these parameters which demonstrates the robustness of our analysis. One first result is that we need to fix temperature to reach values,  $R^{30\%}_{mean}$$<$0.3. Moreover the lowest values of $R^{30\%}_{mean}$ are found for T = 10~K. In Fig.~\ref{modelos-Tg-1}, we show the corner diagrams when fixing the gas temperature. 
At low temperature, T$\sim$10~K, CS and H$_2$S are the best proxies of  D$_{\rm S}$. In the panels including these species, the points corresponding to different values of  D$_{\rm S}$  appear grouped in clearly distinguishable bands. Moreover,  the bands  are almost orthogonal to the X(H$_2$S) axis suggesting that the abundance of H$_2$S is  an excellent tracer of  D$_{\rm S}$ with little dependence on the other parameters. This can be easily interpreted in terms of sulfur chemistry.
At such a low temperature, solid H$_2$S is the main sulfur reservoir, and its abundance in gas-phase is only dependent on the amount of sulfur atoms in the ice and the efficiency of non-thermal desorption mechanisms \citep{Navarro-Almaida2020}. The ubiquitous molecule CS is not such a good tracer of  D$_{\rm S}$. Values of X(CS)$\geq$10$^{-6}$, can only be achieved with low sulfur depletion (see Fig.~\ref{modelos-Tg-1}). However, the situation is less clear for X(CS)$<$10$^{-8}$ that can be explained assuming the three values of D$_{\rm S}$. These low values of X(CS) are the most common in the cold interstellar medium \citep{Agundez2013}, which hinders the determination of D$_{\rm S}$ on the basis of only this compound. 

Fixing the values of density is also useful, especially for n$_{\rm H}$$<$10$^5$~cm$^{-3}$ with $R^{30\%}_{mean}$$<$0.4. The corner diagrams in Fig.~\ref{modelos-nH} show that, fixing density, model clouds are distinguishable, although there is always some overlap among them. The overlap increases with increasing density, which implies that the estimated value of D$_{\rm S}$ is less reliable at high densities. 

As mentioned above, the cosmic-ray molecular hydrogen ionization rate, $\zeta_{H_2}$,  is a key parameter for the chemistry in dark clouds since it is governing the gas ionization degree. Fig.~\ref{modelos-crir} shows the corner diagrams for fixing the values of $\zeta_{H_2}$.  There is an important overlap between the models corresponding to the different values of D$_{\rm S}$ in all cases, challenging its estimation if we only fix this parameter. The same behavior is found when we fix the chemical time (see Fig.~\ref{modelos-age}).This is consistent with the high values of $R$ and  $R^{30\%}$ shown in Table~\ref{Table:R}.Therefore, fixing $\zeta_{H_2}$ and/or  $t$ does not help to determine D$_{\rm S}$. 

According to these results,  it would be desirable to constrain the gas temperature and density in order to have a good estimate of D$_{\rm S}$ in our sample. As explained in Sect.~\ref{method}, we assume that gas and dust are thermalized for our fittings, and use the dust temperature derived from Herschel maps to constrain T. We also constrain density using the values obtained by \citet{Rodriguez-Baras2021} through a multitransitional study of CS and its isotopologues. The  parameters $\zeta_{H_2}$ and t will be fitted together with D$_{\rm S}$ using our molecular database. It is well known that the HCO$^+$/CO and HCS$^+$/CS abundance ratios are good tracers 
of  $\zeta_{H_2}$ (see e.g., \citealp{Fuente2016}). The chemical age is better constrained with the CO abundance. 
This molecule is frozen on dust grains for high densities (n$_{\rm H}$$>$10$^4$ cm$^{-3}$) and dust 
temperatures $<$ 16 K \citep{Wakelam2021}. Since CO depletion depends on the local volume density, we can obtain a 
reliable value of the chemical age as long as the density is known.

Our exploratory study also shows that it is always easier to discern between D$_{\rm S}$ = 187 and D$_{\rm S}$ = 10,  than between D$_{\rm S}$ = 10 and D$_{\rm S}$ = 1. This fact should be taken into account when interpreting our fits.

\section{Methodology}
\label{method}

Our goal is to determine D$_{\rm S}$ in each of the positions observed within GEMS. Our large molecular database with 244 positions allows us to explore possible trends of D$_{\rm S}$ with the local physical parameters (n, T) and/or environment. We fit the abundances of CO, HCO$^+$, HCN, HNC, CS, HCS$^+$, H$_2$S, SO, and OCS as derived by  \citet{Rodriguez-Baras2021} with the output obtained with the grid of chemical models shown in Table~\ref{Table:grid}. We adopted $\chi_{UV}$= 5 for Taurus and $\chi_{UV}$=25 for Perseus. These values were derived by \citet{Fuente2019} and \citet{Navarro-Almaida2020} using the analytic expression obtained by \citet{Hocuk2017} and the Herschel dust temperature images. In Orion, we adopted $\chi_{UV}$=50, similar to that derived  in the
photon-dominated region associated with the Horsehead nebula \citep{Pety2005, Goicoechea2006, Gerin2009, Riviere-Marichalar2019}.  It should be noticed that some compounds are not detected toward the whole sample of 244 positions. In these cases, we perform the fitting using the detected species as long as the number of species is higher than 3, and contains at least one sulfur-bearing species. 

During the fitting process, D$_{\rm S}$ ,  $\zeta (H_2)$, and $t$ are allowed to freely vary among the range of discrete values adopted in the grid. However, we impose some restrictions to the values of A$_{\rm V}$, n(H$_2$) and T. The visual extinction is assumed to be in the range $A_{\rm V} \times 0.5$ and  $A_{\rm V} \times 0.5 + 2$, where A$_{\rm V}$ is the value obtained from {\em Herschel} data  \citep{Malinen2012, Palmeirim2013, Lombardi2014, Zari2016}, and the factor 0.5 accounts for the fact that the cloud is expected to be illuminated from the back and the front. Therefore, at the center of the molecular cloud, the effective visual extinction would be half of the total along the line of sight. The density is assumed to be in the range from 0.5 $\times$ n(H$_2$) to 5 $\times$ n(H$_2$), being the values of n(H$_2$), those obtained by \citet{Rodriguez-Baras2021} . The gas kinetic temperature is allowed to vary between T$_d$ and T$_d +$ 5~K. . 

Finally, we need to select a parameter to describe the goodness of each model in describing the observational results. The standard $\chi^2$ is not adequate to estimate errors when the molecular abundances included in the fit differ by several orders of magnitude. In order to find the "best-fitting" model at each position, we use the parameter $D_{iff}$ that is defined by,

\begin{equation}
D_{diff} = 1/n_{obs} \times \Sigma_i [log_{10} (X_{mod}^i) -  log_{10} (X_{obs}^i)]^2
\end{equation}

\noindent
where $n_{obs}$ is the number of species detected at each position, $X_{mod}^i$ is the model predicted abundance for the species $i$, and $X_{obs}^i$ is the abundance of the species $i$ derived from GEMS observations. This parameter is the square of the {\it Disagreement parameter} that has been previously used in astrochemistry by different authors (see e.g., \citealp{Wakelam2006,Vastel2018}).

\section{Results}
\label{Sect:results}
 
\begin{figure*}
\includegraphics[angle=0,scale=.65]{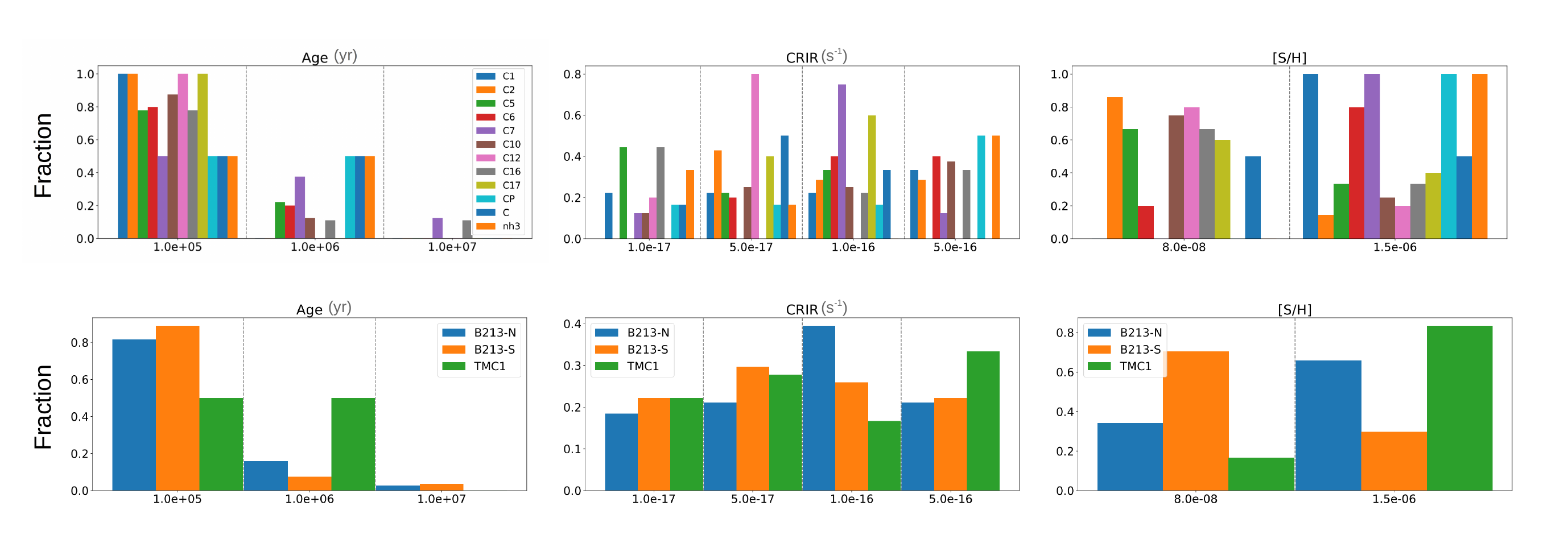}
\caption{Statistics of our molecular fitting in Taurus. The histograms show the fraction of positions that have been fitted with a given value of Age (right), $\zeta_{H_2}$ (center), and [S/H]). The positions are grouped by cut in the upper panels (see Table~\ref{Table: GEMS sample}), and by cloud in the lower panels. It should be noticed that none of the Taurus positions have been fitted with [S/H]=1.5$\times$10$^{-5}$.}
\label{Taurus-1}
\end{figure*}

\begin{figure*}
\includegraphics[angle=0,scale=.65]{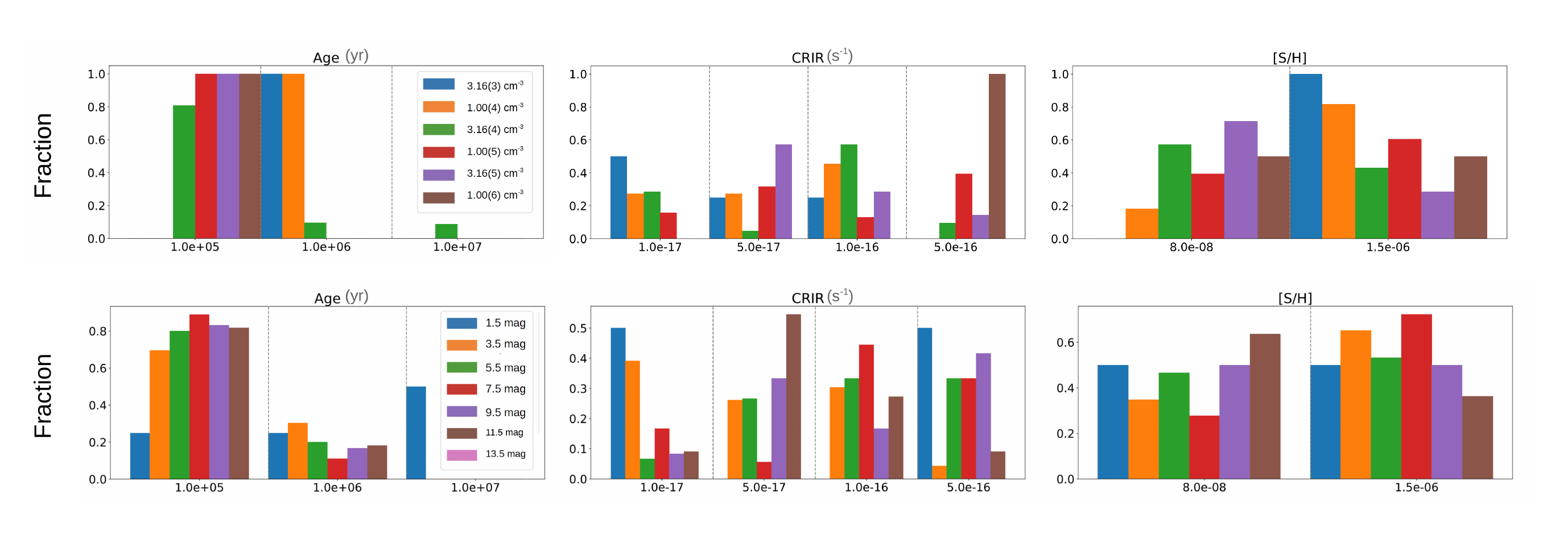}
\caption{Same as Fig.~\ref{Taurus-1}, but in the upper row, the Taurus positions have been separated in bins of density.  In the bottom panels, the positions have been distributed in bins of visual extinction.}
\label{Taurus-2}
\end{figure*}

\begin{figure*}

\includegraphics[angle=0,scale=.65]{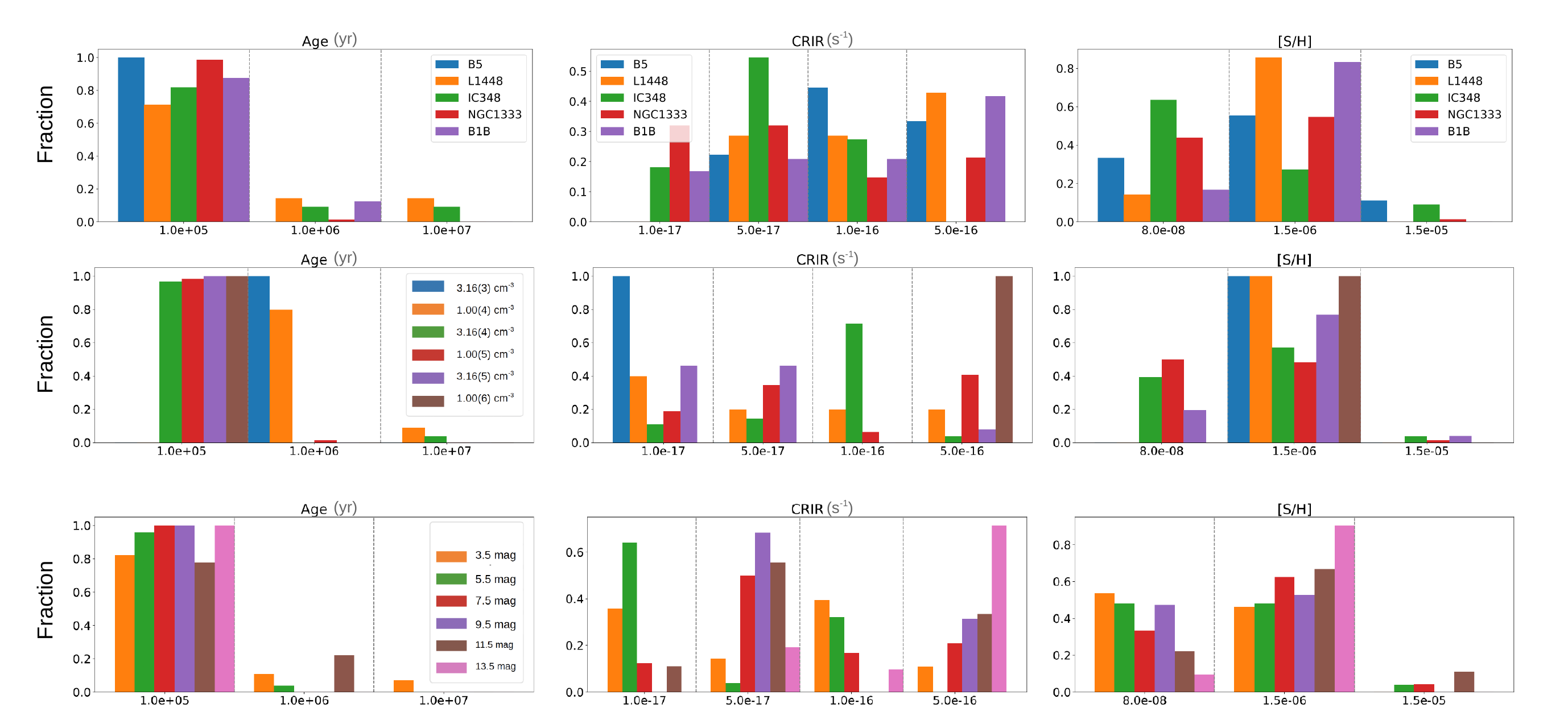}
\caption{Statistics of our molecular fitting in Perseus. The upper rows shows the histograms with the positions separated by cloud. In the middle row,  the positions have been distributed in bins of density, and in the bottom row, in bins of visual extinction.}
\label{Perseus-total}
\end{figure*}

\begin{figure*}
\includegraphics[angle=0,scale=.65]{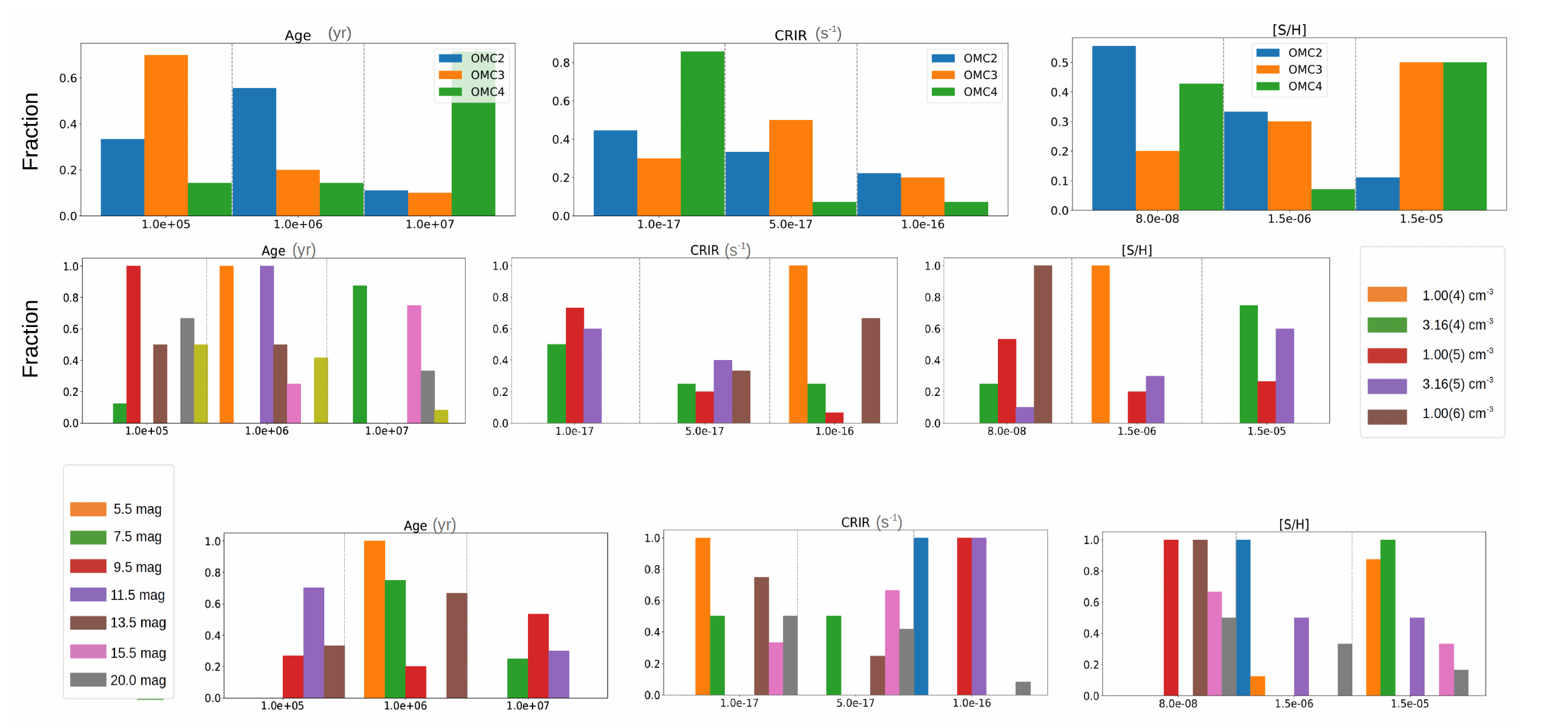}
\caption{Statistics of our molecular fitting in Orion. The upper rows shows the histograms with the positions separated by cloud. In the middle row,  the positions have been distributed in bins of density, and in the bottom row, in bins of visual extinction.}
\label{Orion-1}
\end{figure*}

\subsection{Taurus}
The Taurus molecular cloud is considered as a prototype of low-mass star-forming region.  At a distance of 145 pc \citep{Yan2019}, it has been extensively studied at infrared and millimeter wavelengths \citep{Cernicharo1987, Onishi1996, Narayanan2008, Cam1999, Padoan2002,Andre2010, Schmalzl2010}. Within GEMS, we have studied two well-known filaments:  TMC~1 and B~213/L1495. 

In TMC~1, we observed three cuts along the cores TMC~1-CP, TMC~1-NH3, and TMC~1-C, respectively. 
The positions TMC~1-CP and TMC~1-NH3  (the cyanopolyynes and ammonia emission peaks) are generally adopted as templates of carbon- and oxygen-rich starless cores to compare with chemical codes \citep[e.g.,][]{Feher2016, Gratier2016, Agundez2013}. Less studied from the chemical point of view, TMC~1-C has been identified as an accreting starless core \citep{Schnee2007, Schnee2010}. \citet{Fuente2019} carried out a complete analysis of the data associated with these cores to derive the gas ionization degree and elemental abundances in the TMC~1 translucent cloud. They concluded that the chemistry of the translucent filament is described assuming $\zeta_{H_2}$=(0.5$-$1.8) $\times$ 10$^{-16}$ s$^{-1}$ and  [S/H]=(0.4$-$2.2)$\times$10$^{-6}$. 
 
B~213/L~1495 is a prominent filament in Taurus that has been widely studied in the (sub)mm range \citep{Palmeirim2013, Hacar2013, Marsh2014,Tafalla2015, Bracco2017, Shimajiry2019}. 
The morphology of the map with striations perpendicular to the filament suggests that the filament is accreting material from its surroundings \citep{Goldsmith2008, Palmeirim2013}. \citet{Shimajiri2019} proposed that this active star-forming filament was initially formed by large-scale compression of HI gas, and is now growing in mass due to the gravitational accretion of ambient molecular gas.  A chain of starless cores \citep{Hacar2013} and protostars  \citep{Luhman2009, Rebull2010, Davis2010} is observed along the filament.   
Within GEMS, we observed 9 cuts along clumps \#1, \# 2,  \#5,  \#6,  \#7,  \#10,  \#12,  \#16, and  \#17 (core numbers from the catalog of \citealp{Hacar2013}). Cuts  \#1, \# 2,  \#5,  \#6,  and \#7 are located in the most active star-forming region (hereafter, B~213-N) where the density of protostars is higher. Cuts  \#10,  \#12,  \#16, and  \#17 are located in the quiescent part  (hereafter, B~213-S). Previous works based on CH$_3$OH data suggest the existence of chemical differentiation between B~213-N and  B~213-S \citep{Spezzano2022, Punanova2022}. In the cores located in B~213-N, the methanol peak is spatially coincident with a visual extinction peak, while the contrary behavior is found toward the south. It has been proposed that this behavior could be due to the irradiation on the cores due to nearby protostars which accelerate energetic particles along their outflows. A recent study of the H$_2$CS deuterated compounds \citep{Esplugues2022}, showed that the chemical age of the cores in B~213-N is higher than that of the cores in B~213-S. A combination of environmental and dynamical effects are therefore needed to account for the observed chemical variations.

The GEMS data provide new insights into the variations of chemical age, cosmic-ray ionization rate, and sulfur depletion among the different regions of Taurus. Our approach consists of fitting the abundances of nine species, CO, HCO$^+$, HCN, HNC, CS, HCS$^+$, H$_2$S, SO, and OCS toward 84 positions of Taurus distributed as shown in Table~\ref{Table: GEMS sample}. The results are shown in Fig.~\ref{Taurus-1} and Fig.~\ref{Taurus-2}.  Essentially all the positions in B~213 are best fitted assuming early time chemistry (t=0.1 Myr) which is consistent with the idea that this filament is still accreting material from the surroundings, keeping the gas chemistry far from the steady state. In TMC~1, we have some dispersion in the values of the chemical age with the positions located at A$_V$ $<$ 8~mag being better fitted with t = 1~Myr while those at higher visual extinction being better fitted with  t  = 0.1~Myr. We also find a large dispersion in the values of $\zeta_{H_2}$. In spite of this, we can find some trends. Most positions in TMC~1 and B~213-N are fitted with  $\zeta_{H_2}$$>$ 10$^{-16}$ s$^{-1}$, while B~213-S is better described with  $\zeta_{H_2}$$<$ 10$^{-16}$ s$^{-1}$, consistent with the idea of a harsh environment in B~213-N due to the presence of young (proto-)stars. A high value of  $\zeta_{H_2}$ has also been argued to explain the richness of TMC~1 in complex carbon chains \citep{Agundez2013}. 

Regarding the sulfur elemental abundance, we find that D$_{\rm S}$=1 in TMC~1 and and B~213-N, while  D$_{\rm S}$=187 in B~213-S. 
Based on observations of NS, \citet{Hily-Blant2022} proposed that the sulfur depletion is higher in the more evolved starless cores such as L1544, compared with others at an earlier evolutionary stage. In Fig.~\ref{Taurus-2} we explore the possible dependence of sulfur depletion with density and visual extinction.
There is no clear trend with the visual extinction.  However, we detect a trend of D$_{\rm S}$ with density, with the densest cores showing higher sulfur depletion. The density is expected to increase along the starless core evolution to form a  collapsing core. This would indicate that more evolved cores would present higher values of sulfur depletion in agreement with the results of \citet{Hily-Blant2022}. \citet{Esplugues2022} found that together with density, higher values of sulfur depletion favors the formation of singly and doubly deuterated H$_2$CS. 

\textsc{Nautilus 1.1} gas-grain code includes the freeze out of molecules onto grain surfaces, which is known to increase with density when the grain temperature is below the evaporation temperature of a given species. This means that an additional process which removes sulfur from the volatile chemistry is needed to explain the increase of the depletion with density. On the other hand, as explained in Sect.~\ref{Sect:exploration}, the reliability of our calculation of D$_{\rm S}$ decreases in dense regions. Therefore, we need to be cautious with this result.

\subsection{Perseus}
Located at a distance of 310 pc \citep{Ortiz-Leon2018}, the Perseus molecular cloud is considered the archetype of intermediate-mass star-forming region. This large molecular cloud complex is associated with two clusters containing pre-main-sequence stars: IC~348, with an estimated age of 2~Myr \citep{Luhman2003}, and NGC~1333, which is thought to be younger than $<$1~Myr  \citep{Lada1996, Wilking2004}. The cloud has been extensively studied using a large variety of techniques and wavelengths \citep{Bachiller1984,Warin1996, Hatchell2005, Ridge2006,  Kirk2006, Enoch2006, Hatchell2007a, Hatchell2007b, Curtis2011, Pineda2008, Pineda2010, Pineda2015, Zari2016, Friesen2017, Hacar2017b}.  In particular, the content in dense cores was described in a series of papers based on continuum maps at 850 and 450 $\mu$m obtained with SCUBA at the JCMT \citep{Hatchell2005, Hatchell2007a, Hatchell2007b}. \citet{Hatchell2007a} classified the 91 dense cores detected using their Spectral Energy Distribution (SED), resulting in 47 starless cores, 34 Class 0, and 22 Class I protostars. Later, \citet{Hatchell2007b} surveyed the outflow activity in the region to gain a deeper insight into the evolutionary stage of the protostars. In contrast to Taurus, a significant fraction of these protostars forming proto-clusters. 

We have observed 11 cuts along starless cores distributed in Barnard 1, IC~348, L~1448, NGC~1333, and B5. The group of cores in IC~348 and NGC~1333 are close to the star clusters and therefore immersed in a harsh environment. The cuts associated with B5 and L~1448 are located in more quiescent regions. The Barnard 1 dark cloud contains several dense cores at different evolutionary stages of the star formation process. While B1-a and B1-c are known to host Class 0 sources associated with high velocity outflows \citep{Hatchell2007b}, the B1-b core appears to be more quiet, although its western edge is interacting with an outflow that is possibly arising from sources B1-a or B1-d, which are both located 1$'$ SW of B1-b. Recently, the B1-b core has been associated with an extremely young Class 0 object, B1b-S, and a First Hydrostatic Core (FSHC) candidate, B1b-N \citep{Gerin2015, Gerin2017, Fuente2017, Marcelino2018}.

We have performed the fitting of the observed molecular abundances using the grid shown in Table~\ref{Table:grid}. Essentially, all positions are better fitted assuming early time chemistry ($\sim$0.1 Myr) (see Fig.~\ref{Perseus-total}). Similarly to the case of Taurus, the regions at low visual extinction are better fitted with $t$ = 1~Myr. Moreover, there is a large dispersion in the values of $\zeta_{H_2}$ without any clear trend with visual extinction and density (see Fig.~\ref{Perseus-total}). Regarding sulfur depletion, most of the positions are fitted with  D$_{\rm S}$=10. Contrary to Taurus, we find some positions toward which the best fit corresponds to undepleted sulfur. These positions are found in IC~348, NGC~1333, and B5. It should be noticed that IC 348 and NGC 1333 are the nearest regions to star clusters. Moreover, NGC~1333 hosts a large population of Class 0 and I protostars. The high values of the sulfur elemental abundance in these regions is very likely related to the star formation activity in the vicinity. Finally, we also show the behavior of the sulfur depletion as a function of the density and visual extinction in Fig.~\ref{Perseus-total}. We do not detect any clear trend with these parameters. This is not unexpected since the highest density and highest visual extinction regions are located in NGC~1333 which is hosting numerous protostars and bipolar outflows that are modifying the molecular chemistry.

\subsection{Orion}
Orion is the nearest massive star-forming cluster to Earth (e.g., \citealp{Hillenbrand1997, Lada2000,  Muench2002, Dario2012, Robberto2013, Zari2019}) and it is located at a distance of $\sim$428 pc \citep{Zucker2019}. Traditionally, the “Orion nebula” refers to the visible part of the region, the HII region, powered by the ionizing radiation of the Trapezium OB association. However, this nebula is part of a much larger complex, referred to as the Orion Molecular Cloud (OMC), that is formed by two giant molecular clouds: Orion A hosting the M42 nebula, and the more quiescent Orion B (see, e.g., \citealp{Pety2017}).

Orion A has been extensively mapped in molecular lines using single-dish telescopes and large millimeter arrays \citep{Hacar2017a, Goicoechea2019, Nakamura2019, Tanabe2019, Ishii2019, Hacar2020, Kirk2017,Hacar2018,Monsch2018,Suri2019, Kong2019}. Different clouds can be identified within Orion A based on  millimeter, submillimeter, and infrared observations. Orion molecular cloud 1 (OMC~1) was identified as the dense gas directly associated with Orion KL \citep{Wilson1970, Zuckerman1973, Liszt1974}. The molecular cloud associated with the HII region M~43  is referred to as OMC~2 \citep{Gatley1974}. OMC~3 is composed by a series of clumps detected in CO emission that are located about  25$'$ to the north of OMC~1 \citep{Kutner1976} . The $^{13}$CO (J =1$\rightarrow$0) observations by \citet{Bally1987} revealed that all these clouds form the so-called integral-shaped filament (ISF) of molecular gas, itself part of a larger filamentary structure extending from north to south over 4$^\circ$. After that, the SCUBA maps at 450 $\mu$m and 850 $\mu$m presented concentrations of submillimeter continuum emission in the southern part of the ISF, which are now referred to as OMC~4 \citep{Johnstone1999} and OMC~5 \citep{Johnstone2006}. 

We have observed three cuts: Ori-C1 in OMC~3, Ori-C2 in OMC~4, and Ori-C3 in OMC~2. These cuts are selected to avoid the (proto-)stars, probing quiescent environments far from the Orion nebula. We have fitted the detected molecular abundances using the grid shown in Table~\ref{Table:grid}, and the results are shown in Fig.~\ref{Orion-1}. This region presents clear differences relative to Taurus and Perseus. Although the cut in Ori-C1 in OMC~3 is fitted with early time chemistry (t = 0.1 Myr), the cut Ori-C2 in OMC~4 is fitted with a chemical age of t =10 Myr, and the cut Ori-C3 in OMC~2, with t =1 Myr. Also contrary to Taurus and Perseus, we obtain a low value of cosmic-ray molecular hydrogen ionization rate, $\zeta_{H_2}$$<$ 10$^{-16}$ s$^{-1}$, in most positions of the three cuts. This is an unexpected result taking into account that high star formation rate in the Orion A. We remind that the cuts are selected in quiescent regions far from the massive protostar cluster. On the other hand, the higher temperatures and UV flux in Orion could reduce the sensitivity of the studied molecular  abundances to the value of  $\zeta_{H_2}$. Finally, a significant fraction of the observations are best fitted assuming D$_{\rm S}$=1. This is a notorious difference with Taurus, and with most of the clouds in Perseus, and supports the role of the environment on the sulfur elemental abundance. As in previous cases, we have explored possible trends of the sulfur elemental abundance with the gas density and the visual extinction. We do not detect any trend with density (see Fig.~\ref{Orion-1}). However, there is a weak trend with visual extinction where the higher values of D$_{\rm S}$ are found at high visual extinctions. It is reasonable that the visual extinction plays a more important role in this giant molecular cloud, which is bathed by a more intense UV radiation field.

\begin{figure*}
\includegraphics[angle=0,scale=.76]{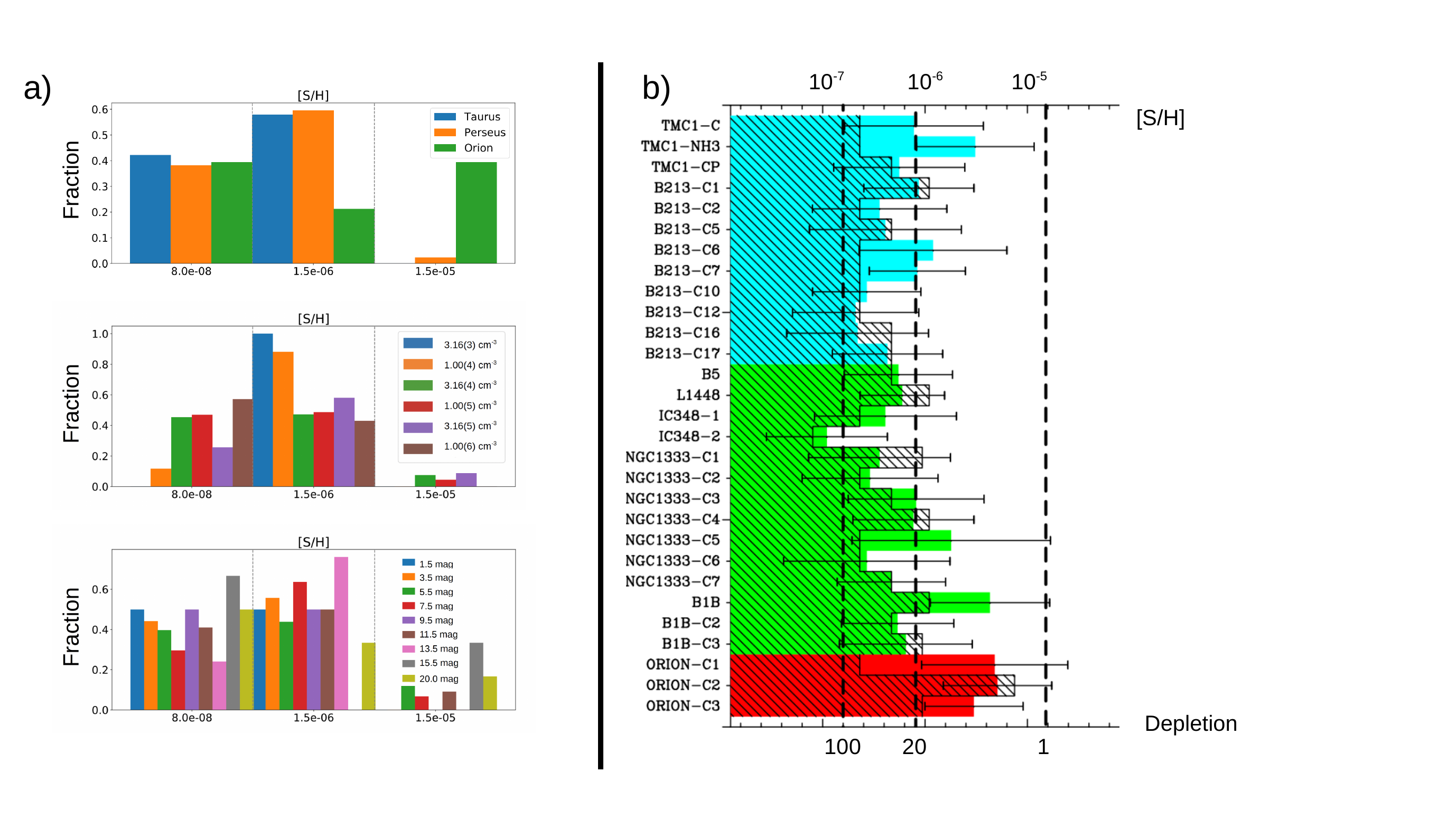}
\caption{Statistics performed with all the positions (Taurus, Perseus, and Orion) of the molecular database. (a) Histograms with the positions ordered by cloud complex (upper panel), in bins of density (middle panel),  and in bins of visual extinction (bottom panel). (b) Mean value of the sulfur elemental abundance derived in each cut. The value of [S/H] toward the high extinction peak is indicated with the dashed area.}
\label{summary}
\end{figure*}

\begin{table*}
\begin{centering}
\caption{Most abundant volatile sulfur-bearing species$^1$}
\label{Table:species}
\begin{tabular}{l | l}
 \hline  \hline
{\bf B213-C12-8 }                                                       &  {\bf B213-C12-1}                                                    \\
n$_{\rm H}$ =  3.16$\times$10$^4$~cm$^{-3}$        &  n$_{\rm H}$ =  3.16$\times$10$^5$~cm$^{-3}$   \\
D$_{\rm S}$  $\sim$ 65                                              &  D$_{\rm S}$  $\sim$  65  \\
97.7 \% of S atoms in refractories                              &  97.7 \% of S atoms refractories \\
1.21 \%of S atoms  in gas                                          &  0.06 \% of S atoms  gas \\
0.32 \% of S atoms in ice                                            &  1.47 \% of S atoms ice \\
{\bf Top-20:}  g-S, m-HS, m-H$_2$S, m-NS, g-CS, m-SO,                          & {\bf Top-20:} m-HS, m-H$_2$S, m-NS, m-SO, m-S, g-S, s-HS,    \\ 
g-SO, m-OCS, g-S$^+$, g-H$_2$S, s-HS, s-H$_2$S, s-NS,                     &  s-H$_2$S, s-NS, g-H$_2$CS, s-SO, g-CS, m-CH$_3$SH, m-OCS,   \\ 
 g-OCS, s-SO, g-H$_2$CS, s-OCS, g-C$_3$S, g-SO$_2$, g-C$_2$S    &  m-CH$_3$S, m-H$_2$CS, g-H$_2$S, s-S, m-NH$_2$CH$_2$SH, m-CH$_2$SH \\
 \hline \hline
{\bf Barnard 1b (+140,0)}                                            &  {\bf Barnard 1b (0,0)}   \\
n$_{\rm H}$ =  3.16$\times$10$^4$~cm$^{-3}$         &  n$_{\rm H}$ =  3.16$\times$10$^5$~cm$^{-3}$   \\
D$_{\rm S}$  $\sim$  8                                               &  D$_{\rm S}$  $\sim$  14  \\
87.3 \% of S atoms  in refractories                             &  92.7 \% of S atoms in refractories \\
9.60 \% of S atoms in gas                                          &  0.22\% of S atoms in gas \\
3.06 \% of S atoms in ice                                           &  7.11\% of S atoms in ice \\
{\bf Top-20:}  g-S, g-CS, m-HS, m-H$_2$S, m-NS,  m-SO,    & {\bf Top-20:} m-HS, m-H$_2$S, m-NS, m-SO, m-S, g-S, s-HS,   \\
 m-OCS, g-S$^+$, g-SO, s-HS,  s-H$_2$S, s-NS,  s-SO,  g-OCS,  &  s-H$_2$S, s-NS, m-CH$_3$SH, m-OCS, m-CH$_3$S, s-SO, g-H$_2$CS, \\
 s-OCS, m-NH$_2$CH$_2$SH, m-SO$_2$, g-C$_3$S, g-H$_2$CS, g-C$_2$S  &  m-NH$_2$CH$_2$SH, m-H$_2$CS,  g-SO, g-CS, g-H$_2$S, m-SO$_2$ \\
\hline \hline
{\bf Ori-C3-11}                                                            &   {\bf Ori-C3-1}        \\
n$_{\rm H}$ =  1.0$\times$10$^5$~cm$^{-3}$         &  n$_{\rm H}$ =  3.16$\times$10$^5$~cm$^{-3}$   \\
D$_{\rm S}$ $\sim$ 2                                                  &  D$_{\rm S}$  $\sim$  16  \\
50.0\% of S atoms in refractories                              & 93.72\% of S atoms in refractories \\
0.29 \% of S atoms in gas                                           &  0.04 \% of S atoms in gas \\
49.71 \% of S atoms in ice                                          &  6.24 \% of S atoms in ice \\
{\bf Top-20:} m-H$_2$S, m-OCS, s-H$_2$S, m-H$_2$S$_3$, m-SO, m-SO$_2$,  & {\bf Top-20:} m-H$_2$S, m-SO, s-H$_2$S, m-HS, m-HSSH, m-NS,  \\
s-OCS, m-HS, s-SO$_2$, s-H$_2$S$_3$, g-S, m-HSSH, m-CS$_2$,                    &  m-CS$_2$, m-HSS, g-S, m-OCS, m-H$_2$S$_3$, m-CH$_3$S, g-H$_2$S, \\
m-HSS, s-SO, g-OCS, m-S, s-HS, g-H$_2$S, s-HSSH                                            &  m-SO$_2$, s-SO, s-HS, m-H$_2$CCS, m-H$_2$C$_3$S, m-S$_8$, s-HSSH\\
\hline \hline
\end{tabular}

\noindent
$^1$ Notation: gas (g-), ice surface (s-) and ice bulk (m-) species.
\end{centering}
\end{table*}

\section{The complete sample: Determining D$_{\rm S}$}
In Sect.~\ref{Sect:results} we have discussed the results of our model fitting considering the three regions, Taurus, Perseus, and Orion, separately. We can also merge the complete set of data in order to obtain an overall view. 
Our first result is that environment is a key parameter to determine sulfur depletion (see Fig.~\ref{summary}a). More than 50\% of the positions in Taurus and Perseus are best fitted with D$_{\rm S}$$\sim$10. Basically, all the other positions need a higher depletion, D$_{\rm S}$$\sim$187. A different behavior is observed in Orion, in which a significant fraction of the positions ($\sim$40\%) can be  fitted with D$_{\rm S}$=1. This suggests that sulfur depletion is dependent on the star formation activity in the environment. It is also interesting to explore possible trends of D$_{\rm S}$ with density and visual extinction. However, we need to put a word of caution in this analysis. Since only $\sim$13\% of the considered positions belong to Orion (see Table~\ref{Table: GEMS sample}), the results are biased to the physical conditions prevailing in Taurus and Perseus. There is a trend of increasing sulfur depletion with density, which suggests a progressive incorporation of sulfur into grains. However, we cannot say that there is a clear correlation between depletion and density. This trend  is dominated by the positions with densities n$_{\rm H}$$\leq$10$^4$~cm$^{-3}$ that present lower values of D$_{\rm S}$.  Since these low-density positions are located in Taurus and Perseus, the statistics is biased. In addition, as discussed in Sect.~ \ref{Sect:exploration}, the uncertainties in the estimated sulfur elemental abundance also increase with density, hindering to establish firm conclusions. No trend is observed in the histograms of D$_{\rm S}$ as a function of visual extinction. As commented above, only in Orion we can find some correlation, and this correlation is missed when merging all positions.

In our first grid, we have only selected three values of D$_{\rm S}$ in order to keep calculation time within reasonable terms. This coarse grid allowed us to consider a wide range of physical parameters but prevented us from determining the value of D$_{\rm S}$ with an accuracy better than a factor of $\sim$10. In a second step, we fixed the values of $t$ and  $\zeta_{H_2}$ according to the results obtained in our first grid, and then ran a second one with finer steps of D$_{\rm S}$=1. In particular, we used $\zeta_{H_2}$ = 10$^{-16}$ s$^{-1}$ for Taurus, $\zeta_{H_2}$ = 5 $\times$ 10$^{-17}$ s$^{-1}$ in Perseus, and $\zeta_{H_2}$ = 10$^{-17}$ s$^{-1}$  for Orion.  The chemical age was fixed to 0.1~Myr in Taurus and Perseus. In the case of Orion where the $t$ is not well determined, we repeated the fitting with an age of 0.1 Myr and 1 Myr, and then selected the chemical age providing the lowest value of $D_{diff}$ for each cut. According to the results, we selected t=0.1~Myr for Ori-C1 (OMC~3), and 1~Myr for Ori-C2 (OMC~4) and Ori-C3 (OMC~2). The value of D$_{\rm S}$ for each cut  was varied from 1 to 187 in multiplicative steps of a factor of
 $\sim$2 (D$_{\rm S}$= 1,2,4,8,16,32,65,187).   

In Table~\ref{Table:summary}, we show a summary of the results thus obtained for the 29 cuts of our sample. The comparison between the 
model-predicted abundances and those derived from observations toward the visual extinction peak in each cut are shown in Fig.~\ref{comparison-1} and \ref{comparison-2}.
As shown in Table~\ref{Table:summary}, the values of $D_{diff}$ vary between 0.1 and 0.5. This means that the abundances of most of the observed molecules are predicted within a factor of 2$-$5. However, there are outliers  with errors of a factor of $\sim$10  toward some positions (see Fig.~\ref{comparison-1} and \ref{comparison-2}).  These outliers are not  necessarily sulfur-bearing species. Indeed we have large errors when fitting the CO, HCN, and HNC abundances in a few locations. These errors cannot be attributed to opacity effects in the case of CO and HCN because, as explained in Sect~\ref{method}, we are using the less abundant isotopologues C$^{18}$O and H$^{13}$CN to estimate their abundances. As commented above, isotopic chemical fractionation might be important for HCN \citep{Loison2020}, but the HCN/H$^{13}$CN does not differ from the assumed value in more than a factor of $\sim$2.  The discrepancy between the observations and model predictions are due to uncertainties in the physical structure and chemistry. We remind that we are using a 0D model that neglects the temperature and density gradients along the line of sight.

Within the group of sulfur-bearing species, we find some systematics in the errors, with the SO abundance being usually over-predicted by the chemical model, while the HCS$^+$ abundance is under-predicted.  This reflects the problem already discussed by \citet{Navarro-Almaida2020} and \citet{Bulut2021} on the difficulty of fitting the abundances of all sulfur-bearing species using the same input parameters. This also explains the discrepancy between the results obtained in this work and previous papers.With a slightly different chemical network, \citet{Vidal2017} concluded that the abundances of sulfur bearing species in TMC~1-CP can be fitted assuming undepleted or low depleted (a factor of $\sim$3) sulfur abundance.  With a chemical network very similar to that used in this paper, \citet{Navarro-Almaida2020} fitted the H$_2$S and emission in B1b and TMC~1 with undepleted sulfur abundance. This fit, however, over-estimated the abundance of CS by factor $\geq$10 along the cores. In a following paper, \citet{Bulut2021} needed to assume a depletion of a factor of 20 to fit the CS and HCS$^+$ abundances in TMC 1-CP, TMC1-NH3, and TMC1-C. In this work, we follow a uniform and systematic method to derive the sulfur depletion in all the GEMS positions. Although our study is limited by the current knowledge of the sulfur chemistry, we are providing a uniform database that allows a reliable comparison between different regions. 
Moreover, we can take advantage of the large number of positions measured in GEMS to obtain average values, and hence minimize possible observational errors.

In Table~\ref{Table:summary} we list the values of [S/H] obtained toward the visual extinction peak, and the mean value of [S/H] along each of the GEMS cuts. These two values intend to be representative of the sulfur elemental abundance in the dense core and the envelope, respectively. This same information is graphically represented in Fig.~\ref{summary}b. The envelopes in all the cuts located Taurus and Perseus, except  TMC1-NH3, and B1b are well reproduced with a depletion of $\geq$20. In Orion, the estimates of [S/H] are compatible with undepleted sulfur within the uncertainties.
The mean value and that toward the extinction peak differ only by a factor of 2$-$3 in most cuts. This is expected behavior taking
into account that our chemical model takes into account the freeze-out of molecules on the grain surfaces in cold and dense regions.
However,  there are some exceptions to this rule: B~213-C6, B~213-C7, TMC~1-NH3, IC~348-C10, NGC1333-C5, B1b, ORI-C1, ORI-C2, and ORI-C3. In these cases, the sulfur elemental abundance in the envelope is a factor of $>$4 higher than that toward the extinction peak. Interestingly, the value of [S/H] toward the extinction peak is similar to those measured in the rest of regions. Although the correspondence  is not perfect, all these cuts are located in regions where molecular chemistry is known to present some peculiarities. The cuts B~213-C6 and B~213-C7 are located in the northern part of the L1495/B~213 region, which is associated with a cluster of (proto-)stars \citep{Hacar2013}. Recent observations of methanol in this region shows that, contrary to the behavior in the more quiescent starless cores in the south, the maximum of the N(CH$_3$OH)/N(C$^{18}$O) ratio is found toward the extinction peak in these cuts \citep{Spezzano2022}. These authors suggest that this behavior is related to the existence of low-mass protostars in the neighborhood. Low-mass stars can accelerate energetic particles, thus increasing the local cosmic-ray flux and the methanol abundance in the densest inner region of the starless core. A high cosmic-ray ionization rate flux has been also proposed to explain the active chemistry in TMC~1-CP and TMC~1-NH3 by several authors \citep{Fuente2019}. Relative to NGC~1333-C5, the cut associated with this core is crossing a cluster of protostars associated with energetic bipolar outflows \citep{Hatchell2007b, Hatchell2009}. 
A different explanation  needs to be found for the cuts studied in Orion since our  fittings suggest that the cosmic-ray ionization rate is low in these regions.
The values derived in Orion are consistent with the scenario of sulfur being undepleted in the envelope of this giant molecular cloud. \citet{Goicoechea2021a} determined that sulfur is undepleted in the photon-dominated surface of OMC1, with a gas-phase carbon to sulfur abundance ratio of $\sim$10. Observations of the 158$\mu$m line of C$^+$  shows that this giant molecular cloud is surrounded by a partially ionized envelope  that extends to the north (OMC~3) and south (OMC~4),  probing that a high UV flux is still found in these suburb regions \citep{Pabst2020}, in line with our results. 

In Table~\ref{Table:species}, we list the 20 most abundant sulfur-bearing species, in order of decreasing abundance, toward three regions that are representative of Taurus, Perseus, and Orion. In regions with n$_{\rm H}$ $>$ 10$^5$ cm$^{-3}$, less than 1 \% of sulfur is in the gas where the most abundant species is atomic sulfur with a fractional abundance $\sim$4$-$10 times larger than the rest of the gaseous compounds.  Contrary to carbon and oxygen, sulfur remains ionized until visual extinctions of $\sim$4 mag. Atomic sulfur is then formed by recombination of S$^+$ and/or dissociative recombinations of HCS$^+$ and CS$^+.$ CS is the most abundant sulfur-bearing molecule at early time but does not become an important sulfur reservoir  as it is CO for carbon. This is because the hydrogenation of CS followed by dissociative recombination of HCS$^+$ produces much more S + CH than H + CS (see discussion in \citealp{Vidal2017}). At later times, in dense clouds, SO is expected to become the most abundant molecule. The amount of SO hence depends a lot on the amount of O$_2$ and OH in molecular clouds. If oxygen is heavily depleted, atomic sulfur will be the most dominant sulfur bearing species even at late times \citep{Fuente2016}.
In the ice, m-H$_2$S, m-HS and m-S (m- indicates ice bulk molecule) are the most abundant ones. A wealth of organo-sulfur species are also found in solid phase as proposed by \citet{Laas2019} with lower abundances. At later times, $\sim$1~Myr, the allotrope m-S$_8$ can form in the ice. As commented above, this stable allotrope can be considered as semi-refractory material. However, current chemical models predict the formation of S$_8$ mainly in dense regions and with  very low abundances, clearly
insufficient to explain the observed sulfur depletion. In the following section, we propose an alternative scenario to efficiently produce allotropes
in the interface between the diffuse and translucent phase where D$_S$$\sim$10 is already observed.

\begin{figure*}
\includegraphics[angle=0,scale=.25]{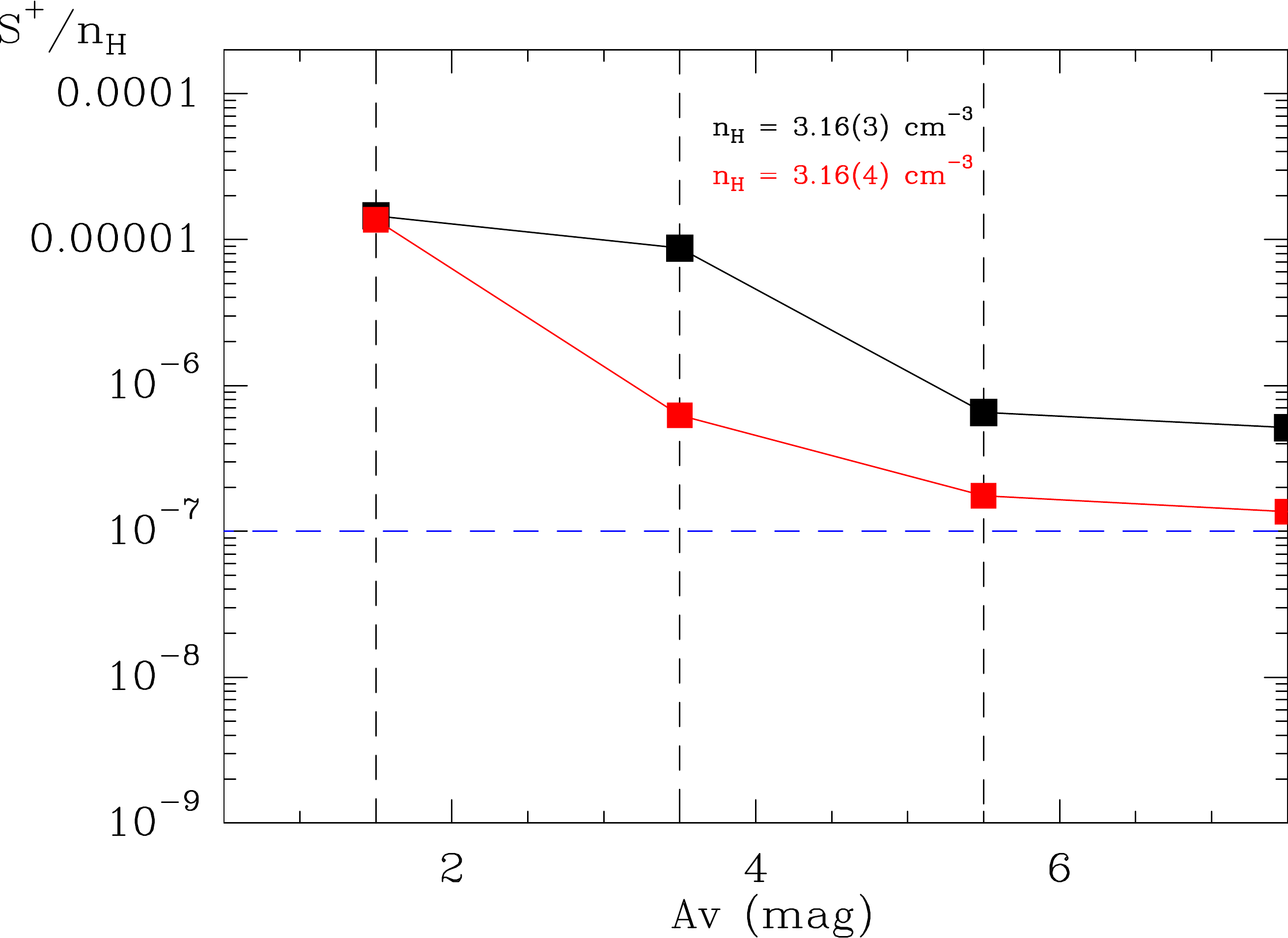}
\includegraphics[angle=0,scale=.19]{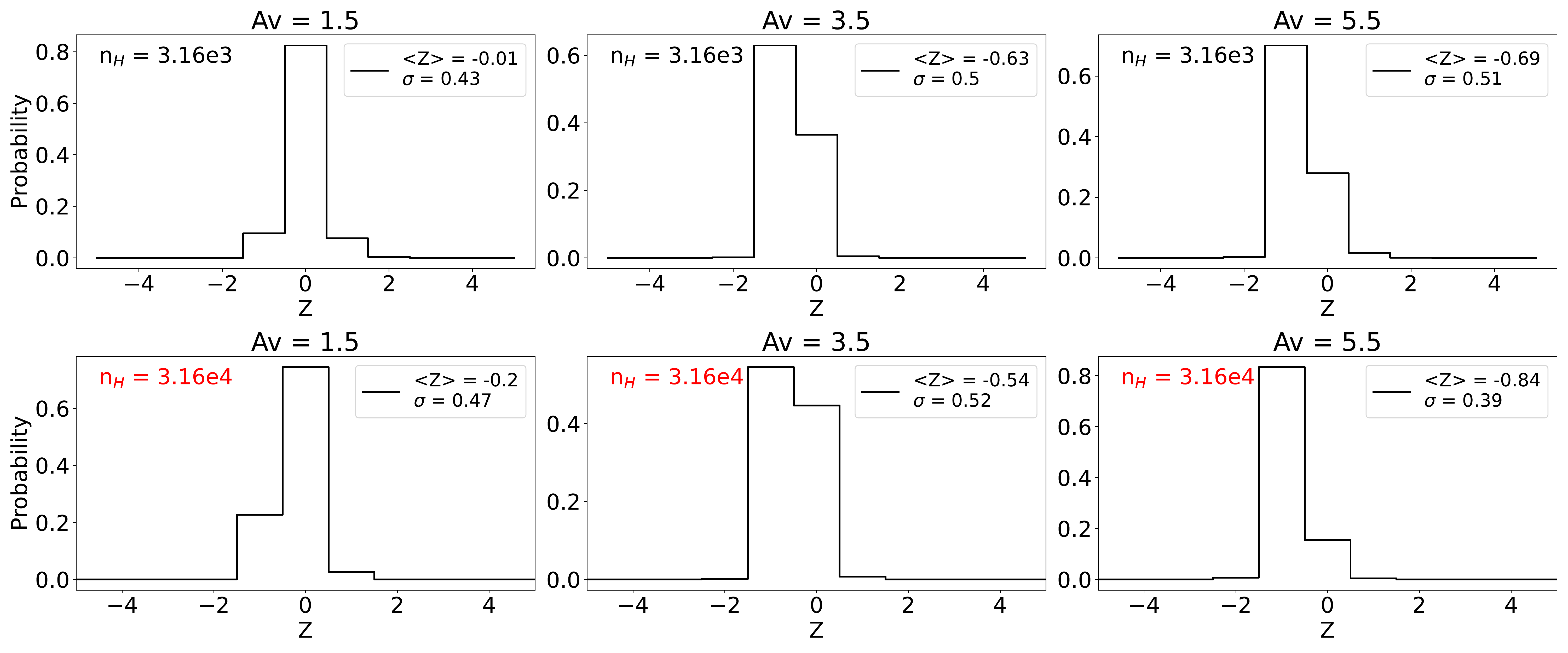}
\caption{Comparison between the spatial distribution of grain charges and sulfur cation in a molecular cloud with similar physical conditions to Taurus. {\it Left:} Abundance of S$^+$ as a function of the visual extinction assuming the physical parameters derived in this work for Taurus ($\chi_{UV}$ = 5, $\zeta_{H_2}$ = 10$^{-16}$ s$^{-1}$, T = 15~K, t = 0.1~Myr). Vertical lines indicate A$_V$ = 1.5, 3.5, and 5.5 mag. The horizontal blue line shows X(S$^+$)=10$^{-7}$. Two values of density are considered, n$_{\rm H}$ = 3.16$\times$10$^3$  cm$^{-3}$ (black) and n$_{\rm H}$ = 3.16$\times$10$^4$  cm$^{-3}$ (red), which are representative of  the translucent and dense phase \citep{Fuente2019}. {\it Right:} Grain charge distribution for silicates with a size of 0.1 $\mu$m  at A$_V$ = 1.5, 3.5, and 5.5 mag assuming the physical conditions described in the left panel. In these calculations we have assumed undepleted sulfur.}
\label{charge-Taurus}
\end{figure*}

\section{Discussion:  The influence of grain charge distribution in sulfur depletion}

In these dense and cold regions, the sum of the observed gas-phase abundances of S-species (the most abundant are SO, SO$_2$, H$_2$S, CS, HCS$^+$, H$_2$CS, C$_2$S and  C$_3$S) only accounts for $<$1\% of the cosmic sulfur abundance \citep{Vastel2018,Fuente2019}.  One could think that most of the sulfur is locked on the icy grain mantles, but surprisingly a similar trend is encountered within the solid phase, where s-OCS \citep{Geballe1985, Palumbo1995} and s-SO$_2$ \citep{Boogert1997} are the  only sulfur-bearing species detected thus far, and only upper limits to the  s-H$_2$S abundance have been derived \citep{Smith1991, Jimenez-Escobar2011}. According to these data, the abundances of the observed icy species account for $<$ 5\% of the total expected sulfur abundance. 
This means that 90\% $-$ 95\% of the sulfur is missing in our counting. It has been suggested that this so-called depleted sulfur may be locked in hitherto undetected reservoirs in gas and icy grain mantles, or as refractory material. In particular, laboratory experiments and theoretical work show that sulfur allotropes, such as S$_8$, could be an important reservoir \citep{Wakelam2004, Jimenez-Escobar2012, Shingledecker2020, Cazaux2022}. The sublimation temperature of this allotrope is $>$ 500 K and can be considered as semi-refractory material.

We have determined the sulfur elemental abundance, that is the amount of sulfur atoms in volatiles in starless cores located in different environments.
An updated chemical network \citep{Laas2019, Navarro-Almaida2020, Bulut2021} has been used to determine the sulfur elemental abundance in the 244 positions forming the GEMS molecular database. We find that sulfur depletion depends on the star formation activity in the neighborhood. While sulfur depletions of a factor of 10$-$20 are needed to explain the observations in  Taurus and Perseus, the abundances of the sulfur-bearing species in the outer parts of Orion are well explained assuming the cosmic value. Within Taurus and Perseus, we detect an additional trend with sulfur depletion increasing with density. We would like to reming that these conclusions are based on the state-of-the-art knowledge of the sulfur chemistry. So what we have determined is the fraction of sulfur that does not participate to the volatile chemistry as we know it nowadays. Still, there could  be volatile species missing from our models and locking the major fraction of sulfur. In the following, we discuss a possible explanation for the lack of sulfur atoms in volatiles, within the state-of-the-art knowledge.

One possibility is that sulfur atoms adsorb on the grain surfaces and form large allotropes in the cloud surface. The evaporation temperature of large allotropes like S$_8$  is of hundreds of K, therefore these compounds would remain on grain surfaces even in the extreme conditions of a hot core. This scenario is supported by recent theoretical Monte Carlo simulations  carried out by \citet{Cazaux2022}, which showed that S$^+$ ions frozen onto grain surfaces could produce large allotropes in a few times 10$^4$ but, in order to promote the formation of allotropes instead of sulfur hydrides, one needs to assume that the sticking coefficient of S$^+$ is higher than that of H.
Based on the treatment developed by \citet{Umebayashi1980} and \citet{Draine1987} for collisions with charged grains,  \citet{Ruffle1999} proposed that the sticking coefficient of positive ions (C$^+$, S$^+$, Na$^+$,...) increases as $A=(1+167/T)$  relative to that of neutrals ($A=1$) in regions where the grains are negatively charged. Contrary to other elements like C$^+$, sulfur would remain ionized in the translucent medium until A$_V$$\sim$ 4$-$7 mag where the grains are expected to be negatively charged, thus increasing its adsorption to the grain surface. With this same approach to describe the adsorption of ions onto grain surfaces, \citet{Laas2019} and \citet{Shingledecker2020} were able to explain gas-phase abundances of most compounds in dark clouds, and predicted that organo-sulfur compounds and S$_8$ could be an important sulfur reservoir in solid phase.
This interpretation is, however, controversial since the sticking efficiency  of positively charged atoms on a negatively charged grain is not fully understood, yet. It would depend on the details of the recombination process  and the possibility of chemisorption of the neutral atom. \citet{Watson1972} concluded for C$^+$ that the rate of sticking could be as much as $\sim$4 times larger than for neutral species in the most favorable case, but also much smaller, $<$1, if the energy liberated during the electron recombination is used for the neutral atom to leave the grain surface. The case of sulfur should be more favorable for accretion than carbon since the energy released during the recombination would be smaller. In the following, we make some simple calculations to test whether there is some correlation between the grain charge distribution and sulfur depletion in our sample. Although not conclusive, this correlation would support the role of the grain charge in sulfur depletion.

Assuming that the main sulfur depletion is due to the accretion of S$^+$ on the negatively charged grain surfaces, sulfur depletion would depend on the amount of S$^+$ available, the grain charge distribution, and the density in the translucent envelope. We have calculated the abundance of S$^+$ and the grain charge distribution as a function of visual extinction in Taurus and Orion. The S$^+$ abundance has been calculated using our chemical network and the physical parameters derived in this work and undepleted sulfur abundance. For Taurus, we have adopted 
$\zeta_{H_2}$ = 10$^{-16}$ s$^{-1}$, G$_0$=9 in unit of Habing field ($\chi$=5 in unit of Draine field), T=15~K, and two values of density, n$_{\rm H}$=3.16$\times$10$^3$ cm$^{-3}$ and  n$_{\rm H}$=3.16$\times$10$^4$ cm$^{-3}$. These values are representative of the translucent and dense gas phases in TMC 1 \citep{Fuente2019}. 
All the sulfur is in the form of S$^+$ at low visual extinctions,  A$_V$ $\leq$1.5 mag (see Fig.~\ref{charge-Taurus}). At higher visual extinctions, the S$^+$ abundance decreases but remains $>$10$^{-6}$ until A$_V$$\sim$ 3.5$-$5.5 mag. 

\begin{figure*}
\includegraphics[angle=0,scale=.25]{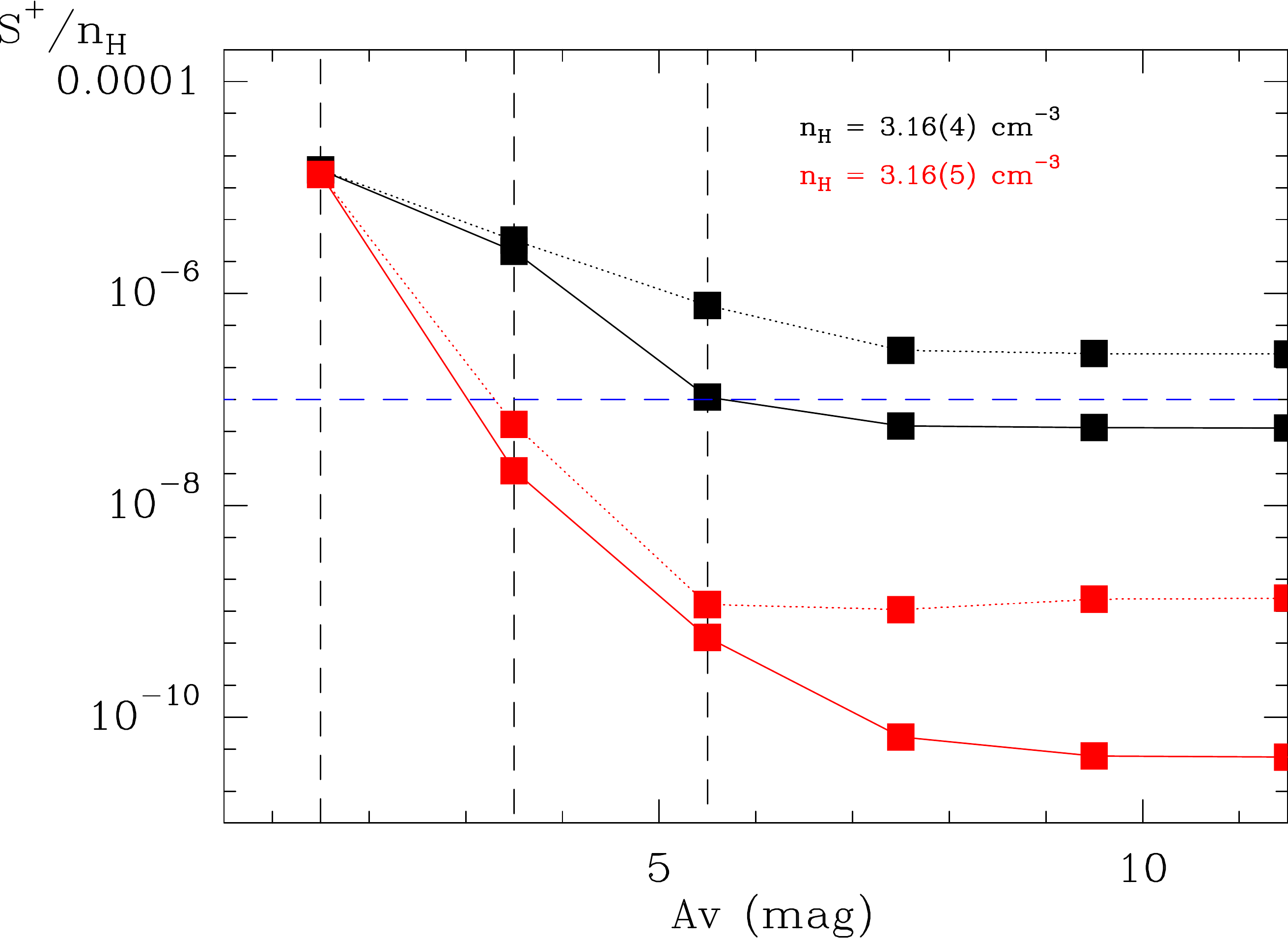}
\includegraphics[angle=0,scale=.19]{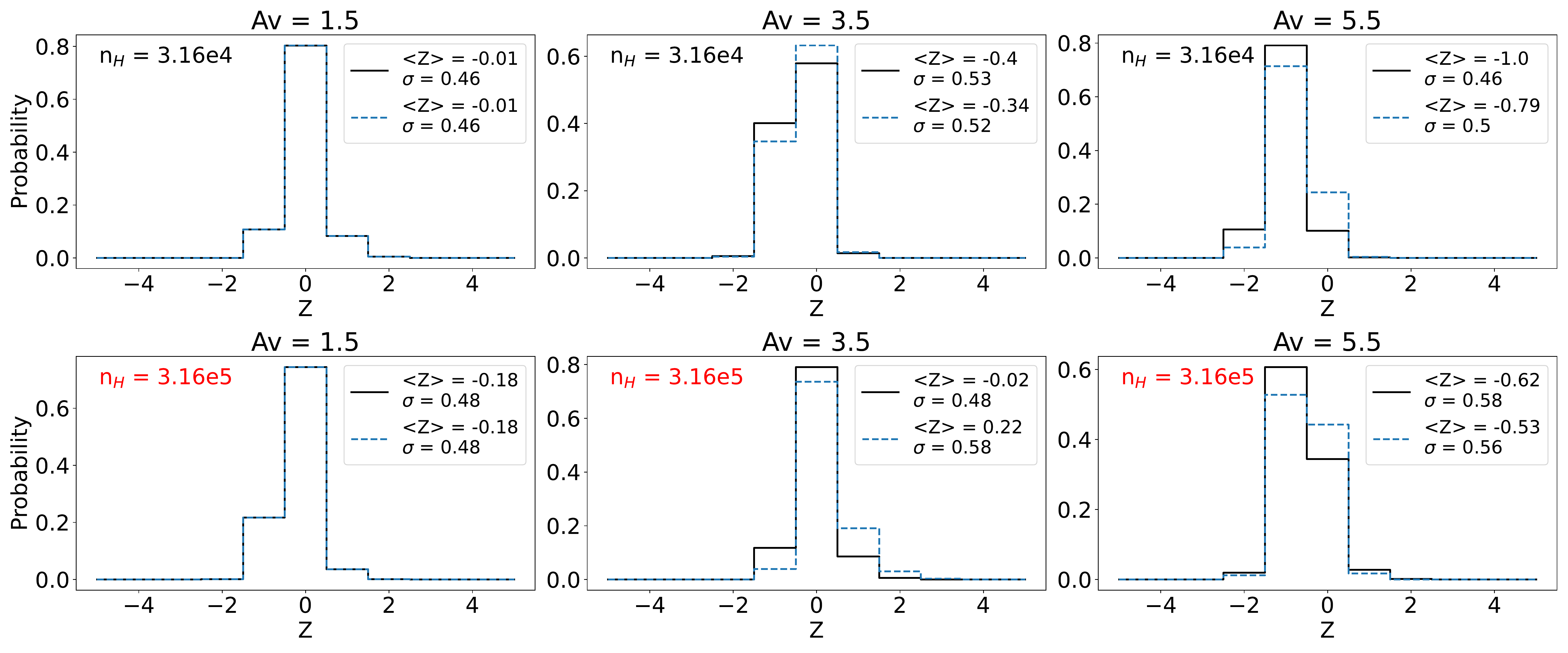}
\caption{Comparison between the spatial distribution of grain charges and sulfur cation in a molecular cloud with similar physical conditions to Orion. {\it Left:} Abundance of S$^+$ as a function of the visual extinction assuming the physical parameters derived for Orion  ($\chi_{UV}$ = 50, T = 25~K) and t = 0.1 Myr. We consider two values of density, n$_{\rm H}$ = 3.16$\times$10$^4$ cm$^{-3}$ (black) and n$_{\rm H}$ = 3.16$\times$10$^5$ cm$^{-3}$ (red). Continuous black and dashed blue lines show the results for $\zeta_{H_2}$ = 10$^{-17}$ s$^{-1}$ and  $\zeta_{H_2}$ = 10$^{-16}$ s$^{-1}$, respectively. Vertical lines indicate A$_V$ = 1.5, 3.5, and 5.5 mag. The horizontal blue line shows X(S$^+$)=10$^{-7}$. {\it Right:} Grain charge distribution for silicates with a size of 0.1 $\mu$m  at A$_V$ = 1.5, 3.5, and 5.5 mag assuming the physical conditions described in the left panel. In these calculations we have assumed undepleted sulfur. }
\label{charge-Orion-1d5}
\end{figure*}

\begin{figure*}
\includegraphics[angle=0,scale=.25]{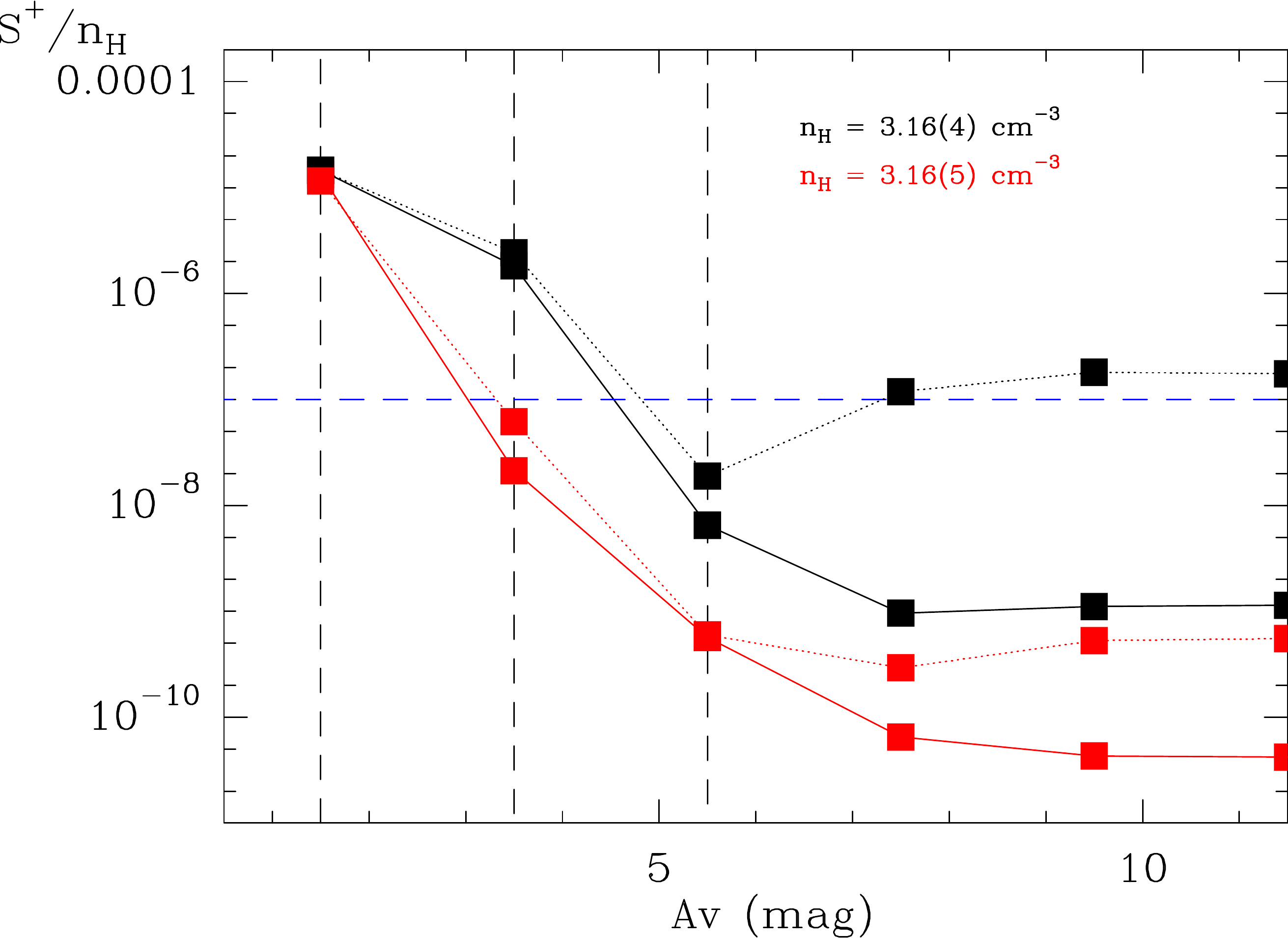}
\includegraphics[angle=0,scale=.19]{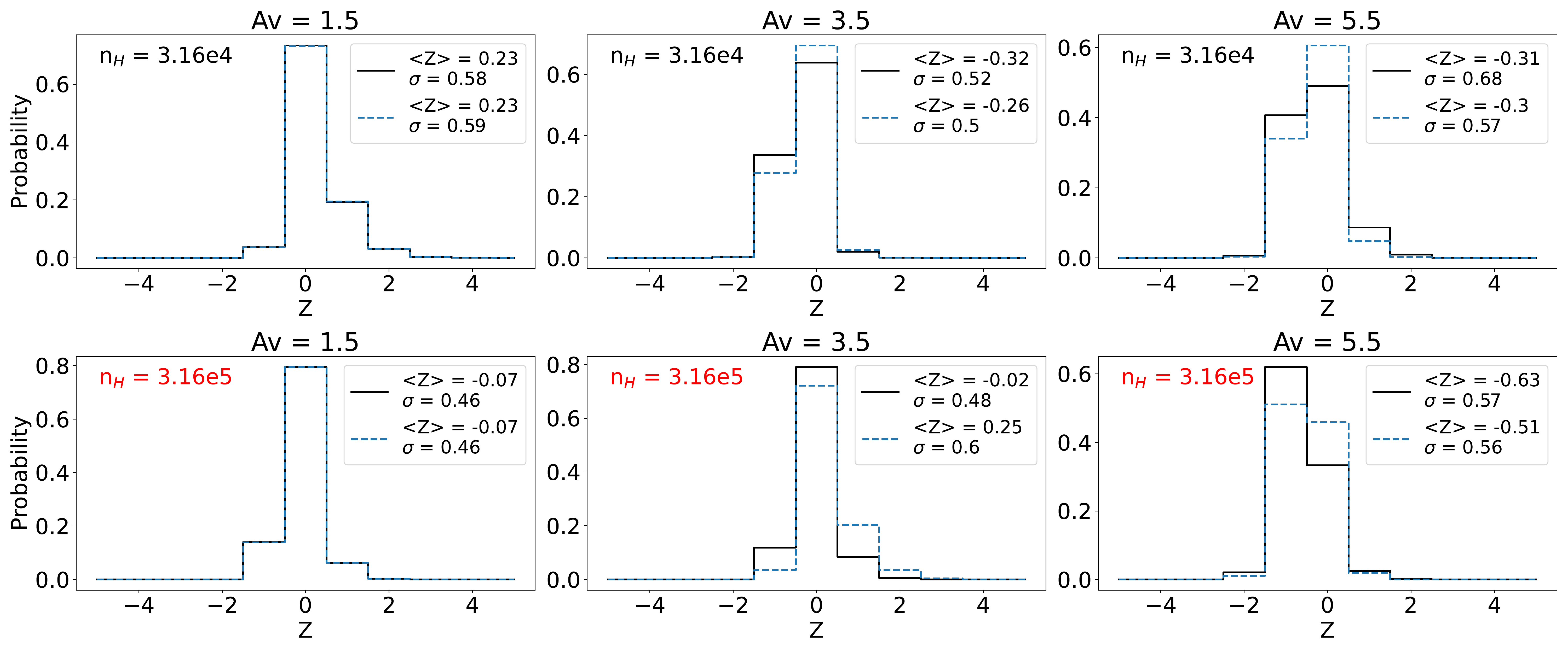}
\caption{Same as Fig.~\ref{charge-Orion-1d5}, but for t=1~Myr.}
\label{charge-Orion-1d6}
\end{figure*}

The case of Orion is more complex, since our fitting revealed different physical conditions in each of the three cuts observed. While the majority of the positions in OMC~4 are fitted with $\zeta_{H_2}$ = 10$^{-17}$ s$^{-1}$,  more than 40\% of the positions in OMC~3 are better fitted with  $\zeta_{H_2}$ = 5 $\times$ 10$^{-17}$ s$^{-1}$ (see Fig.~\ref{Orion-1}). The same problem appears with the chemical age, while early-time chemistry provides the best fit in OMC~3, late-time chemistry better fits the positions in OMC~2 and OMC~4.  It seems clear, however, that the incident UV flux as well as the mean density is higher in all the cuts of this massive
star-forming region \citep{Rodriguez-Baras2021}. Therefore, we fixed the values of  G$_0$=90 ($\chi$=50) and T=25~K,
and considered four sets of physical conditions in our calculations: 1) $\zeta_{H_2}$ = 10$^{-17}$ s$^{-1}$, n$_{\rm H}$=3.16$\times$10$^4$ cm$^{-3}$;  
2) $\zeta_{H_2}$ = 10$^{-17}$ s$^{-1}$, n$_{\rm H}$=3.16$\times$10$^5$ cm$^{-3}$; 
3) $\zeta_{H_2}$ = 10$^{-16}$ s$^{-1}$, n$_{\rm H}$=3.16$\times$10$^4$ cm$^{-3}$; 
and 4) $\zeta_{H_2}$ = 10$^{-16}$ s$^{-1}$, n$_{\rm H}$=3.16$\times$10$^5$ cm$^{-3}$. Moreover, we repeated the calculations for
t = 0.1 Myr  and t = 1 Myr.
Assuming high density, n$_{\rm H}$=3.16$\times$10$^5$ cm$^{-3}$, the S$^+$ abundance is low, X(S$^+$)<10$^{-7}$, even at a low extinction of  A$_V$$\sim$ 3.5 mag, regardless of the other  parameters  (see Fig.~\ref{charge-Orion-1d5} and  Fig.~\ref{charge-Orion-1d6}). In the case of moderate density, n$_{\rm H}$=3.16$\times$10$^4$ cm$^{-3}$ , we obtain high S$^+$ abundance only in the case of  $\zeta_{H_2}$  = 10$^{-16}$ s$^{-1}$ and early-time chemistry. 

The other essential ingredient in our interpretation is the grain charge. The charging of dust grains in the interstellar medium is governed by the equilibrium among different processes.
On the one hand, a dust grain may acquire a positive charge due to photoelectric emission  of electrons induced either by UV radiation \citep{Draine1978} or by fluorescence of H$_2$ \citep{Ivlev2015} depending on the intensity of the local radiation field and on the gas density; additional positive charging of the grain is enhanced by the accretion of free ions in the plasma \citep{Spitzer1941}, although this phenomenon plays a minor role in dense environments where the mobility of the species is drastically reduced. On the other hand, negative grain charging can be produced by the accretion of free electrons in the plasma \citep{Spitzer1941} independently on their origin,
although the contribution of secondary electrons arising from the cosmic-ray population \citep{Draine1979} is of minor importance \citep{Ivlev2015}. For this
work, we have calculated the grain charge distributions in a similar way to \citet{Ibanez2019} for a population of silicate grains of
0.1 $\mu$m; more details on the computation of the distribution can be found in Appendix \ref{app:grain_charging}.
Assuming the same gas densities and radiation field intensities as above, and the abundances for ions and electron density predicted
by our chemical calculations, we obtain the grain charge distributions  shown in Fig.~\ref{charge-Taurus}, Fig.~\ref{charge-Orion-1d5}, and Fig.~\ref{charge-Orion-1d6}. The grain charge distributions have been calculated for A$_V$= 1.5 mag, 3.5 mag, and 5.5 mag.
In all cases, the grain net charge centroid is close to $\sim$0 or positive in the outer part of the molecular cloud (A$_V$ $\sim$ 1.5 mag), and changes to negative at higher extinction. In the case of Taurus, the net charge of grains is negative for A$_V$ $\sim$ 3.5 mag where X(S$^+$)$>$10$^{-6}$, thus promoting sulfur depletion and the formation of sulfur chains as described by \citet{Cazaux2022}.
On the contrary, in Orion the centroid of the grain charge distribution remains close to $\sim$0 until A$_V$=5.5 mag where the abundance of S$^+$ is already very low, especially for n(H$_2$)$\sim$10$^5$~cm$^{-3}$. This low abundance of negatively charged grains  in the cloud envelope could contribute to maintain most of the sulfur atoms in gas phase. These calculations are very simple as long as only one grain size and
composition are taken into account.  One could think that  the inclusion of the grain charges in chemical models would also affect other important species as those formed from C$^+$. However, the situation could be different for carbon. Because of its lower ionization potential, sulfur remains ionized until visual extinctions of $\sim$4 mag where grains are negatively charged under dark cloud conditions. However, carbon would be neutral in this region, thus avoiding an enhanced sticking probability. 
Even though, a more rigorous treatment is necessary to fully disentangle the role of the grain charge in the gas chemistry. 

Finally, we cannot discard the effect that the shocks produced by the expansion of the HII region and those associated with the bipolar outflows of the population of young stellar objects in the region could produce in the sulfur elemental abundance in very active star-forming regions such as Orion A \citep{Gustafsson2003, Colgan2007, Berne2014, Feddersen2020}.  The sulfur depletion is known to be moderate, of the order of  $\sim$10, in the shocks associated with bipolar outflows driven by low-mass stars \citep{Anderson2013, Holdship2019, Feng2020}, still larger than the almost null depletion we have found in outer envelope of Orion A.  Moreover, our cuts have been selected to avoid the jets and molecular outflows that could affect the molecular chemistry hindering the chemical composition of the pristine gas. However, we cannot discard the possibility that previous shocks have released sulfur from the refractory grain cores, that is expelled and turbulently mixed with the molecular gas, thus enriching the sulfur content in the environment. 

\section{Summary and conclusions}
This work uses the GEMS molecular database to derive the sulfur elemental abundance in a wide sample of starless cores located in the nearby star-forming regions Taurus, Perseus, and Orion. These regions have different degrees of star formation activity, and therefore different physical conditions, providing a possibility to explore the effect of environment.  In order to derive the sulfur elemental abundance we have modeled the abundances of 9 species using a state-of-the-art chemical code with an updated sulfur network. In a first  step, in order to explore a wide range  of  physical conditions, we only considered three values of  [S/H] = 1.5$\times$10$^{-5}$ (D$_{\rm S}$=1),
1.5$\times$10$^{-6}$ ((D$_{\rm S}$=10), and 8$\times$10$^{-8}$ ((D$_{\rm S}$=187). Our results can be summarized as follows:

\begin{itemize}
\item Most of the positions in Taurus are best fitted assuming early-time chemistry (t=0.1 Myr). We find a large dispersion in the values of $\zeta_{H_2}$ with a distribution peaking at $\zeta_{H_2}$$\sim$ 1$\times$10$^{-16}$ s$^{-1}$, and without any clear trend with visual extinction and/or molecular hydrogen density. However, we do find a trend with the environment. Most positions in TMC~1 and B~213-N are fitted with  $\zeta_{H_2}$$>$ 10$^{-16}$ s$^{-1}$, while the positions in B~213-S are fitted with  $\zeta_{H_2}$$<$ 10$^{-16}$ s$^{-1}$, consistent with the idea of a harsh environment in B~213-N due to the presence of young (proto-)stars. Regarding sulfur elemental abundance, we find D$_{\rm S}$=10 in TMC~1 and and B~213-N, and D$_{\rm S}$=187 in B~213-S. In addition to environment, we find some correlation of D$_{\rm S}$ with density, with depletion increasing with increasing density.

\item
Similarly to Taurus, essentially most of the positions in Perseus are better fitted with a chemical age of $t$ = 0.1~Myr. There is also large dispersion in the values of $\zeta_{H_2}$ with a distribution peaking at $\zeta_{H_2}$$\sim$ (5-10)$\times$10$^{-17}$ s$^{-1}$, without any clear trend with visual extinction and density. Regarding the sulfur elemental abundance, most of the positions are fitted with D$_{\rm S}$=10. Contrary to Taurus, we find some positions toward which the best fit corresponds to undepleted sulfur abundance. These positions are found toward IC~348, NGC~1333, and B5. IC 348 and NGC 1333 are the nearest regions to star clusters.  NGC~1333 hosts a large population of Class 0 and I protostars. The low values of sulfur depletion in these regions are very likely related to the high local star formation activity.

\item 
We have observed three cuts in the massive star-forming region Orion A: Ori-C1 in OMC~3, Ori-C2 in OMC~4, and Ori-C3 in OMC~2. Although the number of positions observed in Orion A is low, it seems clear that this region presents clear differences relative to Taurus and Perseus. Although the cut  Ori-C1 in OMC~3 is fitted with early time chemistry (t = 0.1 Myr), the cut Ori-C2 in OMC~4 is fitted with a chemical age of $t$ =10 Myr, and the cut Ori-C3 in OMC~2, with $t$ =1 Myr.  We find also  dispersion in the values of $\zeta_{H_2}$. While the majority of the positions in OMC~4 are fitted with $\zeta_{H_2}$ = 10$^{-17}$ s$^{-1}$,  more than 40\% of the positions in OMC~3 are better fitted with  $\zeta_{H_2}$ = 5 $\times$ 10$^{-17}$ s$^{-1}$. Regarding sulfur, a significant fraction of the positions ($\sim$40\%), especially those at low visual extinction, are best fitted assuming undepleted sulfur abundance. 
\end{itemize}

In order to have a deeper insight into the sulfur elemental abundance in the 29 studied starless cores, we have run a model with a finer grid with the values of  [S/H] varying in steps of a factor of 2. In addition, we fix the values of $t$ and $\zeta_{H_2}$ to diminish possible degeneracies. Furthermore, in each cut we considered separately the value of [S/H] derived in the visual extinction peak from that in the envelope. We found sulfur depletion of a factor $>$20  is always needed toward the visual extinction peaks. However, lower values of sulfur depletion are measured in the envelopes of regions with enhanced star formation activity, and especially in Orion A.  

In addition, we have explored the possibility that the observed variations of sulfur depletion is consequence of the influence of the local physical conditions on the abundance of S$^+$ and the grain charges. In the case of Taurus, we find that grains become negatively charged at a visual extinction of A$_V$$\sim$ 3.5 mag. In this region, the abundance of  S$^+$ might be $>$10$^{-6}$, and the electrostatic attraction between S$^+$ and negatively charged grains increases the accretion rate of S$^+$ on dust, and leads to the formation of S chains \citep{Cazaux2022}, which would  enhance sulfur depletion. This could explain that sulfur depletion is already significant, $\sim$20, in the translucent region of Taurus filaments. In the case of Orion, the net charge of grains is close to 0 in the region where the abundance of S$^+$ is high, which could slow down, or even suppress, the depletion of sulfur in the cloud envelope. Therefore, our calculations suggest that grain charge could play an important role to explain the observed differences in the sulfur depletion. However, we have assumed a single composition (silicates) and size (0.1 $\mu$m) for grains, which is a simple scenario. A full distribution of grain sizes and compositions, as well as possible differences of the grain sizes along the cloud, should be taken into account for a more rigorous approach to quantify this effect. 

Summarizing, our data show that the environment is the driving agent of the sulfur depletion in molecular clouds. The mechanisms responsible for this differentiation 
are not known in full detail. The influence of the grain charges on the chemistry, not considered in most chemical models, is very likely one of the causes. We cannot discard that the shocks associated withr massive star formation could erode the grain cores, thus contributing to enhance the sulfur elemental abundance in the hosting molecular cloud. Additional observations of massive star-forming regions using large interferometers would be desirable to have a global and reliable picture of the sulfur chemistry in these intriguing regions.

\begin{acknowledgements}
We thank the Spanish MICIN for funding support from PID2019-106235GB-I00. I.J.-S. acknowledges support from grant No. PID2019-105552RB-C41 by the Spanish Ministry of Science and Innovation/State Agency of Research MCIN/AEI/10.13039/501100011033.  J.R.G. acknowledges Spanish MICIN for support from PID2019-106110GB-I00. VW acknowledges the CNRS program $"$Physique et Chimie du Milieu Interstellaire$"$ (PCMI) co-funded by the Centre National d’Etudes Spatiales (CNES). L.B.-A. acknowledges the receipt of a Margarita Salas postdoctoral fellowship from Universidad Complutense de Madrid (CT31/21), funded by $"$Ministerio de Universidades$"$ with Next Generation EU funds. D.N-A acknowledges the Fundación Ram\'on Areces for funding support through their international postdoc grant program. The authors also thank the anonymous referee for their interesting suggestions.
\end{acknowledgements}

\bibliography{gems}
\appendix
\section{Additional figures and tables}

\begin{table}
\caption{References for the collisional rate coefficients used for the volume density estimates.}
\label{Table:collisional coefficients}
\centering
\begin{tabular}{lll}\\
\hline\hline
\noalign{\smallskip}
Molecule &  & Reference \\
\hline
\noalign{\smallskip}                          
CS & & \citet{Denis2018}   \\
$^{13}$CO & & \citet{Yang2010} \\
C$^{18}$O  & & \citet{Yang2010} \\
HCO$^+$  & & \citet{Yazidi2014}\\
H$^{13}$CO$^+$  & & \citet{Yazidi2014} \\
HC$^{18}$O$^+$  & & \citet{Flower1999} \\
HCS$^+$ && \citet{Flower1999} \\
H$^{13}$CN && \citet{HernandezVera2014, HernandezVera2017} \\
SO & & \citet{Lique2007} \\
$^{34}$SO & & \citet{Lique2007} \\
HNC & & \citet{Dumouchel2011} \\
     & & \citet{ HernandezVera2014} \\
OCS & & \citet{Green1978} \\
o-H$_{2}$S & & \citet{Dagdigian2020} \\
\hline
\noalign{\smallskip}
\end{tabular}
\end{table}

\begin{table*}
\begin{centering}
\caption{Comparison of the mean value of [S/H] in each GEMS cut with that derived toward the visual extinction peak (coordinates in Table~\ref{Table: GEMS sample}). The parameter D$_{diff}$ is the uncertainty as defined in Sect.~\ref{method}. }
\label{Table:summary}
\begin{tabular}{l l| l| c| c| c| c }
\hline\hline
\multicolumn{1}{c}{Region} & \multicolumn{1}{c}{ Cloud} & \multicolumn{1}{c}{ cut}  &    \multicolumn{1}{c}{ Peak} &
 \multicolumn{1}{c}{ D$_{diff}^{\rm peak}$} &
\multicolumn{1}{c}{Mean} & \multicolumn{1}{c}{ D$_{diff}^{\rm mean}$}    \\ \hline \hline 
Taurus &  TMC~1           &  CP           &   2.3$\times$10$^{-7}$   &  0.39     &  7.8$\times$10$^{-7}$ & 0.46 \\
            &                      &  NH3         &   2.3$\times$10$^{-7}$     &  0.16      &  3.1$\times$10$^{-6}$  & 0.33 \\
            &                      &  C             &   4.7$\times$10$^{-7}$    &  0.48      & 5.6$\times$10$^{-7}$  &  0.41  \\
            & B~213-N    &   C1         &  1.1$\times$10$^{-6}$  &  0.14      & 8.7$\times$10$^{-7}$  &  0.29 \\
            &                         &   C2         &   2.3$\times$10$^{-7}$  &  0.34      & 3.6$\times$10$^{-7}$  &  0.43  \\
            &                         &   C5         &   4.7$\times$10$^{-7}$  &  0.31      & 4.1$\times$10$^{-7}$  & 0.55  \\
            &                         &   C6         &   2.3$\times$10$^{-7}$  &  0.27      & 1.2$\times$10$^{-6}$  & 0.52 \\
            &                         &   C7         &  2.3$\times$10$^{-7}$   &  0.25      & 8.4$\times$10$^{-7}$  &  0.22  \\   
                    
            & B~213-S  &   C10      &  2.3$\times$10$^{-7}$    &  0.32     & 2.7$\times$10$^{-7}$   & 0.28 \\
            &                         &   C12      &  2.3$\times$10$^{-7}$    &   0.41    & 2.1$\times$10$^{-7}$  & 0.38 \\
            &                         &   C16      &  4.7$\times$10$^{-7}$    &   0.38    & 2.2$\times$10$^{-7}$  & 0.48  \\            
            &                         &   C17      &  4.7$\times$10$^{-7}$     &   0.14    & 4.3$\times$10$^{-7}$  & 0.29 \\ \hline      
Perseus  &   B5 (\#79)       &    C1      &   4.7$\times$10$^{-7}$  &    0.18  & 5.5$\times$10$^{-7}$   &  0.28 \\
               &  L1448             &    C1      &  1.1$\times$10$^{-6}$   &    0.16  & 6.0$\times$10$^{-7}$   &  0.17 \\
               &  IC348             &    C1      &   2.3$\times$10$^{-7}$  &    0.15  & 4.1$\times$10$^{-7}$  & 0.48   \\
               &                       &    C10     &   8.0$\times$10$^{-8}$  &   0.20  &  1.1$\times$10$^{-7}$  & 0.35  \\            
               &  NGC 1333    &    C1      &  9.4$\times$10$^{-7}$    &   0.48  &  3.6$\times$10$^{-7}$  & 0.48 \\
               &                       &    C2      &  2.3$\times$10$^{-7}$    &  0.28   &  2.9$\times$10$^{-7}$  & 0.44  \\
               &                       &    C3      &  4.7$\times$10$^{-7}$    &  0.16   &  8.2$\times$10$^{-7}$  & 0.44 \\
               &                       &    C4      &  1.1$\times$10$^{-6}$   &   0.09   & 7.7$\times$10$^{-7}$  & 0.35  \\
               &                       &    C5      &  2.3$\times$10$^{-7}$   &  0.23    & 1.8$\times$10$^{-6}$    &  0.94 \\
               &                       &    C6      &  2.3$\times$10$^{-7}$  &  0.23     & 2.7$\times$10$^{-7}$    &  0.66 \\
               &                       &    C7      &  4.7$\times$10$^{-7}$  &  0.28     & 4.7$\times$10$^{-7}$    &  0.28 \\ 
               
               &  B1b               &    C1      &  1.1$\times$10$^{-6}$  &   0.43  &  4.3$\times$10$^{-6}$   &  0.34 \\
               &                       &    C2      &  4.7$\times$10$^{-7}$  &   0.28  &  5.4$\times$10$^{-7}$   &   0.30  \\
              &                        &    C3      &  9.4$\times$10$^{-7}$  &   0.22  &  6.5$\times$10$^{-7}$   &   0.42 \\   \hline
Orion    &        A            &    C1      &  2.3$\times$10$^{-7}$   &  0.51   &   4.8$\times$10$^{-6}$   & 0.51  \\              
             &                      &    C2      &  7.5$\times$10$^{-6}$   &  0.06   &   5.1$\times$10$^{-6}$   & 0.28  \\            
             &                      &    C3      &  9.4$\times$10$^{-7}$   &  0.10   &   3.0$\times$10$^{-6}$   & 0.23   \\                               
\hline \hline
\end{tabular}
\end{centering}
\end{table*}

\begin{figure*}
\includegraphics[angle=0,scale=.7]{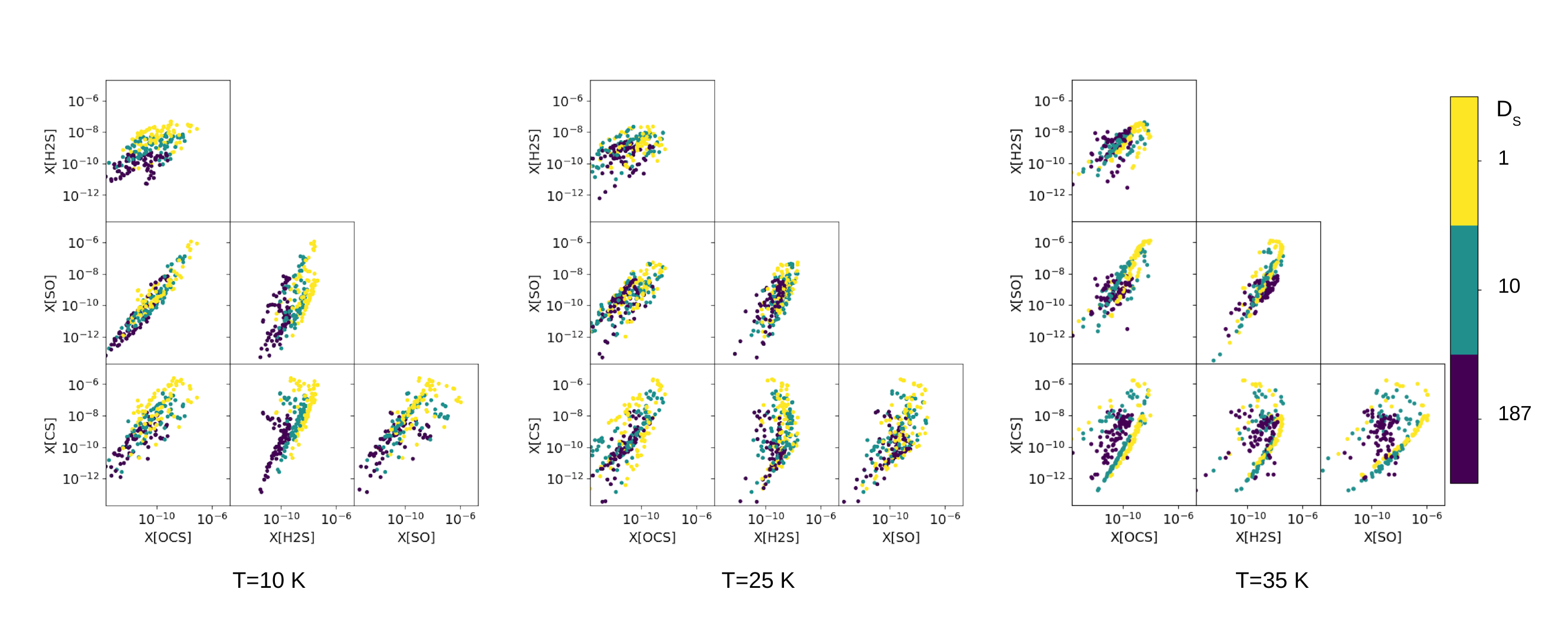}
\caption{Corner diagrams performed by selecting the models with  A$_V$=11.5 mag  from the output of the grid described in Table~\ref{Table:grid} for Taurus. In each panel we show all the models corresponding to a single value of the temperature: $T$ = 10~K (left), 25~K (center), and  35~K (right).}
\label{modelos-Tg-1}
\end{figure*}

\begin{figure*}
\includegraphics[angle=0,scale=.7]{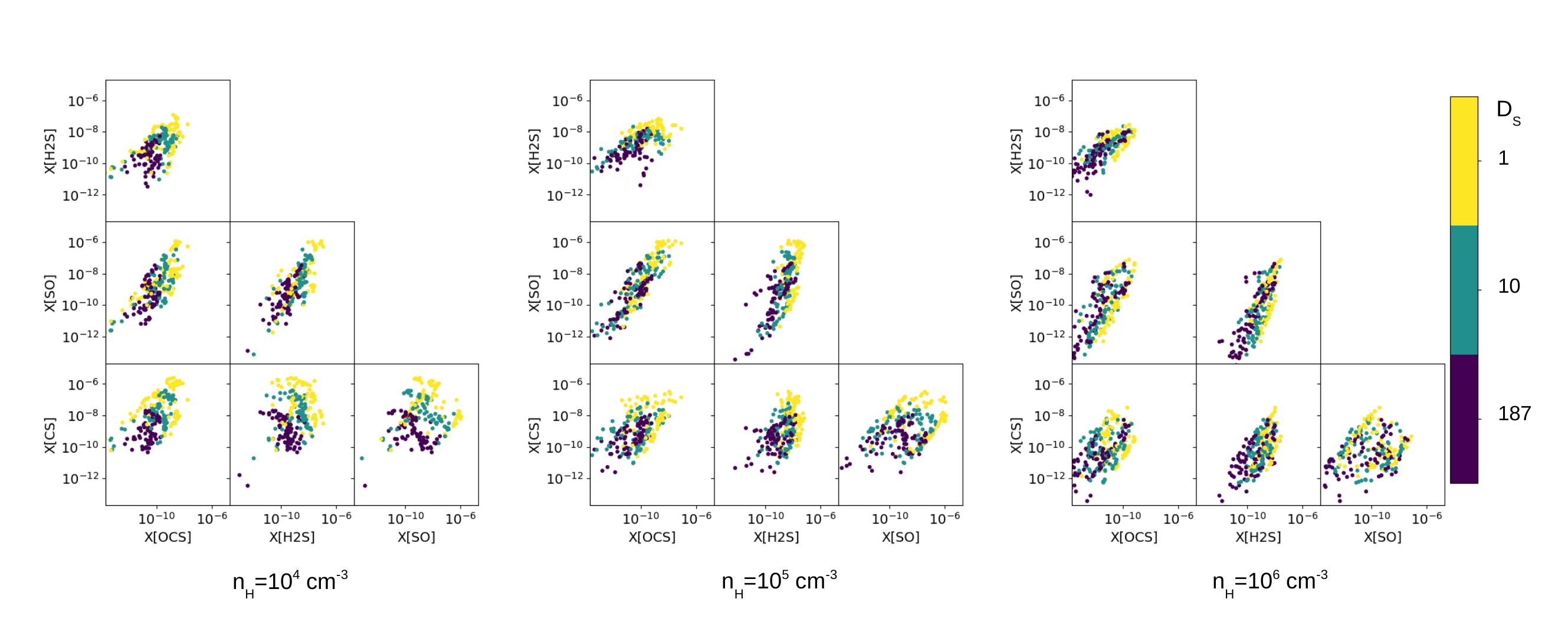}
\caption{Same as Fig.~\ref{modelos-Tg-1}, but in each panel we show all the models sharing the same value of density: $n_{\rm H}$ = 10$^{4}$ cm$^{-3}$ (left), 10$^{5}$ cm$^{-3}$ (center), and 10$^{6}$ cm$^{-3}$ (right).}
\label{modelos-nH}
\end{figure*}

\begin{figure*}
\includegraphics[angle=0,scale=.7]{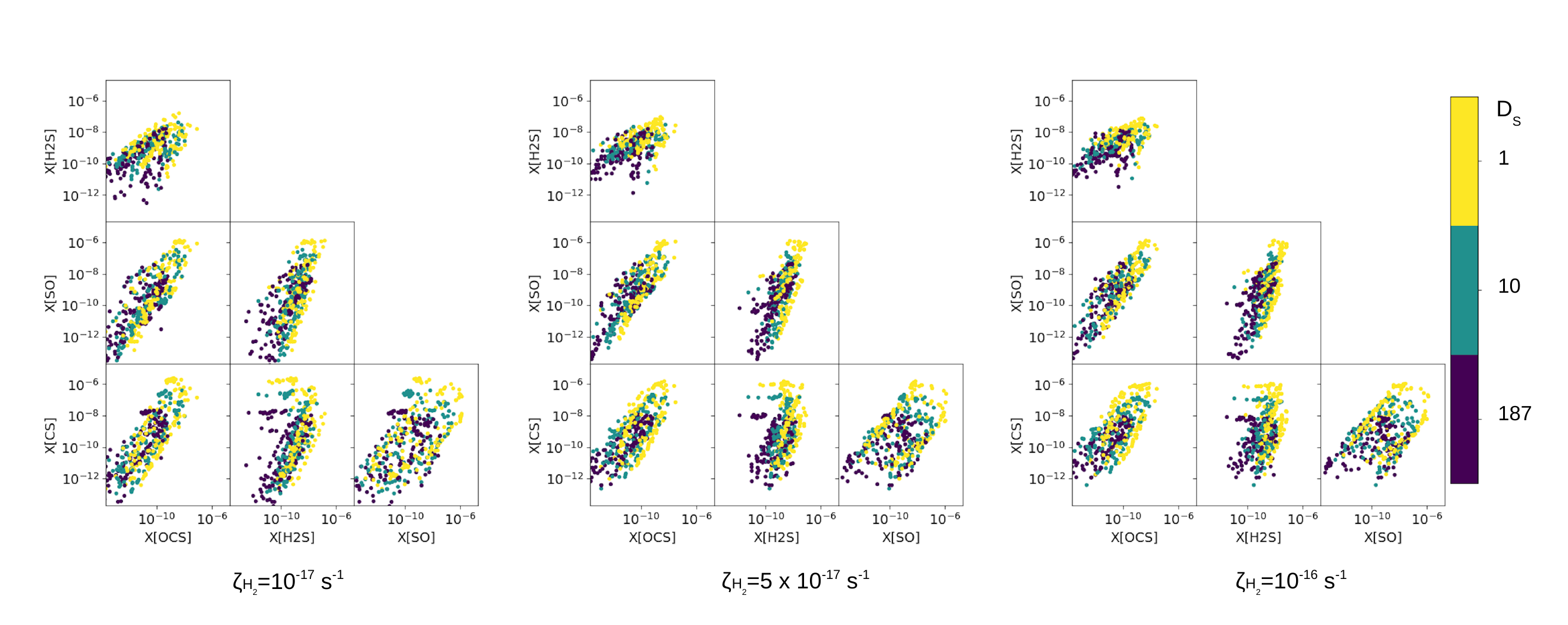}
\caption{Same as Fig.~\ref{modelos-Tg-1}, but in each panel we show all the models sharing the same value of molecular hydrogen cosmic-ray ionization rate:
$\zeta_{H_2}$ = 10$^{-17}$ s$^{-1}$ (left), 5$\times$10$^{-17}$ s$^{-1}$ (center), and 10$^{-16}$ s$^{-1}$ (right).}
\label{modelos-crir}
\end{figure*}

\begin{figure*}
\includegraphics[angle=0,scale=.7]{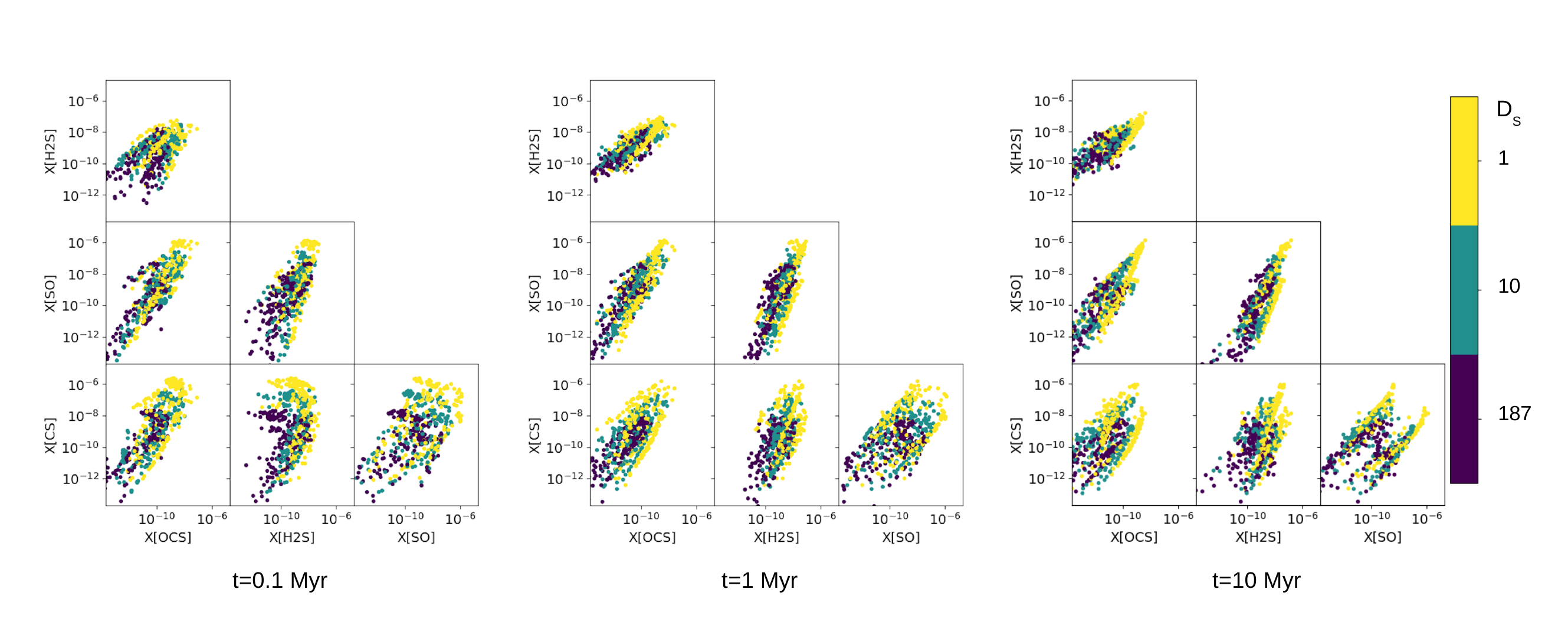}
\caption{Same as Fig.~\ref{modelos-Tg-1}, but in each panel we show all the models sharing the same value of chemical age: $t$= 
0.1~Myr (left), 1~Myr (center), and 10~Myr (right).}
\label{modelos-age}
\end{figure*}

\begin{figure*}
\includegraphics[angle=0,scale=.2]{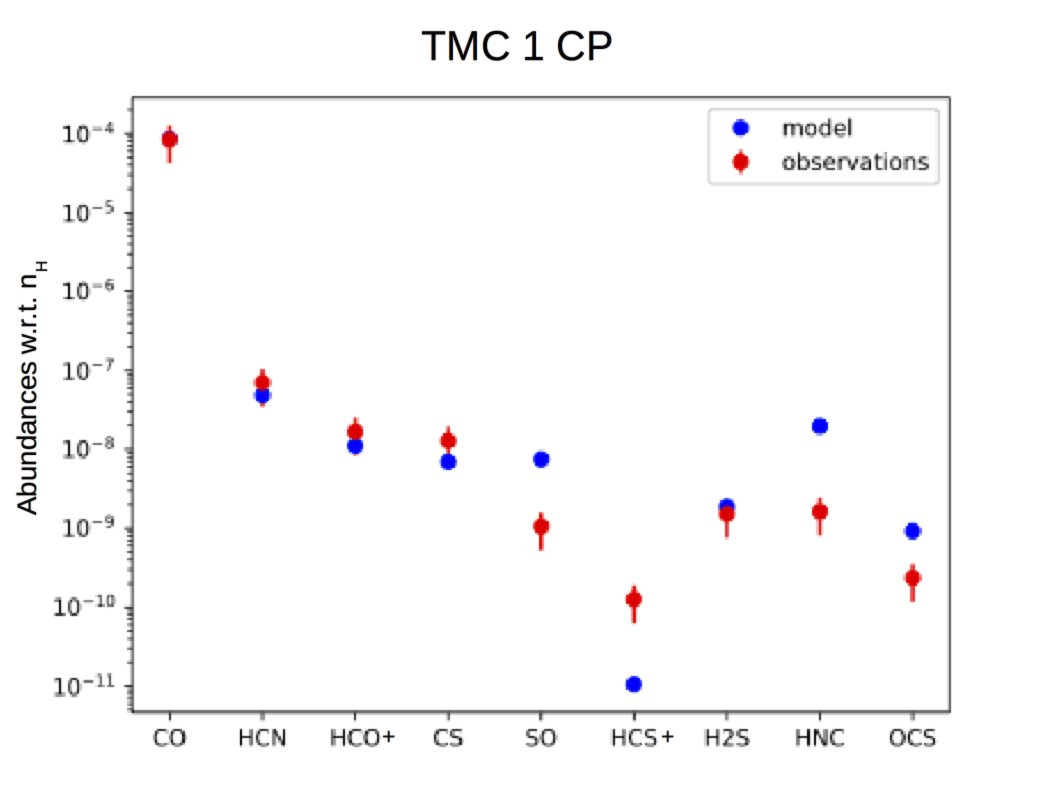}
\includegraphics[angle=0,scale=.2]{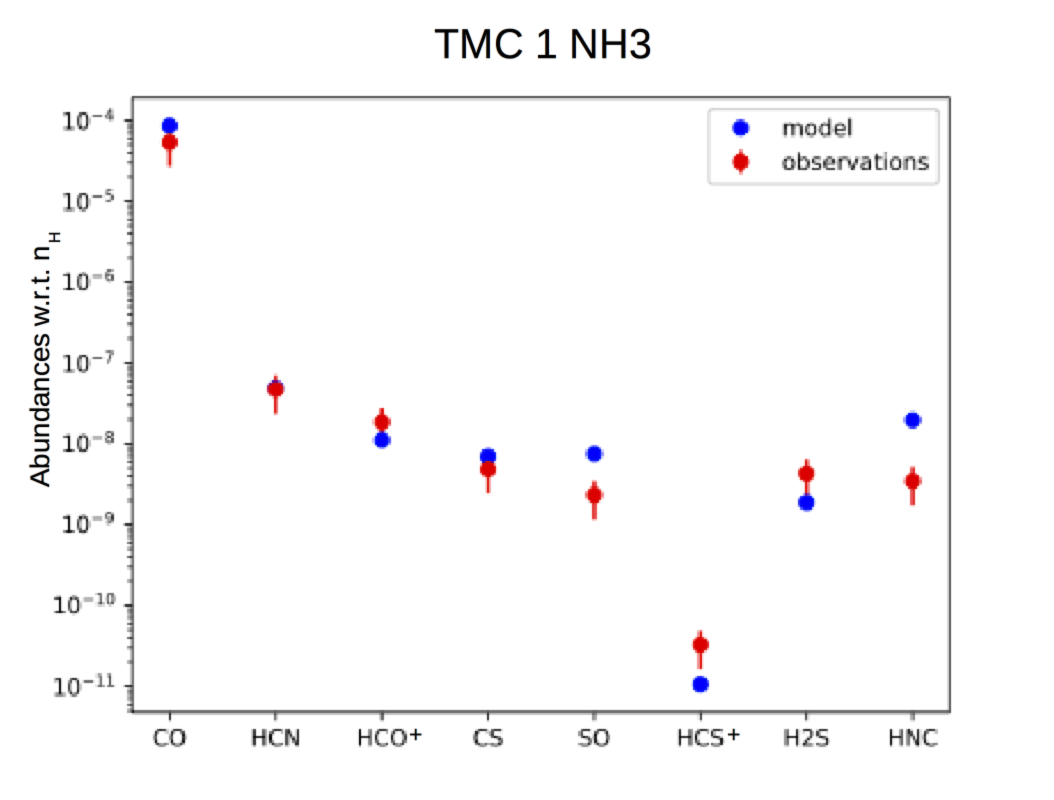}
\includegraphics[angle=0,scale=.2]{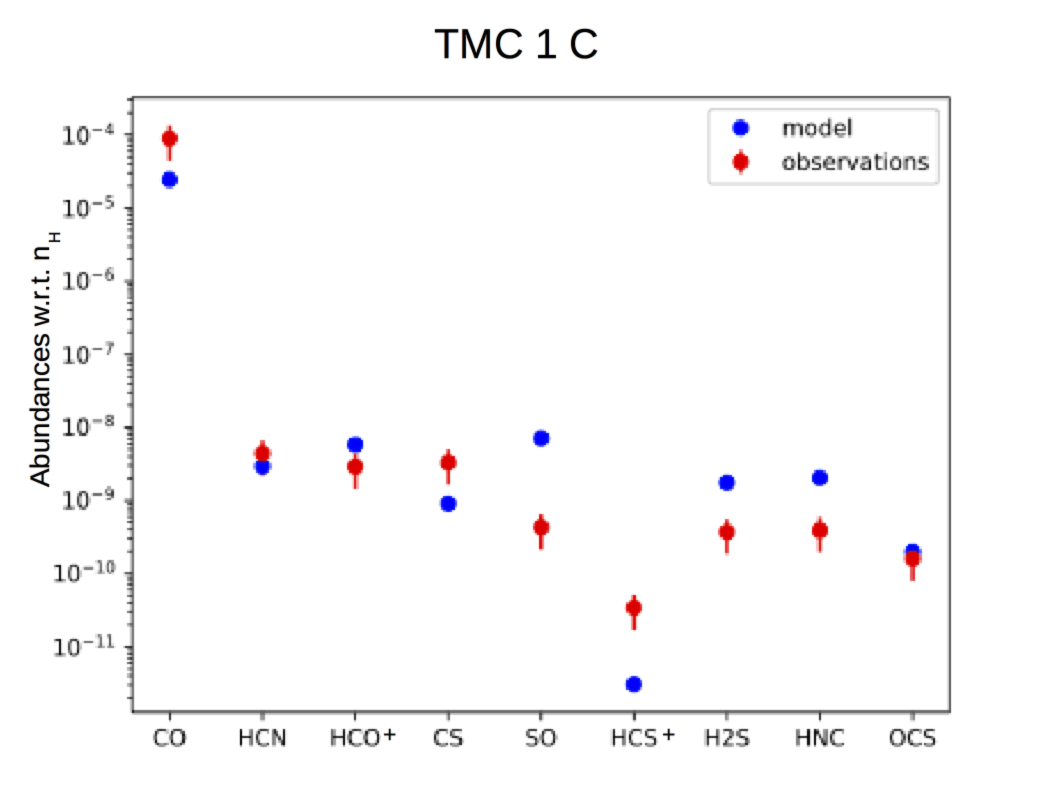}
\includegraphics[angle=0,scale=.2]{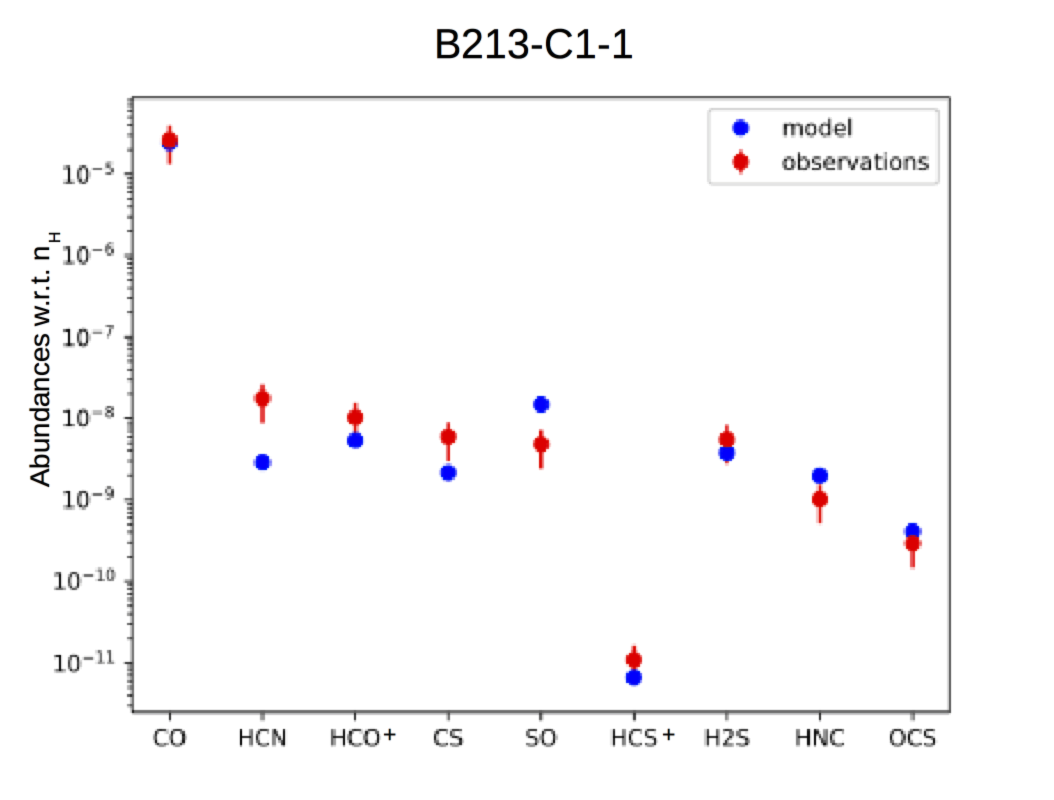}
\includegraphics[angle=0,scale=.2]{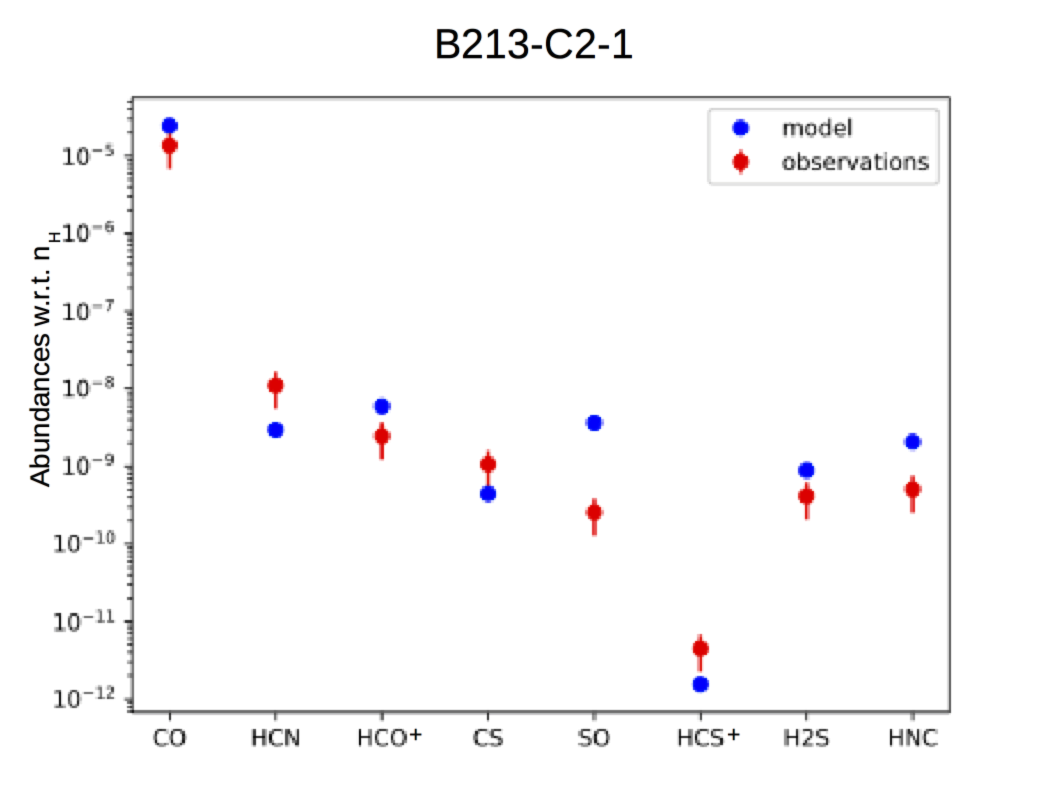}
\includegraphics[angle=0,scale=.2]{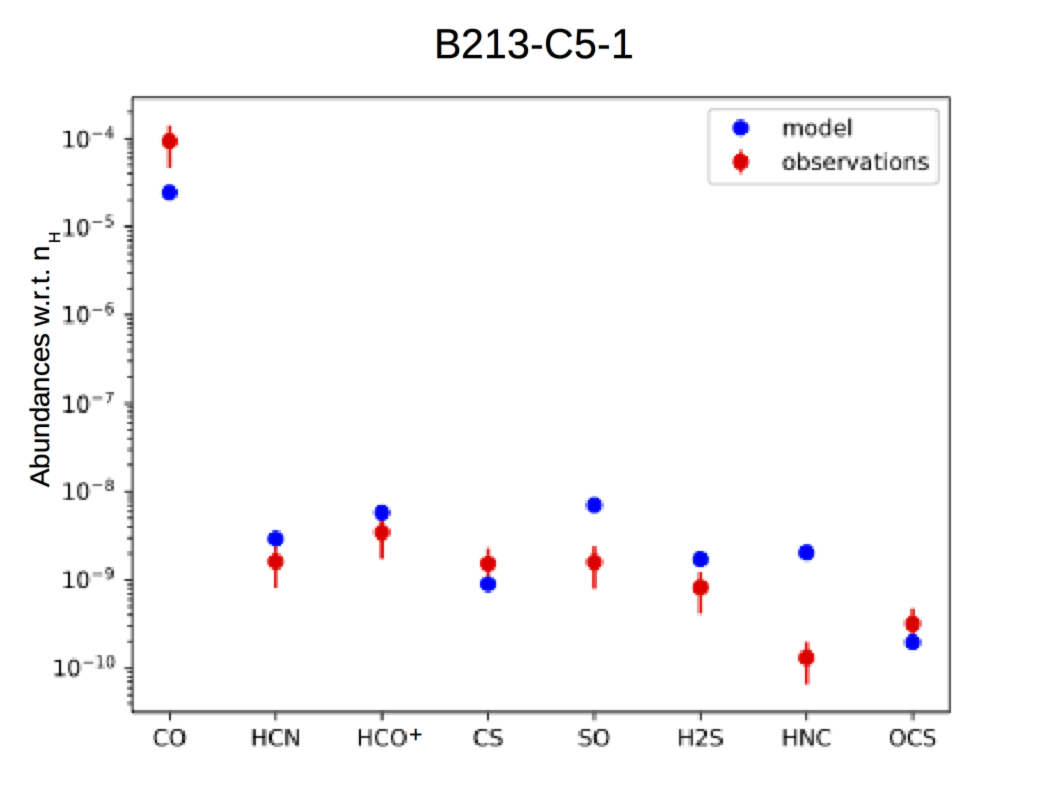}
\includegraphics[angle=0,scale=.2]{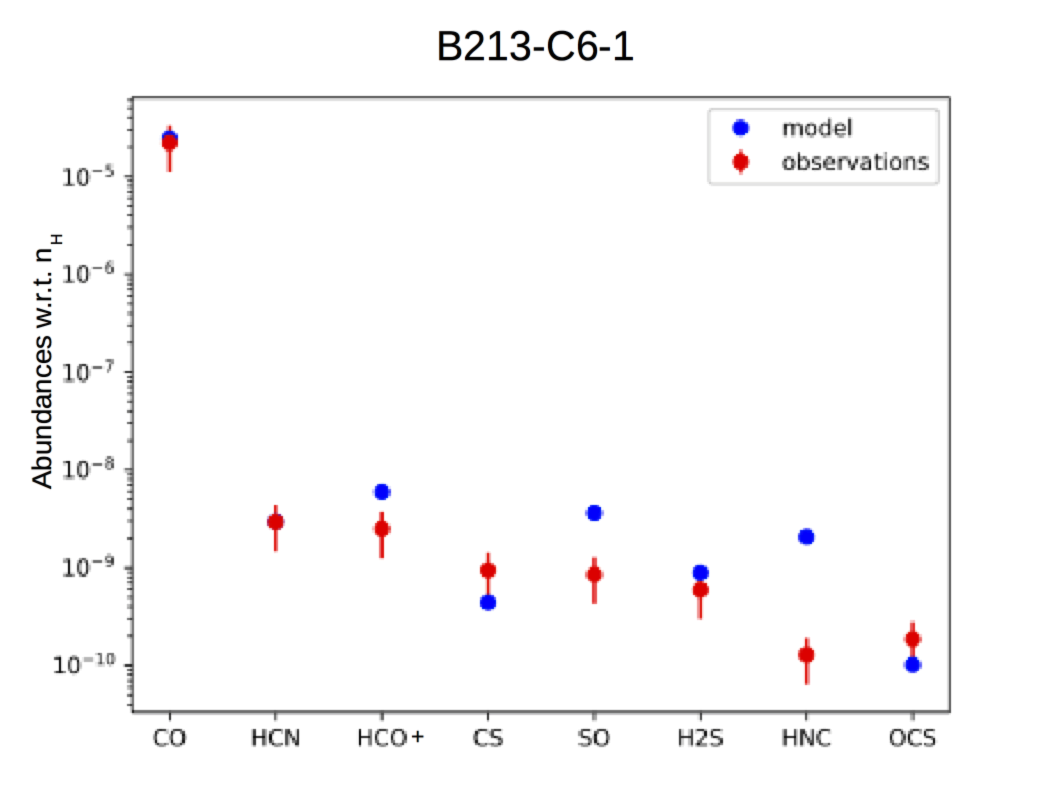}
\includegraphics[angle=0,scale=.2]{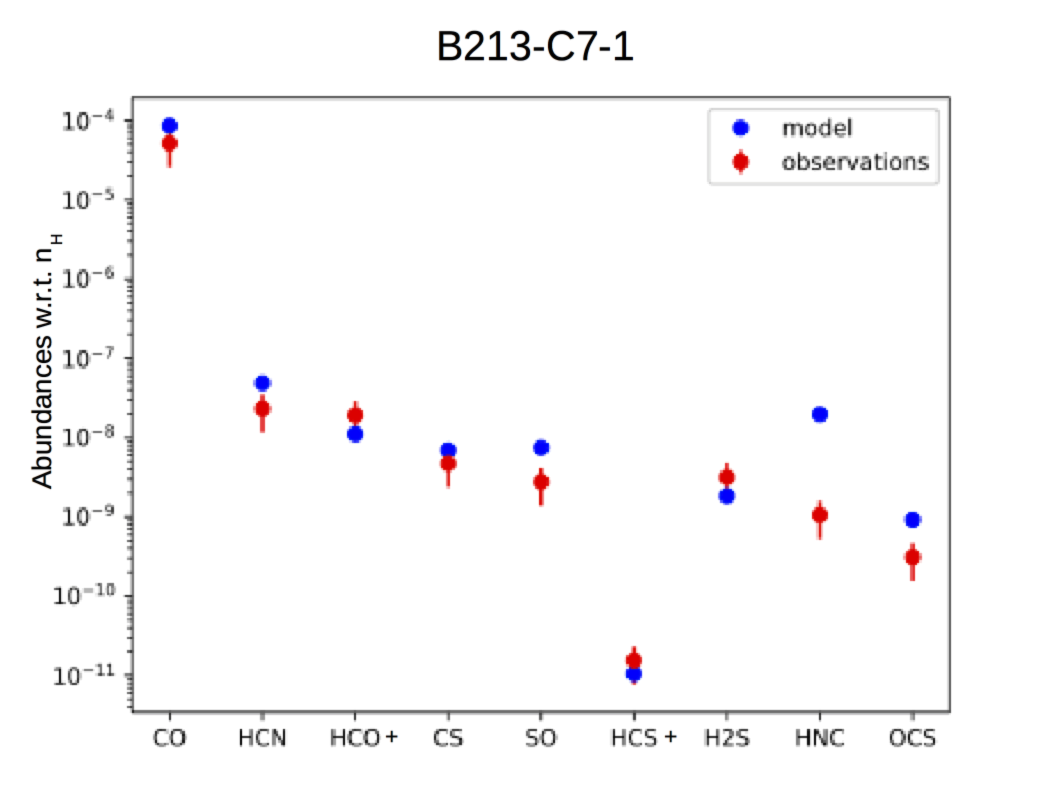}
\includegraphics[angle=0,scale=.2]{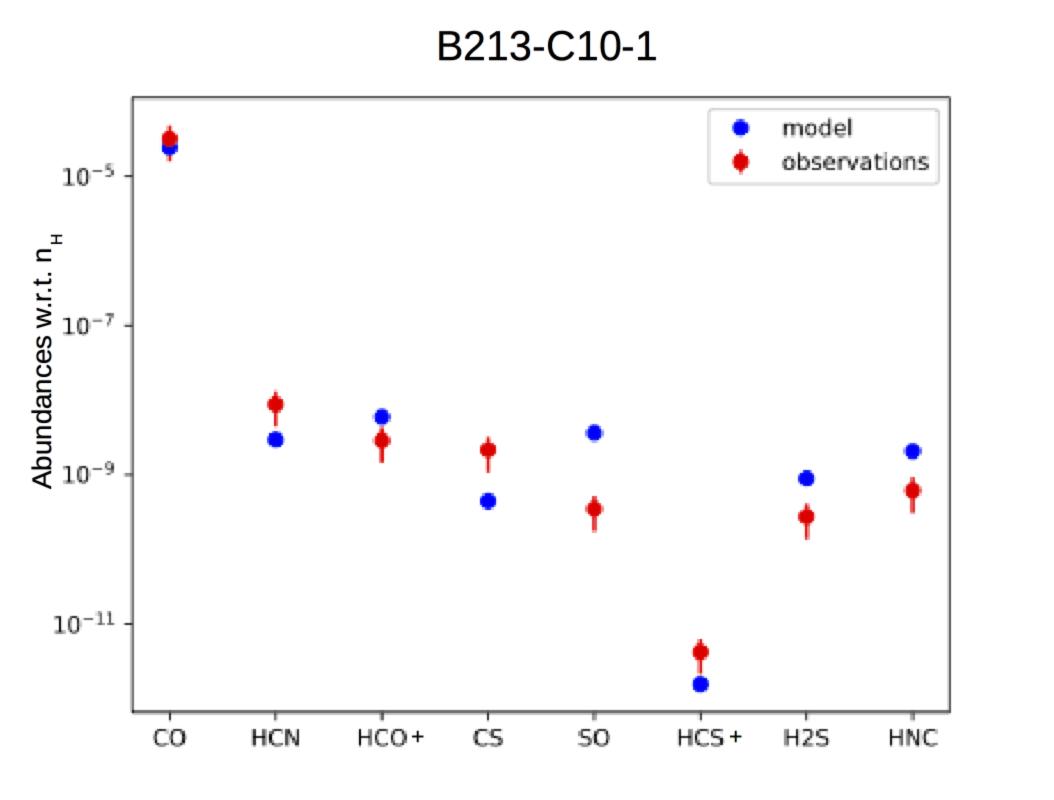}
\includegraphics[angle=0,scale=.2]{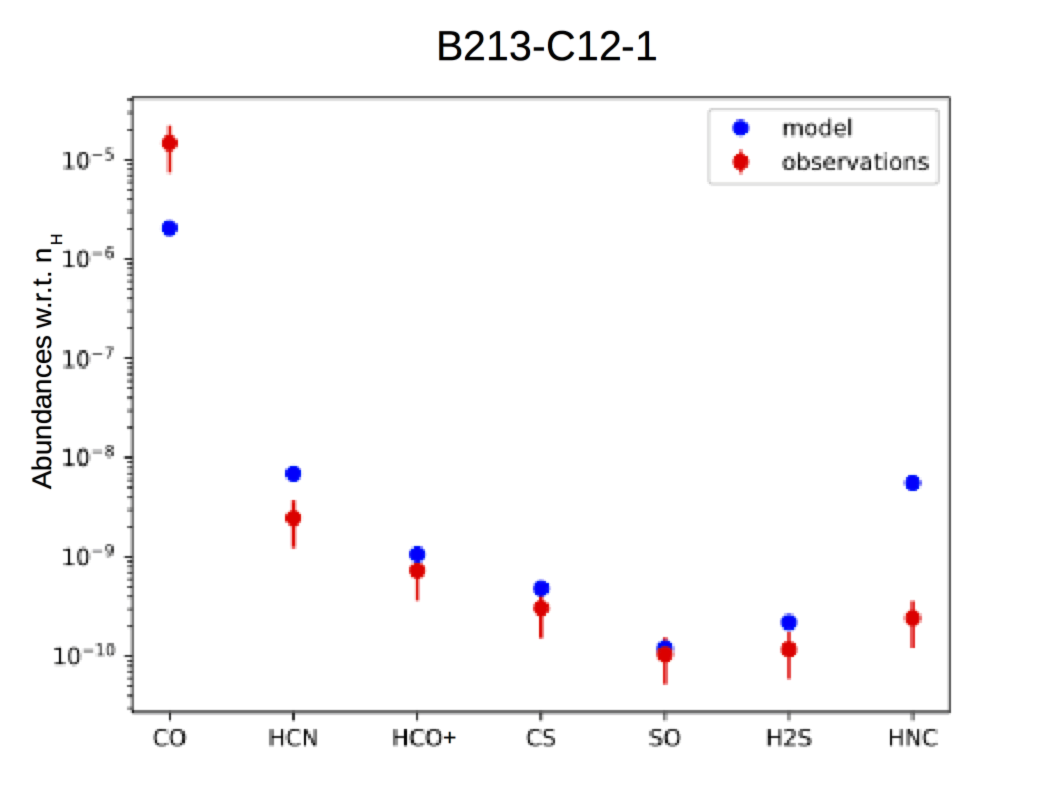}
\includegraphics[angle=0,scale=.2]{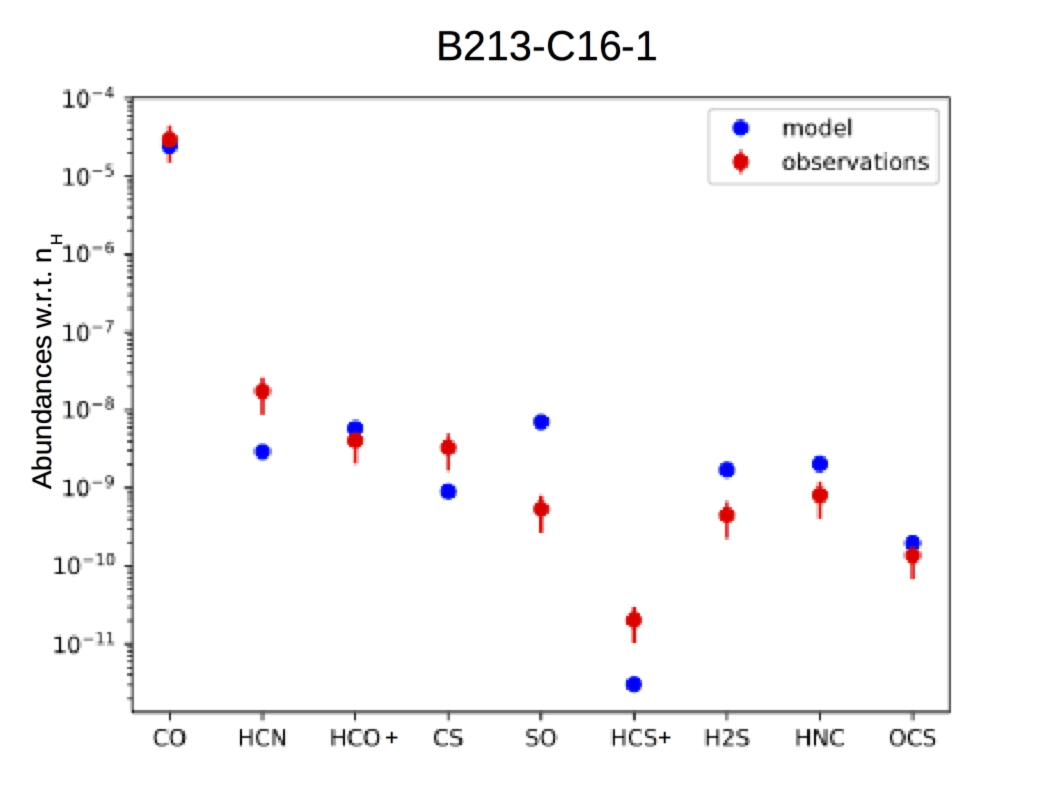}
\includegraphics[angle=0,scale=.2]{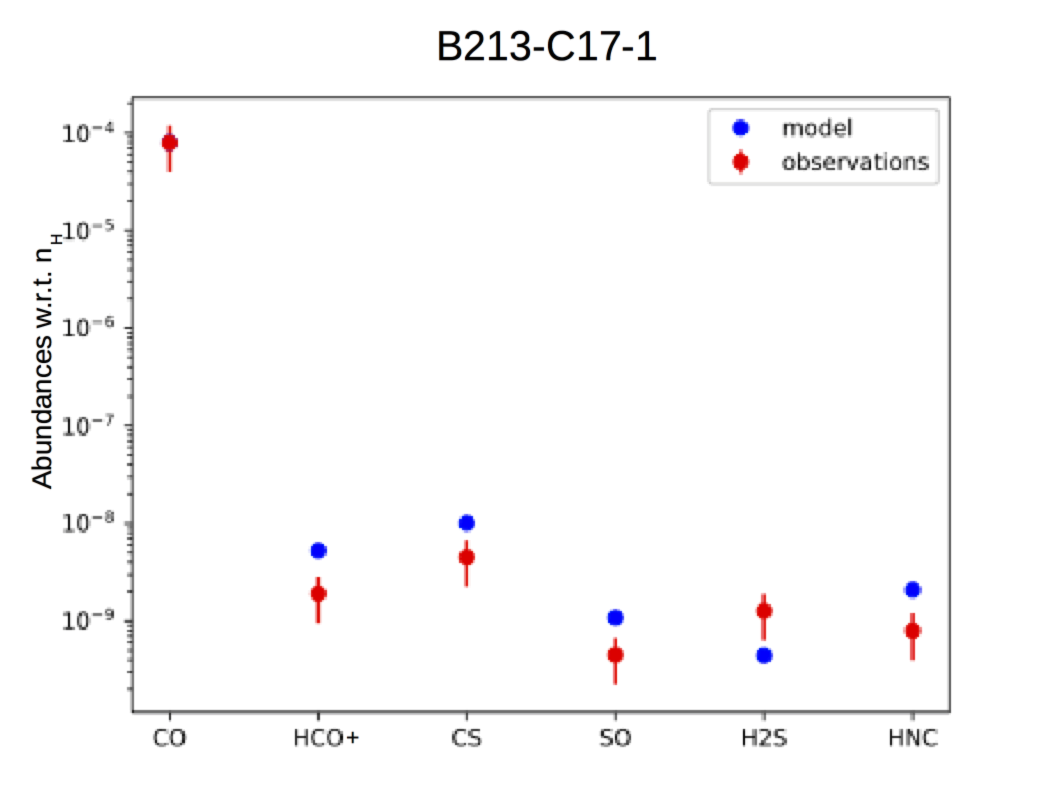}
\caption{Comparison between model predictions and observed abundances toward the position with the highest visual extinction in each of the observed cuts in Taurus.}
\label{comparison-1}
\end{figure*}

\begin{figure*}
\includegraphics[angle=0,scale=.2]{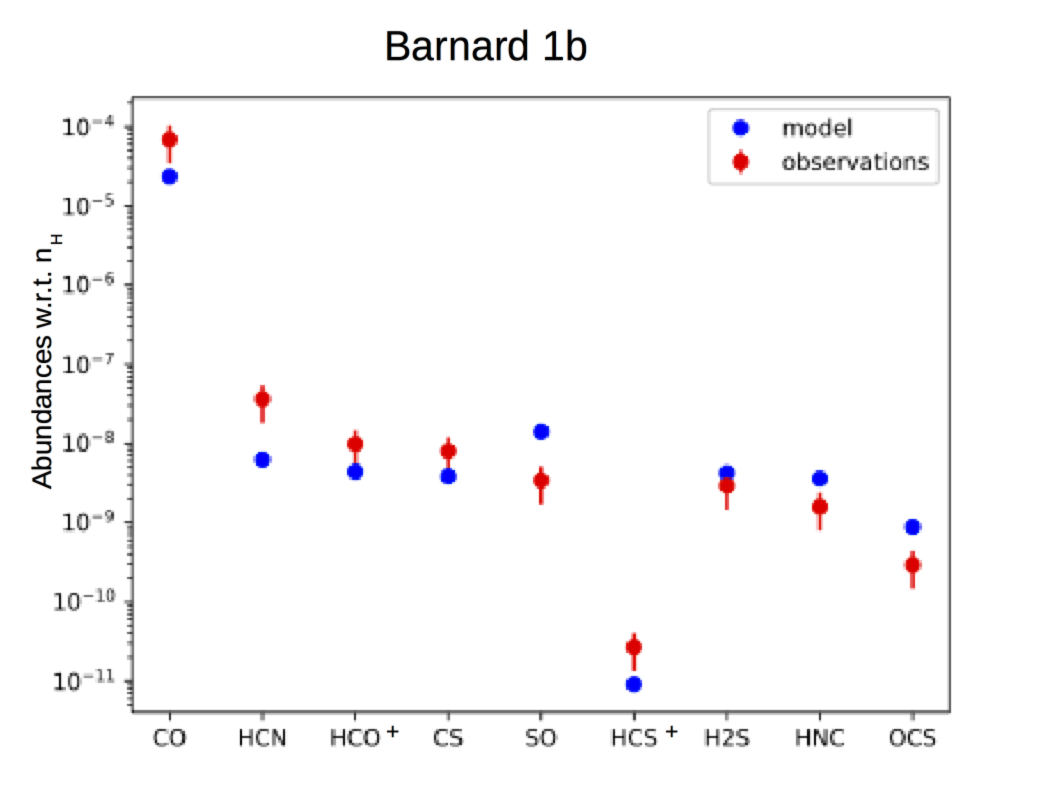}
\includegraphics[angle=0,scale=.2]{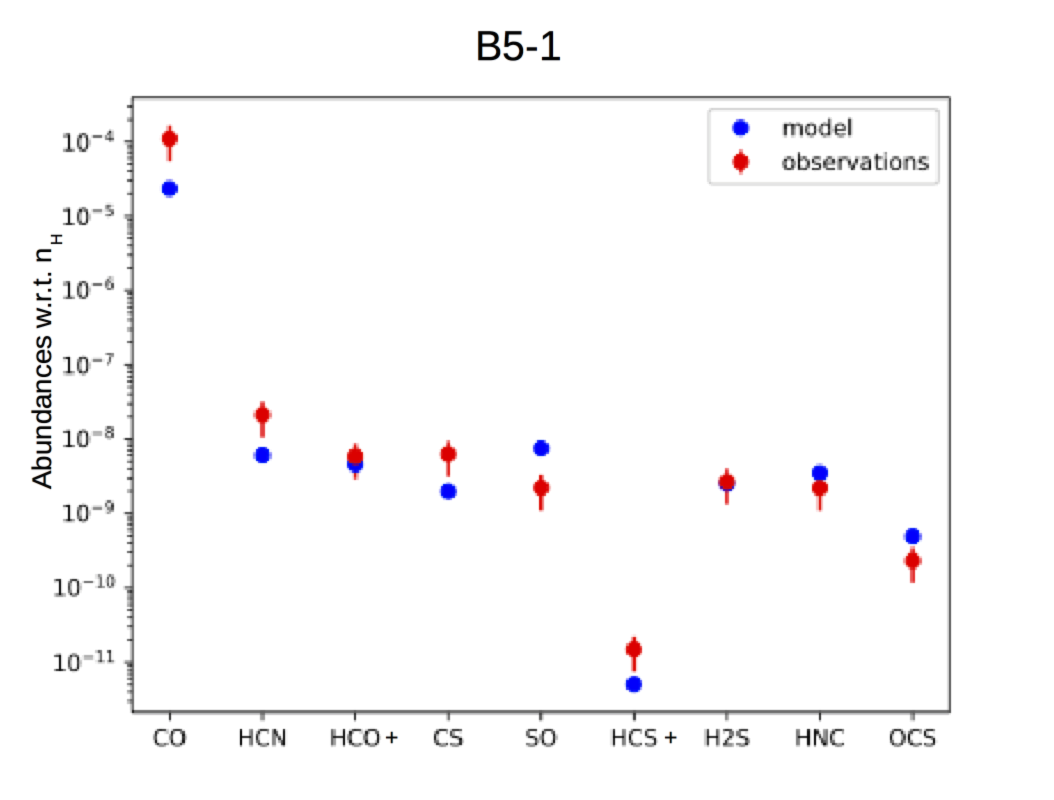}
\includegraphics[angle=0,scale=.2]{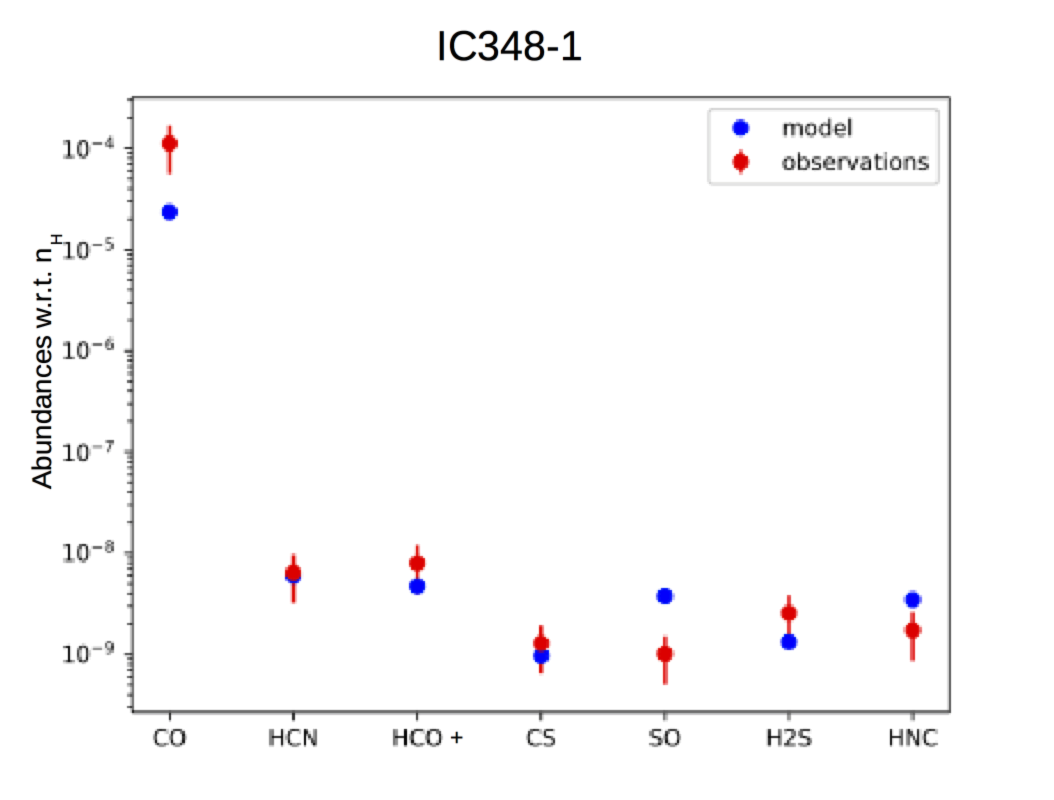}
\includegraphics[angle=0,scale=.2]{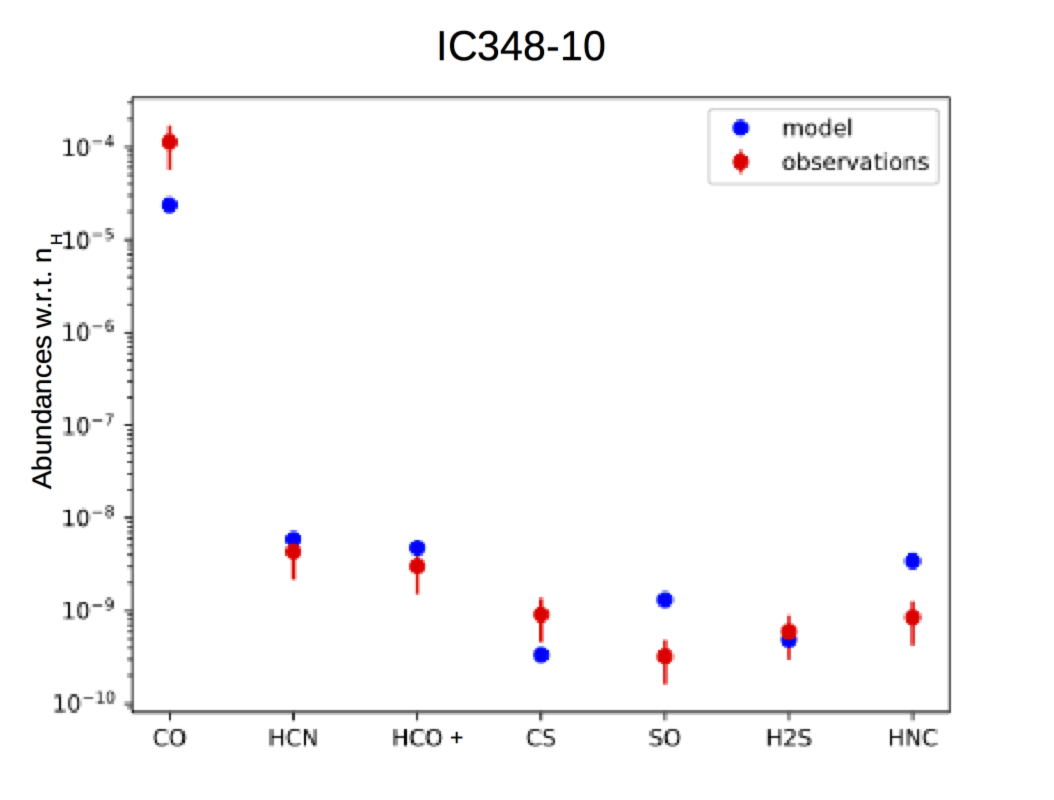}
\includegraphics[angle=0,scale=.2]{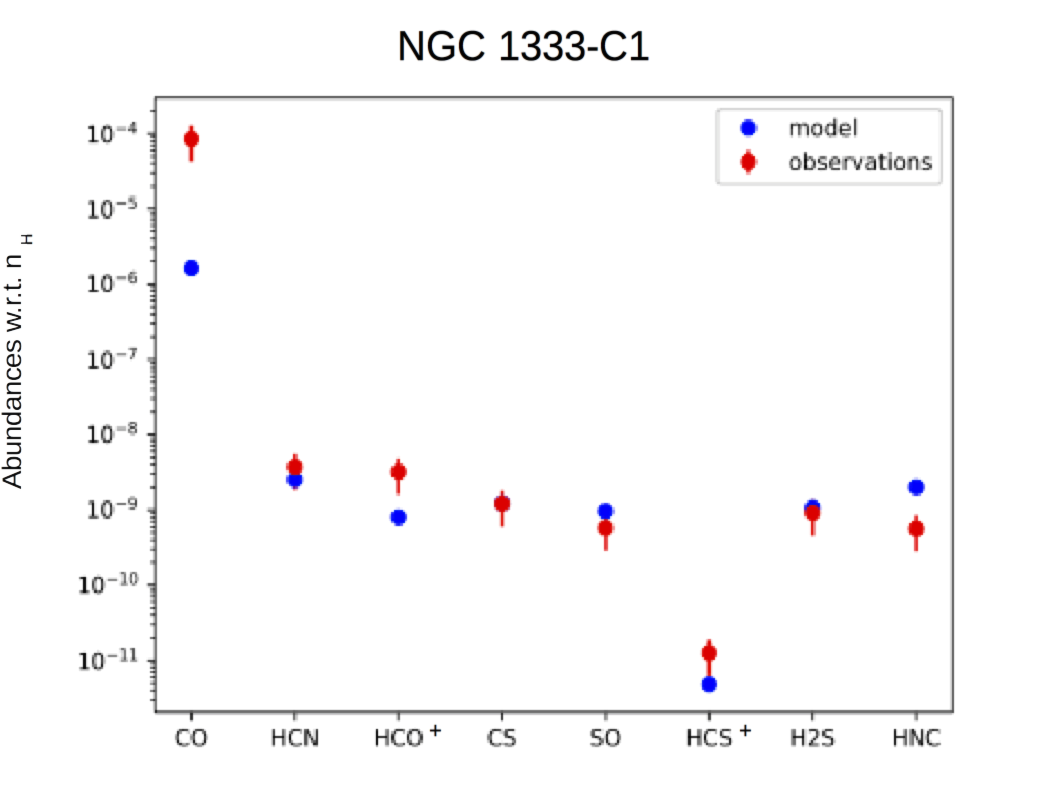}
\includegraphics[angle=0,scale=.2]{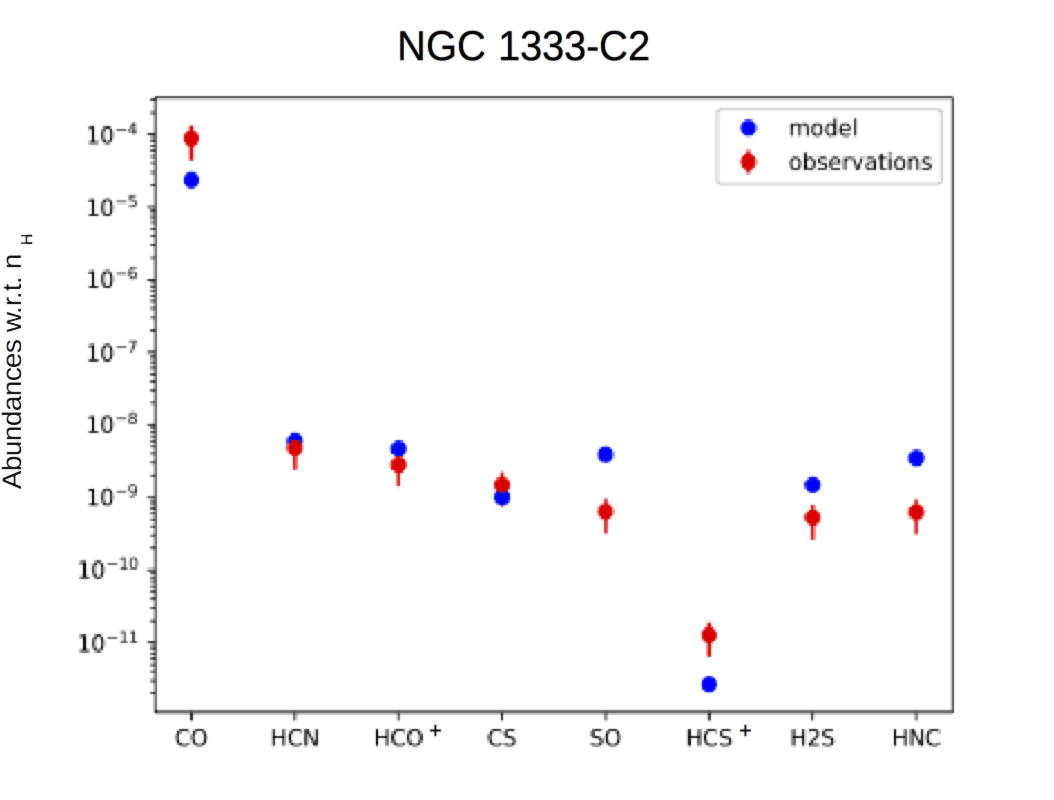}
\includegraphics[angle=0,scale=.2]{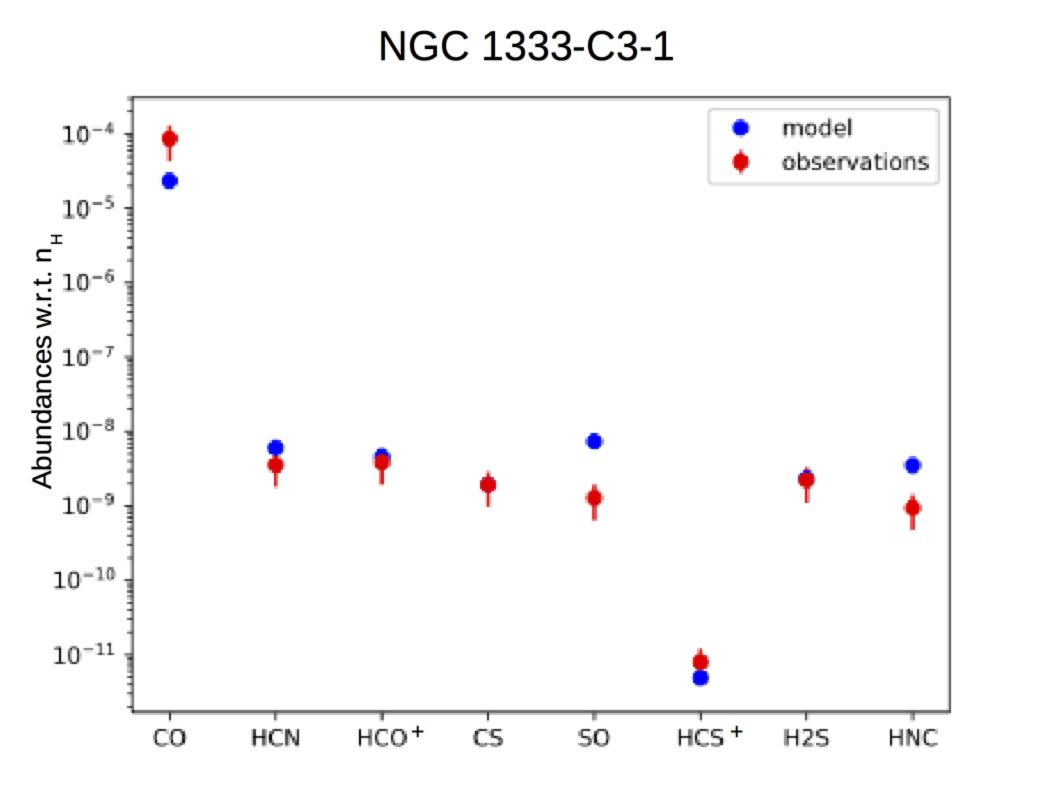}
\includegraphics[angle=0,scale=.2]{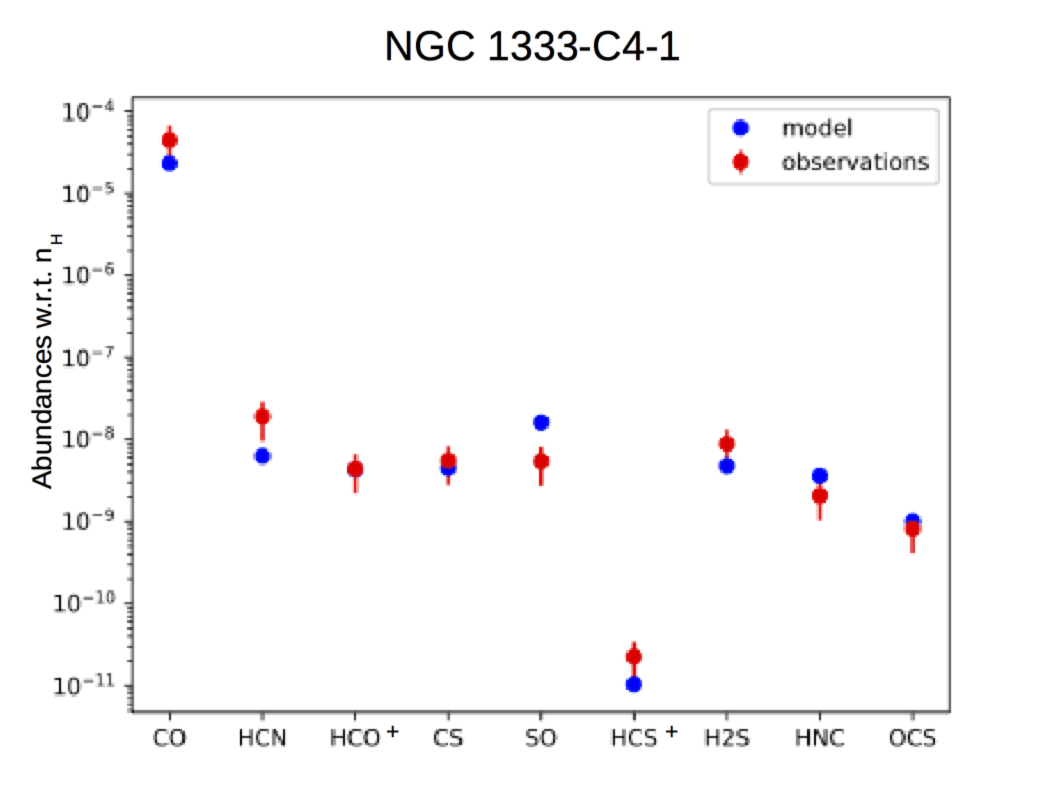}
\includegraphics[angle=0,scale=.2]{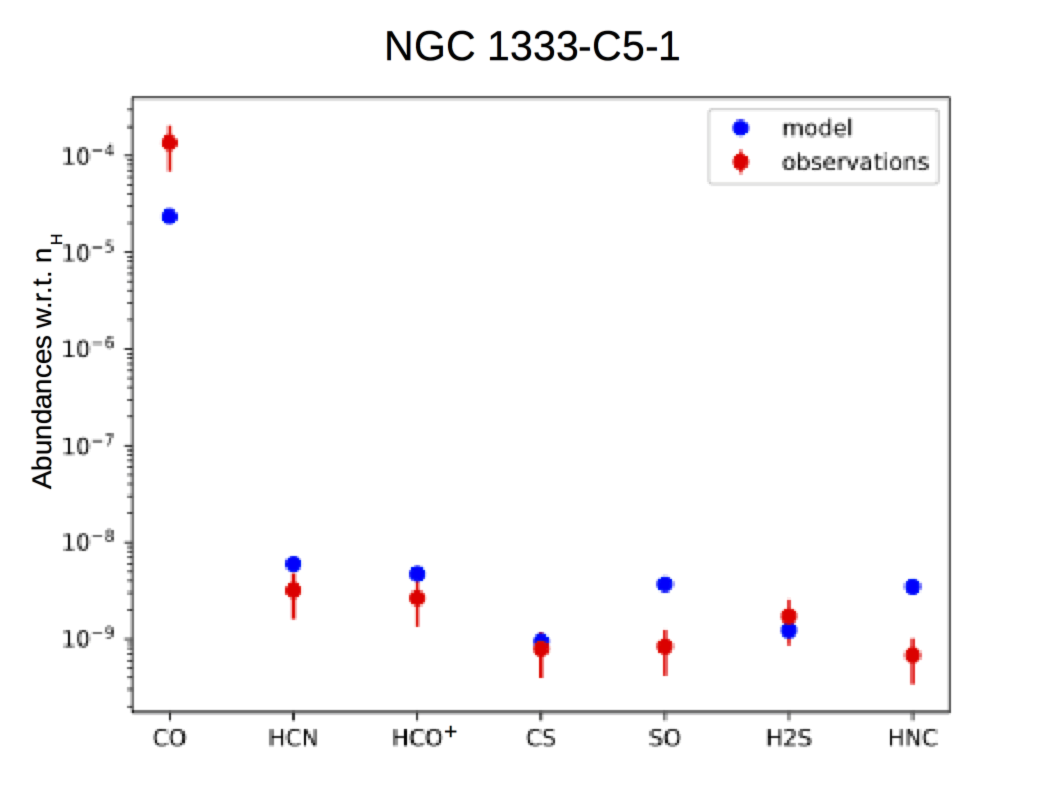}
\includegraphics[angle=0,scale=.2]{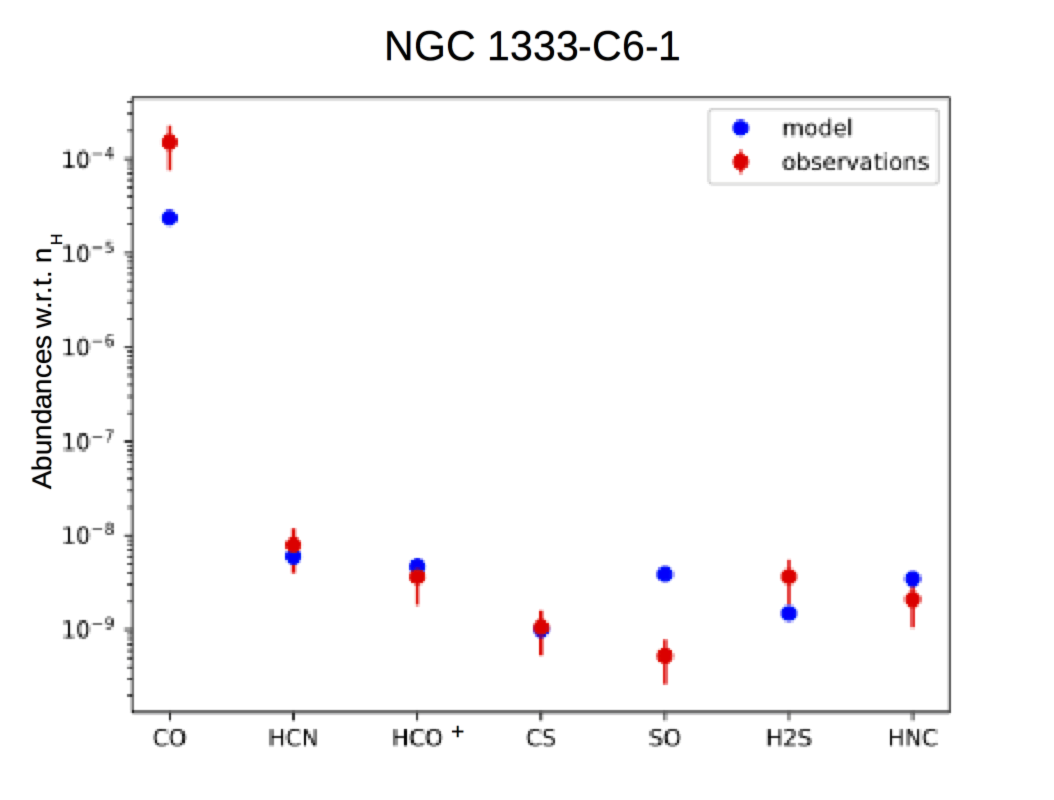}
\includegraphics[angle=0,scale=.2]{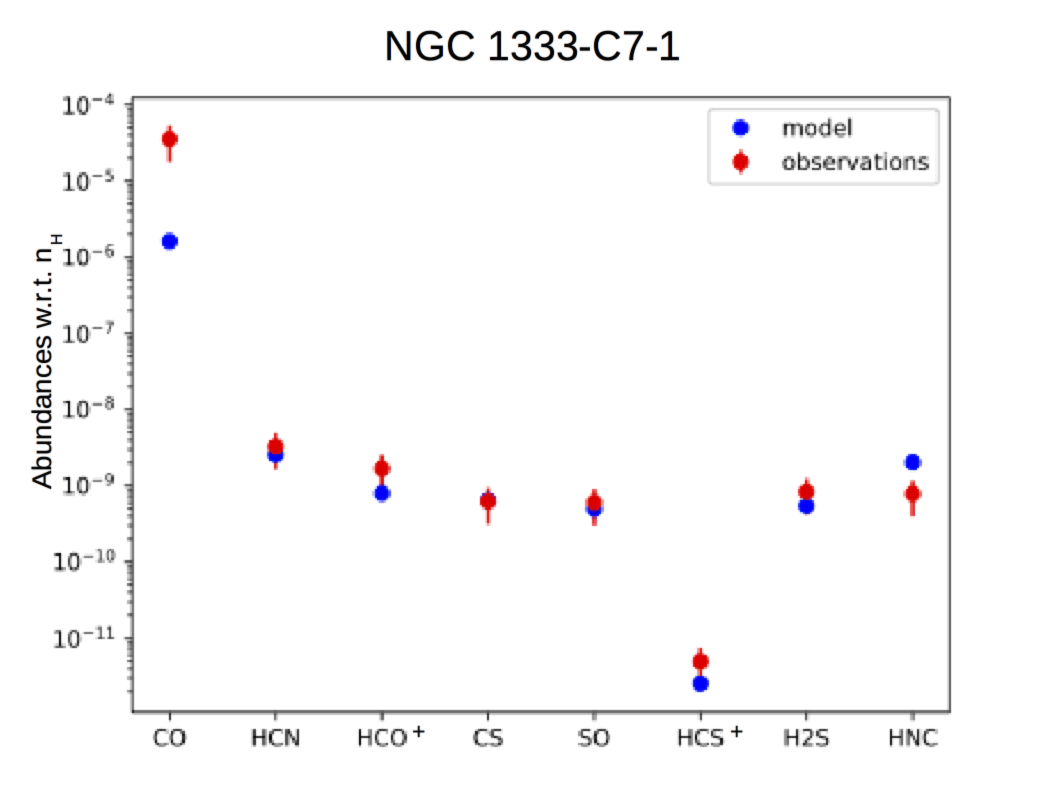}
\includegraphics[angle=0,scale=.2]{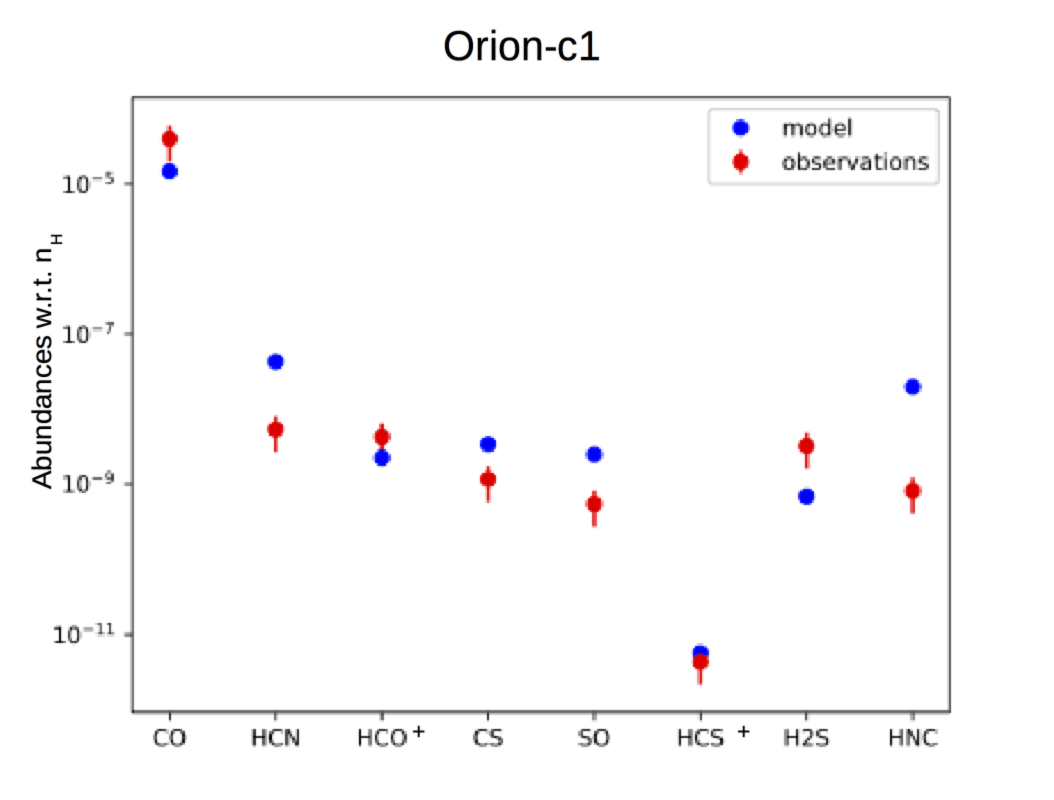}
\includegraphics[angle=0,scale=.2]{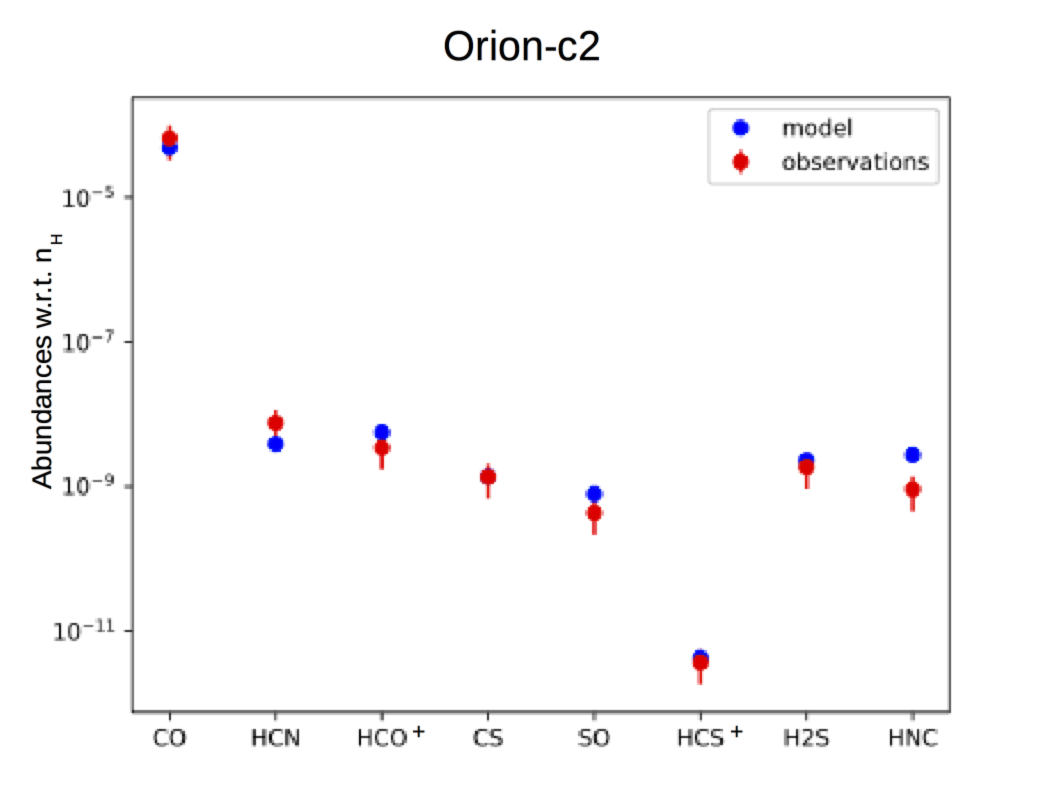}
\includegraphics[angle=0,scale=.2]{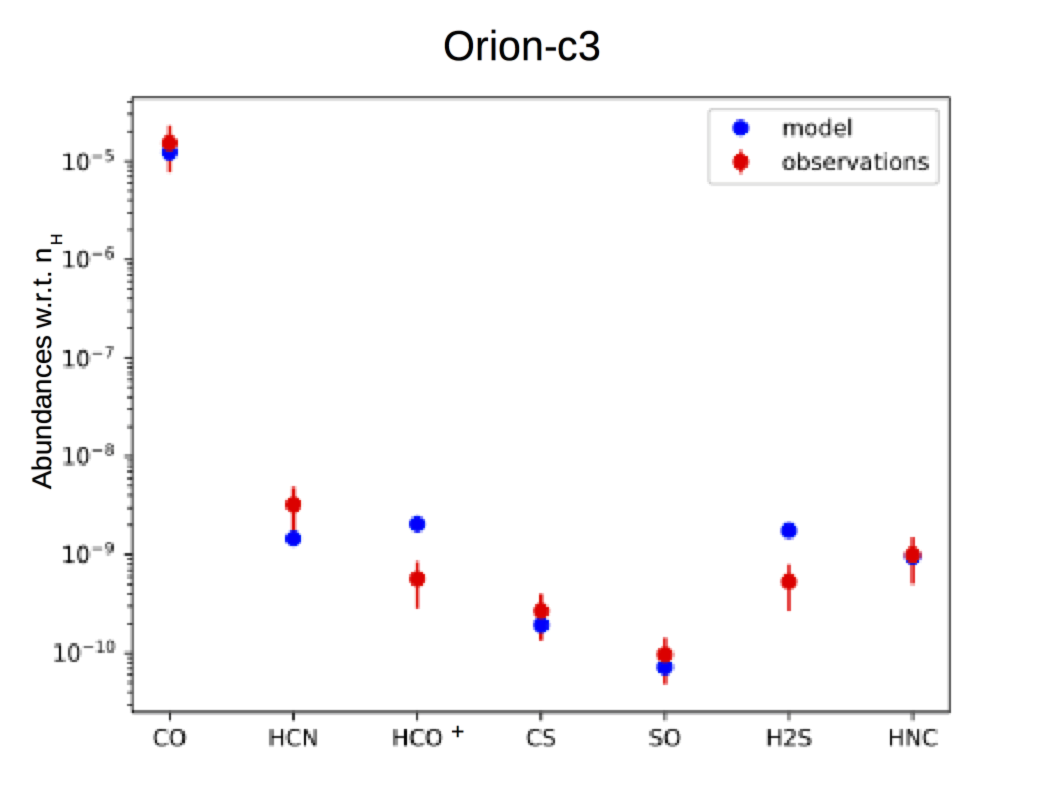}
\caption{Same as Fig~\ref{comparison-1}, but toward the position with the highest visual extinction in each of the observed cuts in Perseus and Orion.}
\label{comparison-2}
\end{figure*}

\newpage
\section{Computation of the grain charge distribution}\label{app:grain_charging}
For the computation of the grain charges we rely on the statistical equilibrium
between positive and negative charging. A detailed explanation on how to compute
the grain charge is out of the scope of this paper, but the recent work by 
\citet{Ibanez2019} can be consulted for more details; we have used this work
as a baseline for our implementation, and in this appendix we present
the main differences between our procedure and that presented in \citet{Ibanez2019}.\par

Positive charging of dust grains is mainly governed by collisions with the ions
in the plasma ($J_{\rm ion}$), and by radiative charging arising from
the interstellar radiation field ($J_{\rm pe}$) or from the cosmic-ray induced
H$_{2}$ fluorescence ($J_{\rm pe, CR}$, \citealp{Prasad1983}). On the other
hand, negative charging of dust grains mainly arise from collisions with
Maxwellian electrons in the plasma ($J_{\rm e}$), since the electron flux induced from
cosmic rays is negligible at these column densities \citep{Ivlev2015}.\par

The main singularity of our computation of grain charges is that we have 
tuned the input quantities to be representative of the two main molecular
complexes studied in this paper: Taurus and Orion. With that purpose, the 
particle number density of the charged species in gas phase has been taken
from the models built with Nautilus. We have considered two densities for each
cloud: $n_{\rm H} = 3.16 \times 10^{3}$ cm$^{-3}$ and 
$n_{\rm H} = 3.16 \times 10^{4}$ cm$^{-3}$ for Taurus, and 
$n_{\rm H} = 3.16 \times 10^{4}$ cm$^{-3}$ and 
$n_{\rm H} = 3.16 \times 10^{5}$ cm$^{-3}$ for Orion; at these densities
the main ionic species are H$^{+}$, He$^{+}$, H$_{2}^{+}$, H$_{3}^{+}$,
N$^{+}$, O$^{+}$, HCO$^{+}$, C$^{+}$, and S$^{+}$. In addition, there are
two parameters related to the cosmic-ray induced ultraviolet flux that we 
have set based on the particular traits of Orion and Taurus: the cosmic-ray ionization rate ($\zeta_{H_2}$) and the slope of the extinction curve at
visible wavelengths ($R_{V}$). These terms influence the final
far-ultraviolet photon flux based on the following relationship derived
by \citet{CecchiPestellini1992}:

\begin{equation}
F_{\rm UV} \simeq 960 \bigg( \frac{1}{1-\omega} \bigg) \bigg( \frac{\zeta}{10^{-17} {\rm s}^{-1}}\bigg)
\bigg( \frac{N_{\rm H_{2}}/A_{V} }{10^{21} {\rm cm}^{-2} {\rm mag}^{-1}}\bigg)
\bigg( \frac{R_{V}} {3.2}\bigg)^{1.5}
\end{equation}

For the dust albedo $\omega$ and the gas-to-extinction ratio $N_{\rm H_{2}}/A_{V}$
we have assumed the same values than \citet{Ibanez2019} (0.5 and 
$1.87\times 10^{21}$ cm$^{-2}$ mag$^{-1}$ respectively). However, based on the
best-fit of the chemical models we have assumed $\zeta_{H_2} = 10^{-16}$ s$^{-1}$
for Taurus, while for Orion we have explored two possible scenarios:
the low-$\zeta$ case suggested by our models ($\zeta_{H_2} = 10^{-17}$ s$^{-1}$)
and a high-$\zeta$ model ($\zeta_{H_2} = 10^{-16}$ s$^{-1}$) 
based on observational evidence of
some massive star-forming regions \citep{Aharonian2019, Padovani2019}.
Finally, we have relied on the values for $R_{V}$ computed by 
\citet{FM2007} based on the full extinction curve of hot stars toward these
complexes. For the observed cuts explored in this paper, we have found
that in Taurus the nearest star (search radius of 15')
with measurements by \citet{FM2007}
is HD 29647 with $R_{V} = 3.46$. For Orion, we find one star for each pointing
in a search radius between 10' and 15': HD 294264 ($R_{V} = 5.48$),
HD 36982 ($R_{V} = 5.60$) and HD 37061 ($R_{V} = 4.55$); as a compromise, we
have taken a characteristic value of $R_{V} = 5.5$ for the whole region, and
we have checked that there are not any appreciable differences between the 
chosen value and that reported by the extinction curves.\par 

The dust charge distribution is then computed by solving the equilibrium
relation:

\begin{equation}
f(Z) [J_{\rm pe}(Z) + J_{\rm pe, CR} + J_{\rm ion}(Z)] = f(Z+1)J_{\rm e}(Z+1)
\end{equation}

The probability distribution is computed between two extreme values $Z_{\rm min}$
and $Z_{\rm max}$ set by the user, for integer values of Z,
and the final curve is normalized such
that:

\begin{equation}
\sum_{Z = Z_{\rm min}}^{Z_{\rm max}} f(Z) = 1
\end{equation}

The python source code is available for download in the public
github repository \url{https://github.com/lbeitia/dust_charge_distribution}.

\end{document}